\documentclass[aip,jcp,numerical,reprint]{revtex4-1}
\usepackage{color}
\usepackage{graphicx}
\usepackage{threeparttable}
\usepackage{longtable}
\usepackage{enumerate}
\usepackage{amsfonts,amssymb}
\usepackage{amsmath}
\usepackage{stmaryrd}
\usepackage{dcolumn}
\usepackage{bm}
\usepackage{bbm}
\usepackage{algpseudocode}
\usepackage{algorithm}
\usepackage[all,cmtip]{xy}
\usepackage{url}

\bibpunct{}{}{,}{s}{}{\textsuperscript{,}}  

\newcommand{\spn}{\mathrm{span}}

\begin{document}

\title{Hilbert space renormalization for the many-electron problem}
\author{Zhendong Li}\email{zhendong@princeton.edu}
\author{Garnet Kin-Lic Chan}\email{gkchan@princeton.edu}
\bigskip
\affiliation{\footnotesize{Department of Chemistry, Frick Laboratory, Princeton University, Princeton, New Jersey 08544, USA}}
\date{\today}

\begin{abstract}
Renormalization is a powerful concept in the many-body problem.
Inspired by the highly successful density matrix renormalization group (DMRG) algorithm,
and the quantum chemical graphical representation of configuration space,
we introduce a new theoretical tool: \textit{Hilbert space renormalization}, to describe many-electron correlations.
While in DMRG, the many-body states in nested Fock subspaces are successively renormalized,
in Hilbert space renormalization, many-body states in nested Hilbert subspaces undergo renormalization.
This provides a new way to classify and combine configurations. The underlying wavefunction ansatz, namely the Hilbert space matrix product state (HS-MPS),
has a very rich and flexible mathematical structure.
It provides low-rank tensor approximations to any configuration interaction (CI) space through restricting either the 'physical indices' or the coupling rules
in the HS-MPS. Alternatively, simply truncating the 'virtual dimension' of the HS-MPS leads  to a family of size-extensive wave function ans\"{a}tze that can be used efficiently in variational calculations.
We make formal and numerical comparisons between the HS-MPS, the traditional Fock-space MPS used in DMRG, and traditional CI approximations.
The analysis and results shed light on fundamental aspects of the efficient representation of many-electron wavefunctions through the renormalization of many-body states.
\end{abstract}
\maketitle

\section{Introduction}

Accurate and efficient solutions of the many-electron problem are central
to all aspects of chemistry and physics\cite{Dirac1929}.
After many decades of investigation\cite{helgaker-molecular-2001},
it has been common to distinguish two kinds of
many-electron correlation: static or strong correlation,
associated with the qualitative physics of the
 valence orbitals; and dynamic correlation, arising from
Coulomb interactions at short range, and associated with a
very large number of high-energy degrees of freedom.


Configuration interaction (CI)\cite{shavitt1998history}, many-body perturbation theory, and
coupled cluster theory\cite{shavitt2009many}, as well as explicit correlation techniques\cite{ten2012explicitly,kong2011explicitly,hattig2011explicitly},
have traditionally presented
highly successful approaches to describe dynamic correlation.
Recent advances in
tensor network states (TNS)\cite{chan2012low,orus2014practical,Legeza2015}
from condensed matter physics and quantum information theory, have further led to efficient representations of strong correlation also.
The most widely used example of this latter class is the
density matrix renormalization group (DMRG) method
\cite{white_density_1992,white_density-matrix_1993,schollwock_density-matrix_2005,schollwock_density-matrix_2011},
which employs the simple one-dimensional TNS, viz., the matrix product state (MPS)\cite{MPS1995}.
DMRG has been successfully applied in quantum chemistry to
compute near-exact many-electron
wavefunctions of several systems with a very large number of
valence quantum degrees of freedom
\cite{white_ab_1999,chan_highly_2002,chan_algorithm_2004,chan_density_2011,
Reiher2011,Sebastian2014,kurashige_multireference_2014,Legeza2015,
olivares-amaya_ab-initio_2015}, such as the oxygen-evolving
complex (Mn$_4$CaO$_5$)\cite{kurashige2013entangled} and
the iron-sulfur clusters\cite{sharma_low-energy_2014}.

Successfully combining these different representations for static and
dynamic correlation is a nontrivial problem, and is an important
research challenge\cite{neuscamman_review_2010,sharma_communication_2014,
saitow_fully_2015,sharma_multireference_2015,Yanai2015}.
For example, TNS methods and
dynamic correlation methods rely on very different physical pictures. TNS approximations parametrize the entanglement between groups of orbitals (referred to as sites), and the
variational objects are states in the local Fock subspaces of
the orbitals, which span the complete set of particle numbers in each subspace.
In contrast, traditional  dynamic correlation methods
are expressed in terms of particle-hole excitations relative to a given $N$-electron
reference. The variational objects or amplitudes are then
of fixed excitation rank, and act on a given number of particles and holes at a time.
Bridging  these two very different parametrizations
is 
a challenge in developing new theories for the electron correlation problem.

To this end, we consider here the possibility to formulate a new kind of TNS. Unlike existing TNS, it will be expressed in a set of nested Hilbert, as opposed to Fock, subspaces.
  The resulting wavefunction is an $N$-electron wavefunction
  directly in Hilbert space, rather than in the Fock-space (occupation number) representation.
In essence, this means that the 'sites' in the TNS will refer to electrons rather than orbitals, and the $N$-electron
wavefunction will be expressed by an interconnected network of $N$ such sites.
At first glance, this does not seem natural,
as fermion antisymmetry precludes a simple factorization of $N$-electron wavefunctions
into a product of single-electron wavefunctions. Indeed, one might
suspect that imposing antisymmetry
could mean that the cost of working with a Hilbert space TNS will be factorial
with respect to the number of electrons $N$. As an example, the antisymmetrized
product of geminals (APG)\cite{Hurley1953,Parr1956,McWeeny1959} wavefunction
can be considered to be a simple Hilbert space TNS where electrons (sites) within pairs are connected by pair expansion coefficients. However, even such a formally simple ansatz is computationally intractable, unless
strong orthogonality constraints are imposed between the geminals. Indeed,
fermion antisymmetry is an important reason why it is easier to formulate TNS
in Fock space, where the antisymmetry
is handled by the anticommuting operator algebra, leaving a simple direct product structure in the state space.

In this work, we will show, however, that the above difficulty of antisymmetry can be circumvented by introducing 'prefix' and 'suffix' constraints for renormalizations in Hilbert space.
This enables a simple composition of two renormalized states, and
more importantly the efficient computation of
operator matrix elements, such as for the overlap or Hamiltonian.
This allows practical algorithms, such as the variational optimization of
wavefunction parameters, to be formulated and applied.
The layout of the remainder of the paper is as follows: The basic concepts
of Fock-space MPS (FS-MPS) are recapitulated in Sec. \ref{sec:FS-MPS},
where the important connection with the graphical
representation of configuration space\cite{Duch1985} is also made.
An alternative view of the same graph naturally
leads to the Hilbert space MPS (HS-MPS) and the necessary
prefix/suffix constraints for renormalizations in Hilbert space
to make the HS-MPS computationally tractable.
The detailed mathematical formulation of the HS-MPS is presented in Sec. \ref{sec:HS-MPS},
along with formal comparisons with FS-MPS and CI (Sec. \ref{sec:formulation}-\ref{sec:restriction}).
We further describe the DMRG-like algorithm for the variational optimization of HS-MPS (Sec. \ref{sec:matrixElements}-
\ref{sec:varopt}), the generalizations to bosons and spin-adaptation of the HS-MPS (Sec. \ref{sec:generalization}),
and connections with other theories (Sec. \ref{sec:connection})
  including
the factorization of multivariate polynomials,
artificial neural networks (ANNs),
and the graphically contracted function (GCF) method of
Shepard et al.\cite{shepard2005general,shepard2006hamiltonian,shepard2014multifacet1,shepard2014multifacet2,gidofalvi2014wave}.
Numerical studies for several typical chemical systems are presented in
Sec. \ref{sec:results}, aiming at a  numerical comparison of HS-MPS and FS-MPS.
Conclusions are given in Sec. \ref{sec:conclusion}.

\section{FS-MPS and its graphical representation}\label{sec:FS-MPS}
Any $N$-electron wavefunction $|\Psi\rangle$ can
be written in the Fock-space (occupation-number) representation\cite{helgaker-molecular-2001} as,
\begin{eqnarray}
|\Psi\rangle=\sum_{n_1n_2\cdots n_K}\Psi^{n_1n_2\cdots n_K}|n_1n_2\cdots n_K\rangle,\label{FSrep}
\end{eqnarray}
where $n_k\in\{0,1\}$ is the occupation number for the $k$-th spin-orbital
and $K$ is the total number of spin-orbitals. The coefficients $\Psi^{n_1n_2\cdots n_K}\in\mathbb{C}^{2^K}$ form a (complex) tensor
of dimension $2^K$, which has the structure
\begin{eqnarray}
\Psi^{n_1n_2\cdots n_K}=
\left\{\begin{array}{ll}
\Psi^{n_1n_2\cdots n_K},& \sum_{k=1}^K n_k=N,\\
0,& \mathrm{otherwise}.
\end{array}\right.\label{FStensor}
\end{eqnarray}
in order to be a well-defined $N$-electron wavefunction.
In the Fock-space representation, the system with $K$ orbitals can be viewed as a spin lattice
with $K$ distinguishable sites.
The FS-MPS (with open-boundary conditions) can be expressed as a chain of tensor products
\begin{eqnarray}
\Psi^{n_1n_2\cdots n_K}
=\sum_{\{\alpha_k\}} A^{n_1}_{\alpha_1}[1]A^{n_2}_{\alpha_1\alpha_2}[2]\cdots A^{n_K}_{\alpha_{K-1}}[K],\label{FSMPS}
\end{eqnarray}
where $A^{n_k}_{\alpha_{k-1}\alpha_k}[k]$ ($1<k<K$) is a rank-3 tensor, and the tensors at
the boundary ($k=1$ and $k=K$) are rank-2 tensors (i.e. matrices).
For simplicity, we will use $A[k]$ to denote both kinds of tensors in the
following discussion.
In MPS terminology, the occupation number $n_k$ is usually referred as the 'physical index', while $\alpha_k$ is
referred as the 'virtual index'. The virtual index can be viewed
to arise from successive singular value decompositions (SVD) of the Fock-space tensor $\Psi^{n_1n_2\cdots n_K}$, i.e.,
\begin{eqnarray}
\Psi^{n_1n_2\cdots n_K}&=&\sum_{\alpha_1}U^{n_1}_{\alpha_1}
s_{\alpha_1}V_{\alpha_1}^{n_2n_3\cdots n_K*}
\triangleq \sum_{\alpha_1}U^{n_1}_{\alpha_1}W_{\alpha_1}^{n_2n_3\cdots n_K}\nonumber\\
&=&\sum_{\alpha_1\alpha_2}U^{n_1}_{\alpha_1}U^{n_2}_{\alpha_1\alpha_2}
W^{n_3\cdots n_K}_{\alpha_2}=\cdots.\label{SVD}
\end{eqnarray}
The dimension of $\alpha_k$ denoted by $D_k$ is called 'bond dimension'
for the $k$-th 'bond' between sites $k$ and $k+1$. The FS-MPS representation \eqref{FSMPS}
can be \emph{exact} as long as $D_k$ is sufficiently large. Given $K$, $N$, and a maximal level of CI excitations,
the minimal values of $D_k$ ($1\le k\le K-1$) for the exact representation
of an arbitrary
wavefunction in the specified configuration space can be determined, see Sec. \ref{sec:dims}.
For convenience, we refer these minimal values as the 'theoretical' bond dimensions.
The minimal theoretical bond dimensions along the MPS chain
increase exponentially for FCI, as $N$ and $K$ increase.
The success of TNS lies in the fact
that physically relevant states live
only in a corner of the full configuration space, where the entanglement
of the states is generally limited by the so-called 'area law'\cite{hastings2007area,eisert2010colloquium}.
This implies that if the wavefunction on the right hand side of Eq. \eqref{FSMPS}
with a finite bond dimension $D$ is taken as an approximate variational ansatz, then
to represent physically relevant states
$D$ needs only to grow as polynomial function of $K$ and $N$ to achieve good accuracy.
Notably, for ground states of one-dimensional gapped systems with Hamiltonians
containing only local interactions, constant $D$ is sufficient to
obtain accurate results for very large $N$ and $K$. This is the underlying
reason why the DMRG algorithm works extremely well in one-dimensional systems.

The MPS representation \eqref{FSMPS} has a rich mathematical structure.
One of the most significant, is that it naturally encodes a
recursive chain of renormalization transformations.
To illustrate this, the wavefunction \eqref{FSMPS} can be
re-expressed in terms of the following renormalized intermediate states,
\begin{eqnarray}
|\alpha_1\rangle&=&\sum_{n_1}|n_1\rangle A^{n_1}_{\alpha_1}[1],\nonumber\\
|\alpha_2\rangle&=&\sum_{\alpha_1n_2}|\alpha_1n_2\rangle A^{n_2}_{\alpha_1\alpha_2}[2],\nonumber\\
&&\cdots\nonumber\\
|\alpha_{K-1}\rangle&=&\sum_{\alpha_{K-2}n_{K-1}}|\alpha_{K-2}n_{K-1}\rangle A^{n_{K-1}}_{\alpha_{K-2}\alpha_{K-1}}[K-1],\nonumber\\
|\alpha_{K}\rangle&=&\sum_{\alpha_{K-1}n_{K}}|\alpha_{K-1}n_{K}\rangle A^{n_{K}}_{\alpha_{K-1}}[K]\equiv|\Psi\rangle.\label{FSChain}
\end{eqnarray}
It is now evident that the set of intermediate states $\{|\alpha_k\rangle\}$ are many-body states in the Fock
space $\mathcal{F}_k$ defined by the
direct product space of the first $k$ orbitals.
From Eq. \eqref{FSChain}, the tensor $A[k]$ can be recognized
as a linear map from the space spanned by renormalized states
to the direct product space, e.g.,
\begin{eqnarray}
A[k]: \spn\{|\alpha_k\rangle\}\mapsto
\spn\{|\alpha_{k-1}\rangle\}\otimes\spn\{|n_k\rangle\}.\label{Ak}
\end{eqnarray}
From this perspective, this mapping is a many-body analog of the contraction of basis
functions at the one-particle level\cite{DunningChapter}, which maps a large
underlying set of primitives to a smaller set of contracted functions.
Without  loss of generality, the set of renormalized states in Eq. \eqref{FSChain} can be made orthonormal.
This orthonormal set will be denoted by $\{|l_k\rangle\}$ and the corresponding MPS
is then usually referred to as being in 'left canonical form'\cite{chan_density_2011,schollwock_density-matrix_2011}.
Likewise, the renormalized intermediate states can be defined in 'right canonical form' denoted by $\{|r_k\rangle\}$ ,
where the renormalization process starts from the last site and proceeds to the first site.
The FS-MPS can also be expressed in a basis of mixed forms, e.g.,
\begin{eqnarray}
|\Psi\rangle&=&\sum_{l_{k-1}n_k r_k}|l_{k-1}n_k r_{k}\rangle A^{n_k}_{l_{k-1}r_k}[k],\nonumber\\
|l_{k-1}\rangle&=&
\sum_{n_1\cdots n_{k-1}}
(A^{n_1}[1]\cdots A^{n_{k-1}}[k-1])_{l_{k-1}}|n_1\cdots n_{k-1}\rangle\nonumber\\
&=&
\sum_{l_{k-2}n_{k-1}}|l_{k-2}n_{k-1}\rangle
A^{n_{k-1}}_{l_{k-2}l_{k-1}}[k-1],\nonumber\\
|r_{k}\rangle&=&
\sum_{n_{k+1}\cdots n_{K}}
(A^{n_{k+1}}[k+1]\cdots A^{n_{K}}[K])_{r_k}|n_{k+1}\cdots n_{K}\rangle\nonumber\\
&=&
\sum_{n_{k+1}r_{k+1}}
A^{n_{k+1}}_{r_{k}r_{k+1}}[k+1]|n_{k+1}r_{k+1}\rangle.\label{Mixed}
\end{eqnarray}
Such a mixed canonical form is particularly useful in numerical optimizations of MPS,
since the basis $\{|l_{k-1}n_k r_{k}\rangle\}$ is orthonormal for each $k$.

\begin{figure}
\begin{tabular}{c}
{\resizebox{!}{0.18\textheight}{\includegraphics{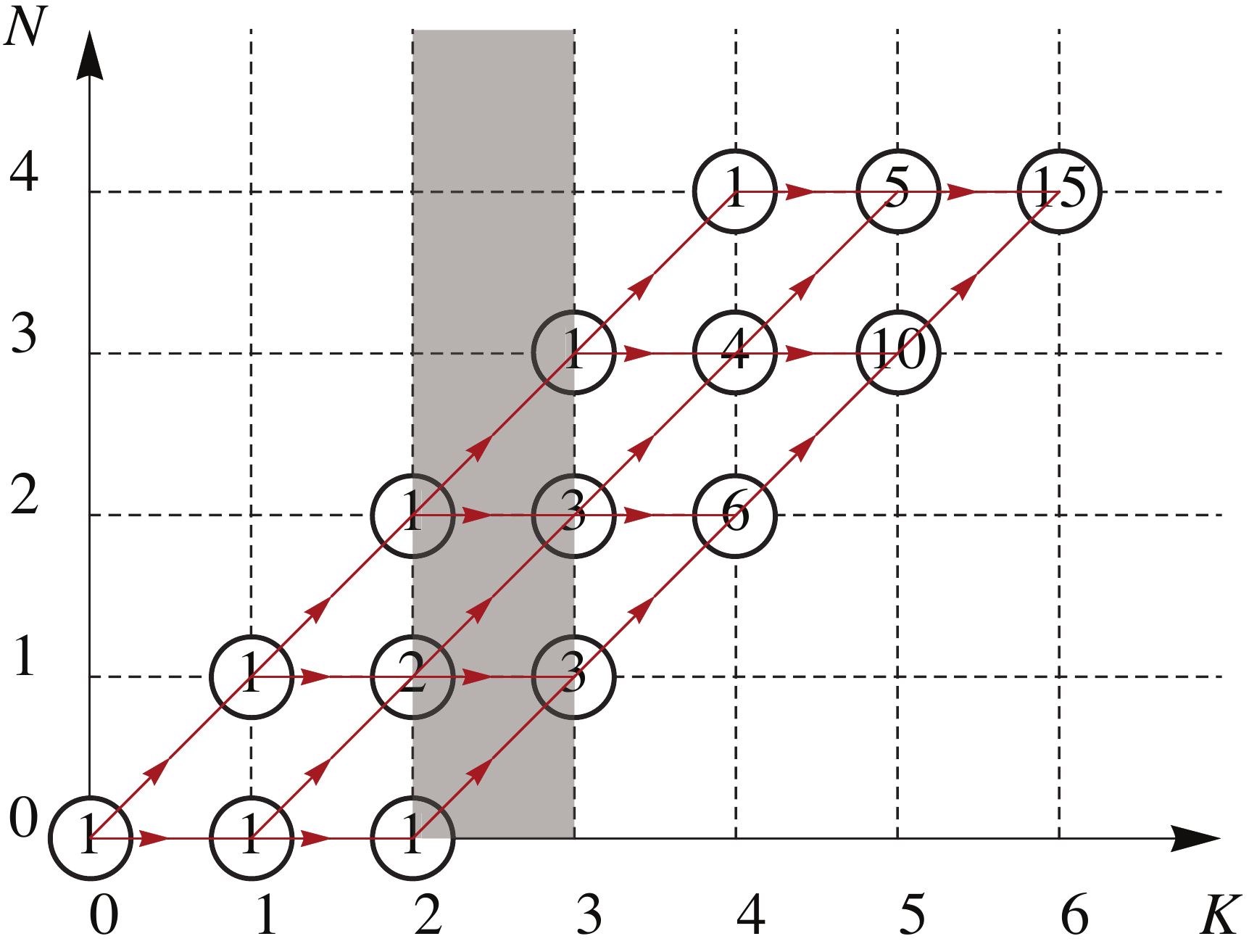}}}\\
(a) \\
{\resizebox{!}{0.25\textheight}{\includegraphics{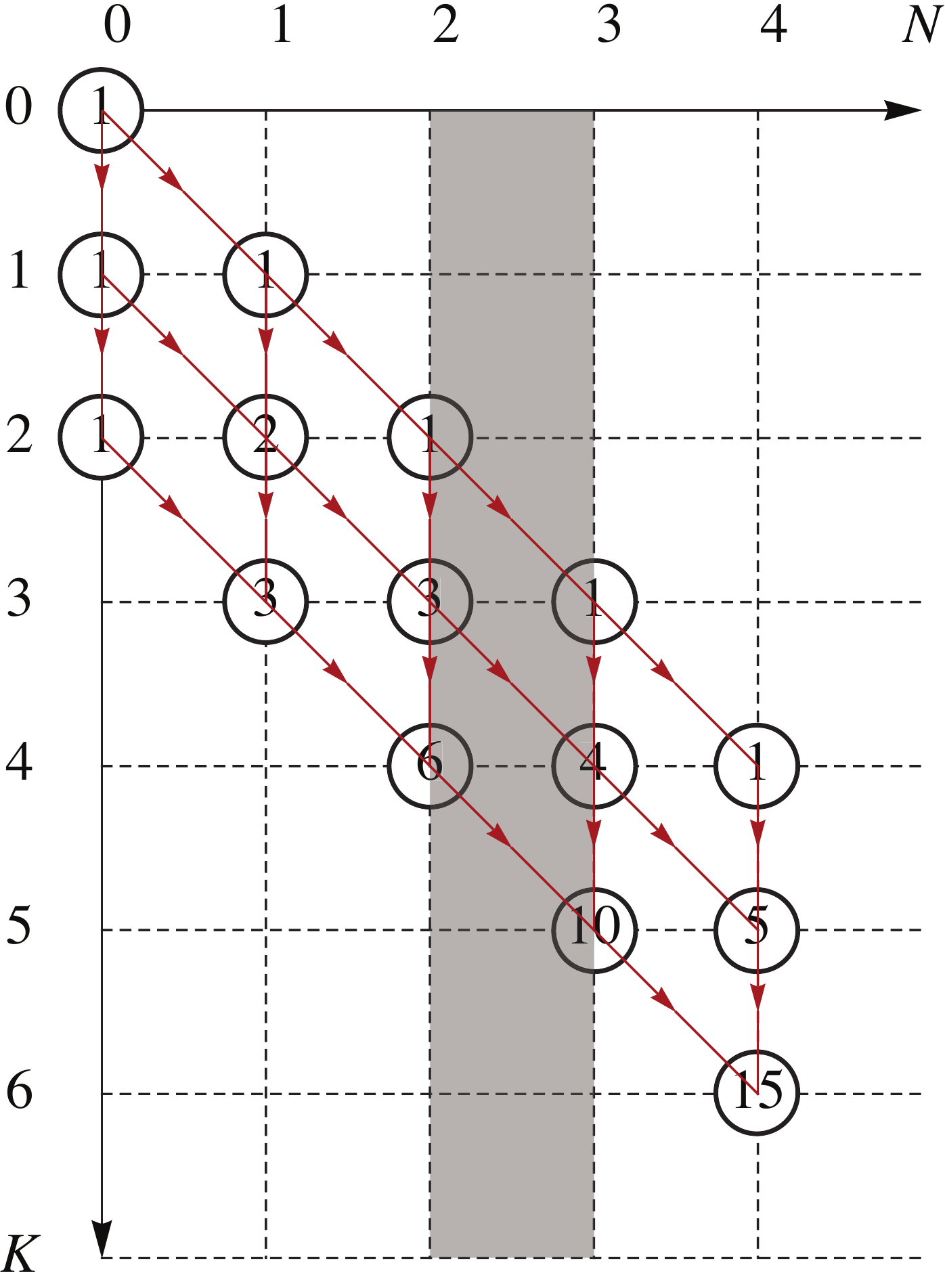}}}\\
(b) \\
{\resizebox{!}{0.25\textheight}{\includegraphics{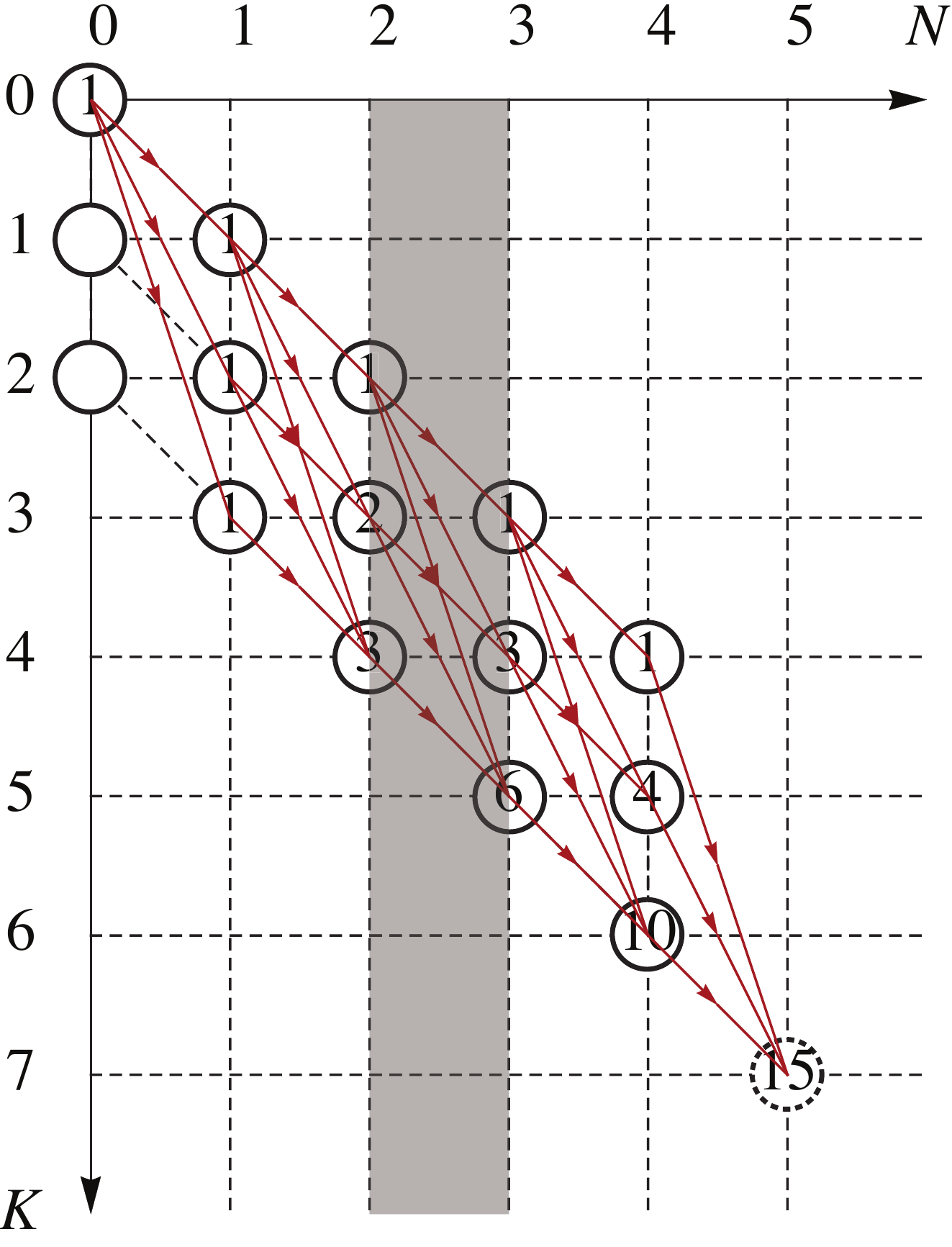}}}\\
(c) \\
\end{tabular}
\caption{Graphical representation of configuration space
and renormalizations: (a) Configuration graph for $(K,N)=(6,4)$;
(b) Rotated from (a) by 90 degrees; (c) Readjusted configuration graph obtained
by replacing all the 
downwards
arrows in (b).
The shaded regions correspond to the tensor in Eq. \eqref{Ak}.}\label{fig:rotate}
\end{figure}

For FS-MPS with  particle number symmetry, by which we mean that the intermediate
states $\{|\alpha_k\rangle\}$ are required to be eigenfunctions of particle number operators
$\mathcal{N}_k\triangleq\sum_{l=1}^k a_l^\dagger a_l$,
the above recursive structure of FS-MPS can be visualized with the
help of the graphical representation of the configuration space of determinants\cite{Duch1985}. The similar graphical representation
in terms of configuration state functions rather than determinants, viz., the Shavitt graph in the Graphical Unitary Group Approach (GUGA)\cite{paldus1974group,shavitt1977,shavitt1978matrix} for CI, has also been used to motivate
  the construction of the graphical contracted function by Shepard and coworkers~\cite{shepard2014multifacet1}, a kind of FS-MPS as discussed in Sec.~\ref{sec:connection}, although without the renormalization
  interpretation taken here. Figure \ref{fig:rotate}(a) illustrates the configuration graph for $(K,N)=(6,4)$.
Such graph is usually employed in CI algorithms, where any Slater determinant can be
assigned a unique address based on the path from the origin $(0,0)$ to the destination $(K,N)$
on the graph. We note that the construction of such a graph is very much the same as
the recursive construction in Eq. \eqref{FSChain}.
For instance, starting from the origin, which is the vacuum (with no orbitals
and no electron) the first orbital can be added to create
new states. Depending on whether it is occupied ($n_1=1$) or unoccupied ($n_1=0$),
either a new state, shown in the circle $(K,N)=(1,1)$, can be created,
or the state remains a vacuum state $(K,N)=(1,0)$ in the Fock space $\mathcal{F}_1$.
Thus, the two circles in the layer with $K=1$ comprise the Fock space $\mathcal{F}_1$.
The stepwise construction can proceed until the last orbital $K$ is reached, which creates the whole
Fock space. Therefore, the circle in the configuration graph with coordinate
$(K,N)$ represents the Hilbert space $\mathcal{H}_N$, and the value in the circle shows
the dimension of this space, given by the binomial coefficient $C_{K}^N$. The set of
all circles in the same layer $K$ comprise the Fock space $\mathcal{F}_K=\bigoplus_{N=0}^{2^K}\mathcal{H}_N$.
If only the states in the sector $(K,N)$ are of interest, then there is a
parallelogram that restricts the possible intermediate states, see Figure \ref{fig:rotate}(a).
States outside of this parallelogram are irrelevant to the study of
the target sector. From the same kind of recursive construction, one can immediately
recognize that $A[k]$ \eqref{Ak} is the mapping from states in one
layer to states in another layer, and its information is all contained
in the shaded region between two layers if no truncation is made, e.g., see Figure \ref{fig:rotate}(a) for $k=3$.
Whenever there is a truncation in the renormalization process, the number of
states in the corresponding circle that can enter the next layer gets reduced, for example
so as to avoid the exponential growth in the number of configurations.
The ability to constrain the growth of configuration space is the very reason why renormalization is so powerful
when combined with the DMRG algorithm to variationally optimize the
intermediate states in Eq. \eqref{FSChain}, or equivalently
the contraction coefficients in Eq. \eqref{Ak}.

\section{Hilbert space MPS}\label{sec:HS-MPS}


\subsection{Formulation}\label{sec:formulation}
Having established the connection between FS-MPS and the
graphical representation of configuration space, one may naturally wonder
what is obtained if the renormalization is performed
along the axis of particle numbers instead of orbitals.
For this purpose, we can obtain a different view of Figure \ref{fig:rotate}(a)  by rotating it to Figure \ref{fig:rotate}(b). (This is actually the more usual way
to draw the configuration graph\cite{Duch1985}). We can then apply the same
interpretation to the shaded region as a renormalization process.

It is easily identified that states in the same layer
are now within the same $N$-electron Hilbert space.
However, there is a crucial difference
between Figures 1(a) and 1(b): While in the former case
circles in the same layer represent {\it different} Hilbert subspaces of a Fock space, in the latter case, downwards arrows between circles in the same layer indicate
that the space of the circle above is a {\it subspace} of the space of the circle below, e.g., $\mathcal{H}_{n}^{k=n}\subset
\mathcal{H}_{n}^{k=n+1}\subset\cdots\subset
\mathcal{H}_{n}^{k=n+K-N}$ for the circles in the $n$-th layer with different
$k$ values. This makes the interpretation of Figure \ref{fig:rotate}(b)
as a renormalization flow less transparent.
To avoid such difficulties, we  write a slightly different graph, where the spaces associated with circles in the same layer are once again distinct,
and obtained by taking the complementary part of the space
to the circles above in Figure 1(b).
That is, from Figure 1(b),
each circle except the top one in the layer is replaced by
the complementary part $\mathcal{P}_{n}^{k=p}$ ($\mathcal{H}_{n}^{k=p+1}=\mathcal{H}_{n}^{k=p}\oplus\mathcal{P}_{n}^{k=p}$),
such that the direct sum of all the circles gives rise to
$\mathcal{H}_{n}^{k=n+K-N}=\bigoplus_{p=n}^{n+K-N}\mathcal{P}_{n}^{k=p}$.
Note that $\mathcal{P}_{n}^{k=p}$ is nothing but the space spanned by
those configurations that come from the previous layers, by occupying the $p$-th orbital
as indicated by the southeasterly arrows in Figure \ref{fig:rotate}(b).
Thus, the dimension of these spaces can be read off from the previous layer in Figure \ref{fig:rotate}(b).
The changes are summarized in Figure \ref{fig:rotate}(c), where the downwards arrows have
been eliminated and new oblique arrows with different slopes appear.
The ending point $(k,n)$ of an arrow indicates that the $k$-th orbital is
added to the $(n-1)$-electron states from the starting point of the arrow, to
form new $n$-electron configurations that share the same last
orbital $k$ (suffix) in their orbital string. Note that a virtual node at $(K+1,N+1)$ (dashed circle) has been added in Figure \ref{fig:rotate}(c)
to accommodate this new convention.

Applying the renormalization interpretation
to Figure \ref{fig:rotate}(c) now leads to a renormalization process
that combines $n$-electron configurations that share the same orbital suffix.
Similarly, in reverse, if the renormalization proceeds from the
virtual node $(K+1,N+1)$ to the node $(0,0)$, then $n$-electron configurations
that share the same orbital prefix will be combined.
We refer to these renormalizations along the axis of particle numbers simply as {\it Hilbert space} renormalizations.
Similar to Eq. \eqref{Mixed}, the left and right renormalized states can be defined recursively as
\begin{eqnarray}
|\Psi\rangle&=&\sum_{l_{i-1}p_{i} r_i}|l_{i-1}p_i r_{i}\rangle A^{p_i}_{l_{i-1}r_i}[i],\nonumber\\
|l_{i-1}\rangle
&=&\sum_{p_1<\cdots<p_{i-1}}
(A^{p_1}[1]\cdots A^{p_{i-1}}[i-1])_{l_{i-1}}|p_1\cdots p_{i-1}\rangle\nonumber\\
&=&
\sum_{l_{i-2}p_{i-1}}|l_{i-2}p_{i-1}\rangle
A^{p_{i-1}}_{l_{i-2}l_{i-1}}[i-1],\nonumber\\
|r_{i}\rangle&=&
\sum_{p_{i+1}<\cdots<p_{N}}
(A^{p_{i+1}}[i+1]\cdots A^{p_{N}}[N])_{r_i}|p_{i+1}\cdots p_{N}\rangle\nonumber\\
&=&
\sum_{p_{i+1}r_{i+1}}
A^{p_{i+1}}_{r_{i}r_{i+1}}[i+1]|p_{i+1}r_{i+1}\rangle.\label{Mixed2}
\end{eqnarray}
where the orbital index $p_{i}\in\{1,\cdots,K\}$ and $|p_1p_2\cdots p_i\rangle$ are $i$-electron Slater determinants.
The summations over $p_{i-1}$ for the states $|l_{i-1}\rangle$ are purely formal, as the index can take only only
 value for each $|l_{i-1}\rangle$, due to the suffix constraint.
This also applies to the prefix $p_{i+1}$ for the states $|r_{i}\rangle$.
Thus, the prefix (suffix) of the left (right) renormalized states can be
viewed as a kind of 'symmetry' index, and a counterpart of the particle number index for renormalized
states in Fock space. Both of them represent the $x$-coordinates of the Hilbert subspaces (circles)
in the respective configuration graphs, see Figures \ref{fig:rotate}(a) and (c).
However, unlike the particle number symmetry in Fock space, it should
be emphasized that the prefix/suffix 'symmetry'
is not a physical symmetry of the Hamiltonian,
e.g., $\langle l_{i}'|H|l_{i}\rangle\ne 0$
for $|l_{i}'\rangle$ and $|l_{i}\rangle$ with different suffixes.

From Eq. \eqref{Mixed2}, a Hilbert-space MPS (HS-MPS) representation
for $N$-electron wavefunctions can be written down as,
\begin{eqnarray}
|\Psi\rangle=\sum_{p_1p_2\cdots p_N}\Psi^{p_1p_2\cdots p_N}|p_1p_2\cdots p_K\rangle,\label{HST}
\end{eqnarray}
where the tensor $\Psi^{p_1p_2\cdots p_N}\in\mathbb{C}^{N^K}$ is given by
\begin{eqnarray}
\Psi^{p_1p_2\cdots p_N}
=\sum_{\{\alpha_i\}} A^{p_1}_{\alpha_1}[1]A^{p_2}_{\alpha_1\alpha_2}[2]\cdots A^{p_N}_{\alpha_{N-1}}[N].\label{HSMPS}
\end{eqnarray}
This is clearly  an analogue of the FS-MPS \eqref{FSMPS} for wavefunctions in the Fock-space representation \eqref{FSrep}.
The relation of $\Psi^{p_1p_2\cdots p_N}$ \eqref{HST} with the FCI vector
is as follow:
Suppose the FCI wavefunction is expressed as
\begin{eqnarray}
|\Psi\rangle=\sum_{p_1<p_2<\cdots < p_N}
\Psi^{(p_1p_2\cdots p_N)}|p_1p_2\cdots p_N\rangle,\label{FCI}
\end{eqnarray}
where $\Psi^{(p_1p_2\cdots p_N)}$ is the FCI vector
with dimension given by the binomial coefficient $C_K^N$, and $(p_1p_2\cdots p_N)$ represents
an ordered set of orbital indices with $p_1<p_2<\cdots < p_N$.
From the way that the
HS-MPS is constructed in Figure \ref{fig:rotate}(c), it is easily seen that $\Psi^{p_1p_2\cdots p_N}$ \eqref{HSMPS} is naturally zero
if the orbital indices do not satisfy $p_1<p_2<\cdots < p_N$. Then,
the FCI wavefunction \eqref{HST} is obtained through the following relation,
\begin{eqnarray}
\Psi^{p_1p_2\cdots p_N}=
\left\{\begin{array}{ll}
\Psi^{(p_1p_2\cdots p_N)},& p_1<p_2<\cdots <p_N,\\
0,& \mathrm{otherwise}.
\end{array}\right.\label{HSrep}
\end{eqnarray}
This gives a 'strictly upper triangular'
tensor representation of the FCI wavefunction.
Such choice differs from the more common
antisymmetric tensor representation in Hilbert space,
where $\Psi^{p_1p_2\cdots p_N}_A
=\frac{1}{\sqrt{N!}}\Psi^{(p_1p_2\cdots p_N)}$ for $p_1<p_2<\cdots <p_N$, and all other entries of the tensor
$\Psi^{p_1p_2\cdots p_N}$ are fixed
by imposing the antisymmetry, e.g,
\begin{eqnarray}
\Psi^{p_1p_2\cdots p_N}_A=-\Psi^{p_2p_1\cdots p_N}_A=\cdots.\label{Antisym}
\end{eqnarray}
The factor $\frac{1}{\sqrt{N!}}$ arises from the normalization
condition. The advantage of the antisymmetric
tensor representation is that the set
of antisymmetric tensors is closed
under the rotation of the single particle basis:
It is straightforward to show that the tensor $\tilde{\Psi}^{p_1p_2\cdots p_N}$,
\begin{eqnarray}
\tilde{\Psi}^{p_1p_2\cdots p_N}
=\sum_{p_1'p_2'\cdots p_N'} \Psi_A^{p_1'p_2'\cdots p_N'}X_{p_1'p_1}X_{p_2'p_2}\cdots X_{p_N'p_N},\label{Trans}
\end{eqnarray}
which represents the same wavefunction $|\Psi\rangle$ when expressed in another one-particle basis obtained from a transformation $X_{pq}$,
is still antisymmetric,
\begin{eqnarray}
\tilde{\Psi}^{p_2p_1\cdots p_N}
&=&
\sum_{p_1'p_2'\cdots p_N'} \Psi_A^{p_2'p_1'\cdots p_N'}X_{p_2'p_2}X_{p_1'p_1}
\cdots X_{p_N'p_N}\nonumber\\
&=&
\sum_{p_1'p_2'\cdots p_N'} -\Psi_A^{p_1'p_2'\cdots p_N'}X_{p_2'p_2}X_{p_1p_1}
\cdots X_{p_N'p_N}\nonumber\\
&=&
\sum_{p_1'p_2'\cdots p_N'} -\Psi_A^{p_1'p_2'\cdots p_N'}X_{p_1'p_1}X_{p_2'p_2}
\cdots X_{p_N'p_N}\nonumber\\
&=&-\tilde{\Psi}^{p_1p_2\cdots p_N},
\end{eqnarray}
as the multiplications of the numbers $X_{pq}$ commute.
It is a particular consequence of the Schur-Weyl duality that relates irreducible
finite-dimensional representations of the general linear and symmetric groups, viz., $GL(K)$ and $S_N$ in our notation\cite{duality}.
Furthermore, if $\Psi^{p_1p_2\cdots p_N}_A$ has an MPS representation like Eq. \eqref{HSMPS} obtained
by applying successive SVD's, then the MPS representation for the
transformed $\tilde{\Psi}^{p_1p_2\cdots p_N}$ can be simply
obtained as
\begin{eqnarray}
\tilde{\Psi}^{p_1p_2\cdots p_N}
&=&\sum_{\{\alpha_i\}} \tilde{A}^{p_1}_{\alpha_1}[1]\tilde{A}^{p_2}_{\alpha_1\alpha_2}[2]\cdots \tilde{A}^{p_N}_{\alpha_{N-1}}[N],\nonumber\\
\tilde{A}^{p_i}[i]&=&
\sum_{p_i'}A^{p_i'}[i]X_{p_i'p_i},\label{HSMPSt}
\end{eqnarray}
which shows that the bond dimensions are not
be altered by the orbital rotation.

Unfortunately, albeit with these nice formal properties,
the antisymmetric tensor representation
turns out not to be a good starting point
for exploring the Hilbert space MPS representation. Because
the antisymmetry generates more nonzero terms in the tensors
than the 'strictly upper-triangular' representation \eqref{HSrep}, this leads to
a significant increase of the bond dimensions (see Sec. \ref{sec:numdecomp} for numerical examples).
The simplest example to reveal this important defect
is to consider the trivial
case $(K,N)=(2,2)$, where the dimension of the
FCI vector space is one and the wavefunction
is simply denoted by $|\Psi\rangle=|12\rangle$.
When mapped into the antisymmetric tensor representation,
the corresponding tensor (just a matrix in this case)
$\Psi^{p_1p_2}_A=\frac{1}{\sqrt{2}}\left[\begin{array}{cc}
0 & 1\\
-1 & 0 \\
\end{array}\right]$ is rank-2,
which increases the complexity. In comparison, in either the Fock-space representation
$\Psi^{n_1n_2}=\left[\begin{array}{cc}
0 & 0\\
0 & 1 \\
\end{array}\right]$ or the Hilbert-space representation \eqref{HSrep}
$\Psi^{p_1p_2}=\left[\begin{array}{cc}
0 & 1\\
0 & 0 \\
\end{array}\right]$, the tensors are rank-1.
In particular, the latter two representations
have the same number of nonzero entries as the
original FCI vector. The disadvantage
of these two representations is that the closed property of the MPS manifold
under orbital rotation no longer holds.
For instance, the transformation $X_{pq}$ in Eq. \eqref{Trans}
will generally bring nonzero
values into the entries of $\Psi^{p_1p_2\cdots p_N}$ violating $p_1<p_2<\cdots <p_N$.
In practice, this means a proper ordering of orbitals needs to be chosen.
But this seems to be a necessary price to pay when exploring the low-rank
structure of fermionic wavefunctions.

\subsection{Particle-hole duality}\label{sec:phtrans}
Before going into a detailed study of various properties of HS-MPS,
we introduce another representation of the wavefunction
through using particle-hole duality.
Within the finite basis scheme, the same $N$-electron wavefunction can also be understood
as a $(K-N)$-hole wavefunction. The FCI expansion \eqref{FCI} can be
written in terms of determinants
of holes $|h_1h_2\cdots h_{K-N}\rangle$, which have a one-to-one correspondences with determinants of electrons $|p_1p_2\cdots p_N\rangle$.
In CI or FS-MPS, this picture change does not lead to any nontrivial
advantage.
However, in the case of Hilbert-space MPS it gives a new MPS representation for the same wavefunction, because the sites become $K-N$ holes instead of $N$ electrons. We refer to this new MPS representation
as the HS-MPS for holes (HS-MPS[h]) with respect to the fully filled state, and the original representation \eqref{HSMPS} as the
HS-MPS for particles (HS-MPS[p]) with respect to the physical vacuum. In variational calculations,
the particle-hole duality implies that rather than minimizing
the energy of the second quantized Hamiltonian
\begin{eqnarray}
H = \sum_{pq}h_{pq}a_p^\dagger a_q + \frac{1}{2}\sum_{pqrs}g_{pq,rs} a_p^\dagger a_q^\dagger a_s a_r \equiv H_N^p,\label{Hnp}
\end{eqnarray}
where $g_{pq,rs} = \langle pq|rs\rangle$ in physicists' notation\cite{helgaker-molecular-2001}, using
the HS-MPS[p] as an ansatz, the HS-MPS[h] can alternatively be employed.
Here, the notation $H_N^p$ means the Hamiltonian is normal ordered with respect to the genuine vacuum
without electrons. To apply exactly the same variational optimization algorithm to
HS-MPS[h], the Hamiltonian $H$ can be simply rewritten in terms of the set of transformed
operators $b_p\triangleq a_p^\dagger$ with respect to the fully filled state as
\begin{eqnarray}
H = \sum_{pq}h_{pq}b_p b^\dagger_q + \frac{1}{2}\sum_{pqrs}\langle pq|rs\rangle b_p b_q b_s^\dagger b_r^\dagger.
\end{eqnarray}
Through Wick's theorem\cite{shavitt2009many}, this can be recast into
a sum of a constant reference energy
for the new vacuum, and a normal ordered Hamiltonian $H_N^h$ with the same mathematical form as $H_N^p$ \eqref{Hnp},
\begin{eqnarray}
H &=& E_{\mathrm{ref}} + H_{N}^h,\nonumber\\
E_{\mathrm{ref}} &=& \sum_{p} h_{pp} +\frac{1}{2}\sum_{pq}\langle pq\|pq\rangle,\nonumber\\
H_{N}^h &=&\sum_{pq}\tilde{h}_{pq} b_p^\dagger b_q + \frac{1}{2}\sum_{pqrs}\tilde{g}_{pq,rs} b_p^\dagger b_q^\dagger b_s b_r, \nonumber\\
\tilde{h}_{pq} &=& -h_{qp} - \sum_s \langle qs\|ps\rangle,\nonumber\\
\tilde{g}_{pq,rs} &=& g_{pq,rs}^* = \langle rs|pq\rangle.
\end{eqnarray}
Unless the bond dimension is sufficiently large to reach the FCI limit, the HS-MPS[p]
and HS-MPS[h] with the same maximal bond dimension $D$ generally parametrize different
manifolds of states. This point will become clear
when examining the theoretical bond dimensions for various CI models
in the next section.

\subsection{Bond dimensions for virtual indices}\label{sec:dims}

In this section, we investigate the minimal bond dimensions $D_k$ to represent
the FCI or truncated CI spaces by FS/HS-MPS, or in other words, the maximum bond dimensions
that are necessary to represent an arbitrary state in a given FCI or truncated CI space.
This analysis will be crucial for understanding the performance of FS/HS-MPS.

The starting point is to examine a bipartition of the sites.
For simplicity, the Fock space case is considered first. Suppose the sites (orbitals)
are (bi)partitioned into two sets and the many-body basis for the left and right Fock spaces are denoted by $\{|L_\beta\rangle\triangleq|n_1\cdots n_k\rangle\}$ and
$\{|R_\gamma\rangle\triangleq|n_{k+1}\cdots n_{K}\rangle\}$, respectively. The Fock space expansion \eqref{FSrep} can then
be rewritten as
\begin{eqnarray}
|\Psi\rangle = \sum_{\beta\gamma}\Psi^{L_\beta R_\gamma}|L_\beta R_\gamma\rangle.\label{Matrix}
\end{eqnarray}
Meanwhile, the form of FS-MPS \eqref{FSMPS} allows the wavefunction to be written
a sum of $D_k$ terms
\begin{eqnarray}
|\Psi\rangle&=&\sum_{\alpha_k}|\alpha_k^L\alpha_k^R\rangle,\nonumber\\
|\alpha_k^L\rangle&\triangleq &\sum_{n_1\cdots n_k}(A^{n_1}[1]\cdots A^{n_{k}}[k])_{\alpha_{k}}|n_1\cdots n_{k}\rangle,\nonumber\\
|\alpha_k^R\rangle&\triangleq &\sum_{n_{k+1}\cdots n_K}
(A^{n_{k+1}}[k+1]\cdots A^{n_{K}}[K])_{\alpha_k}|n_{k+1}\cdots n_{K}\rangle,\label{MPSpartition}
\end{eqnarray}
which is similar to Eq. \eqref{Mixed}. By comparing Eqs. \eqref{Matrix}
with \eqref{MPSpartition}, one can identify the theoretical bond dimension $D_k$ as the rank of
the matrix $\Psi^{L_\beta R_\gamma}$. In particular, if the SVD is applied to $\Psi^{L_\beta R_\gamma}$,
then the resulting decomposition is just the Schmidt decomposition\cite{qcalgebra} with the singular values
characterizing the entanglement between the two Fock subspaces.

\begin{figure}
\begin{tabular}{c}
{\resizebox{!}{0.15\textheight}{\includegraphics{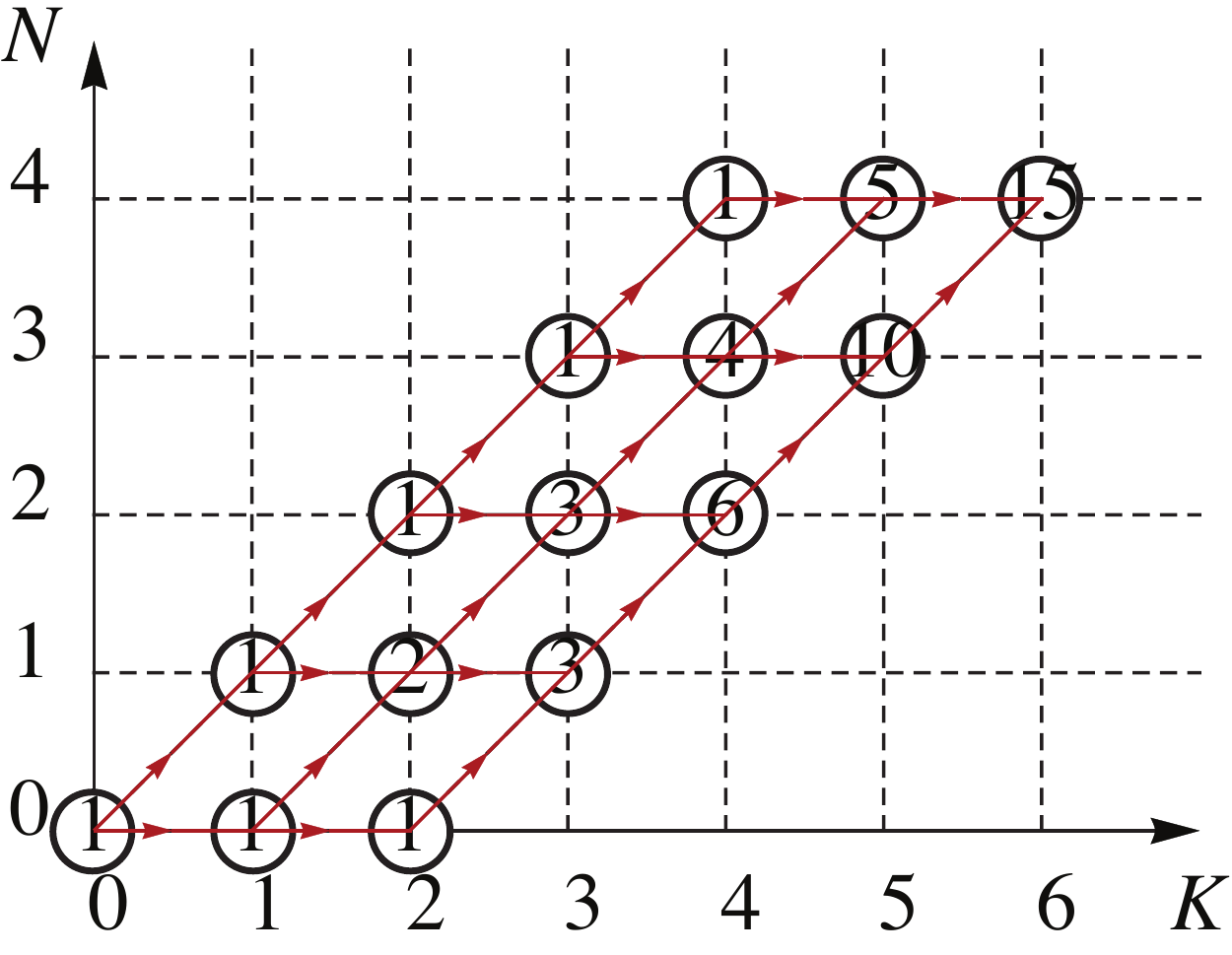}}}\\
(a) \\
{\resizebox{!}{0.15\textheight}{\includegraphics{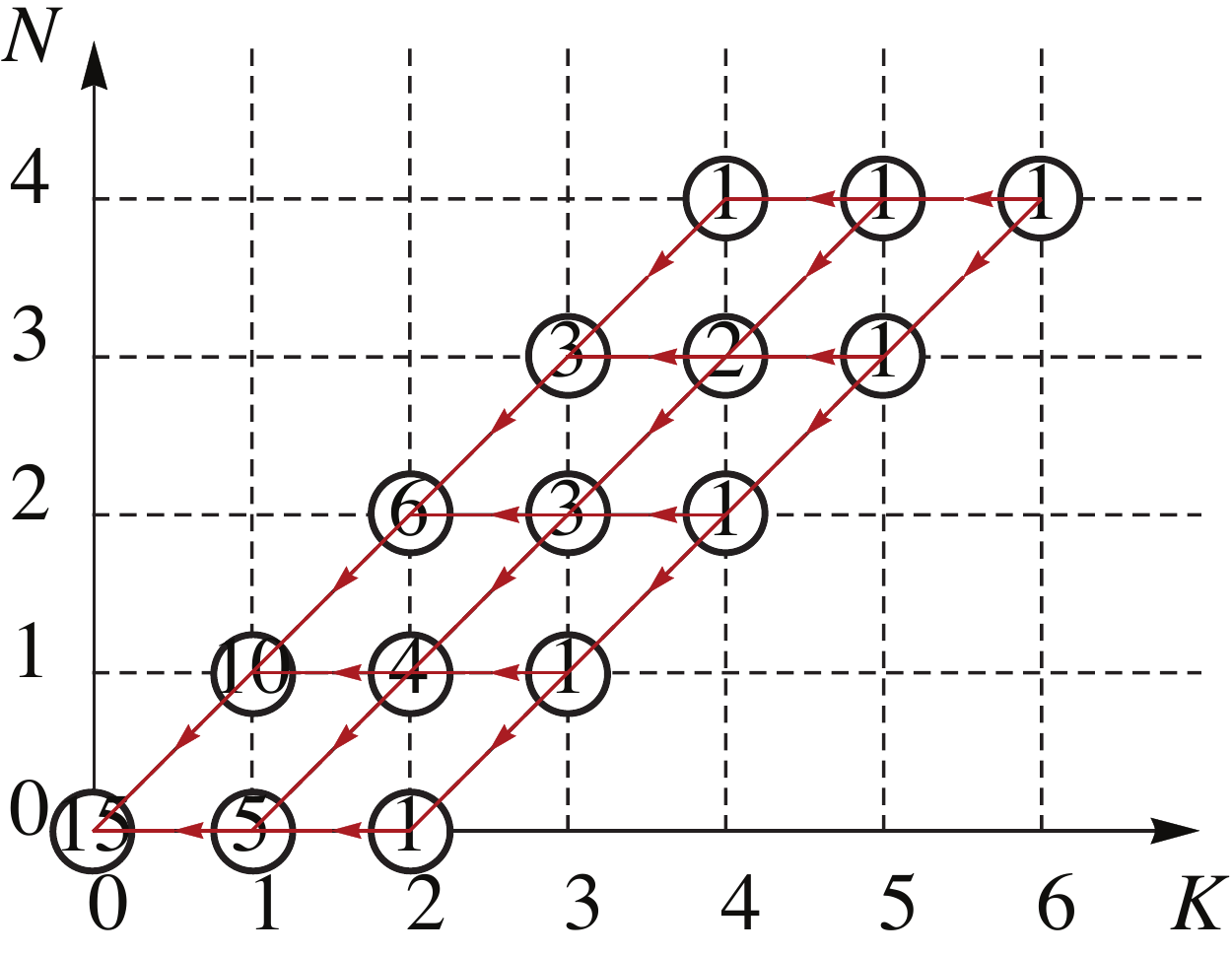}}}\\
(b) \\
{\resizebox{!}{0.2\textheight}{\includegraphics{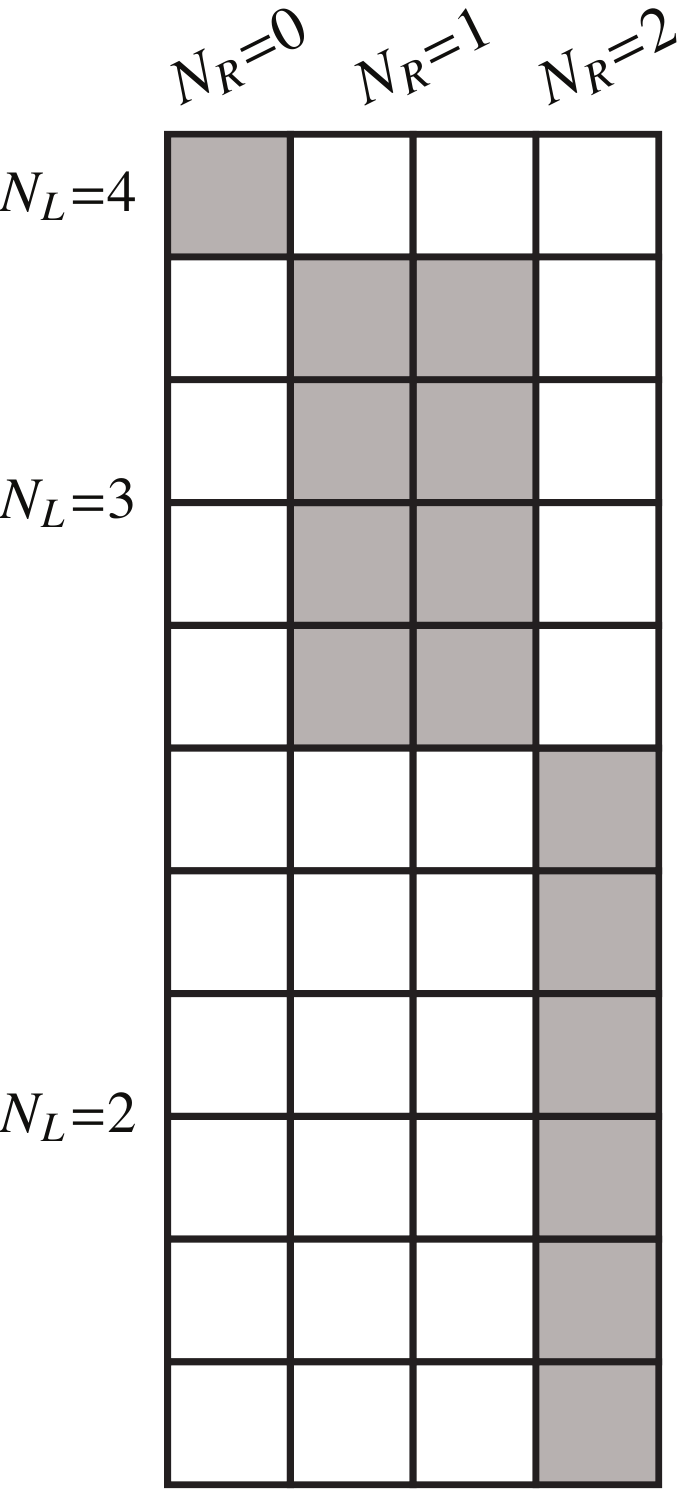}}}\\
(c) \\
{\resizebox{!}{0.15\textheight}{\includegraphics{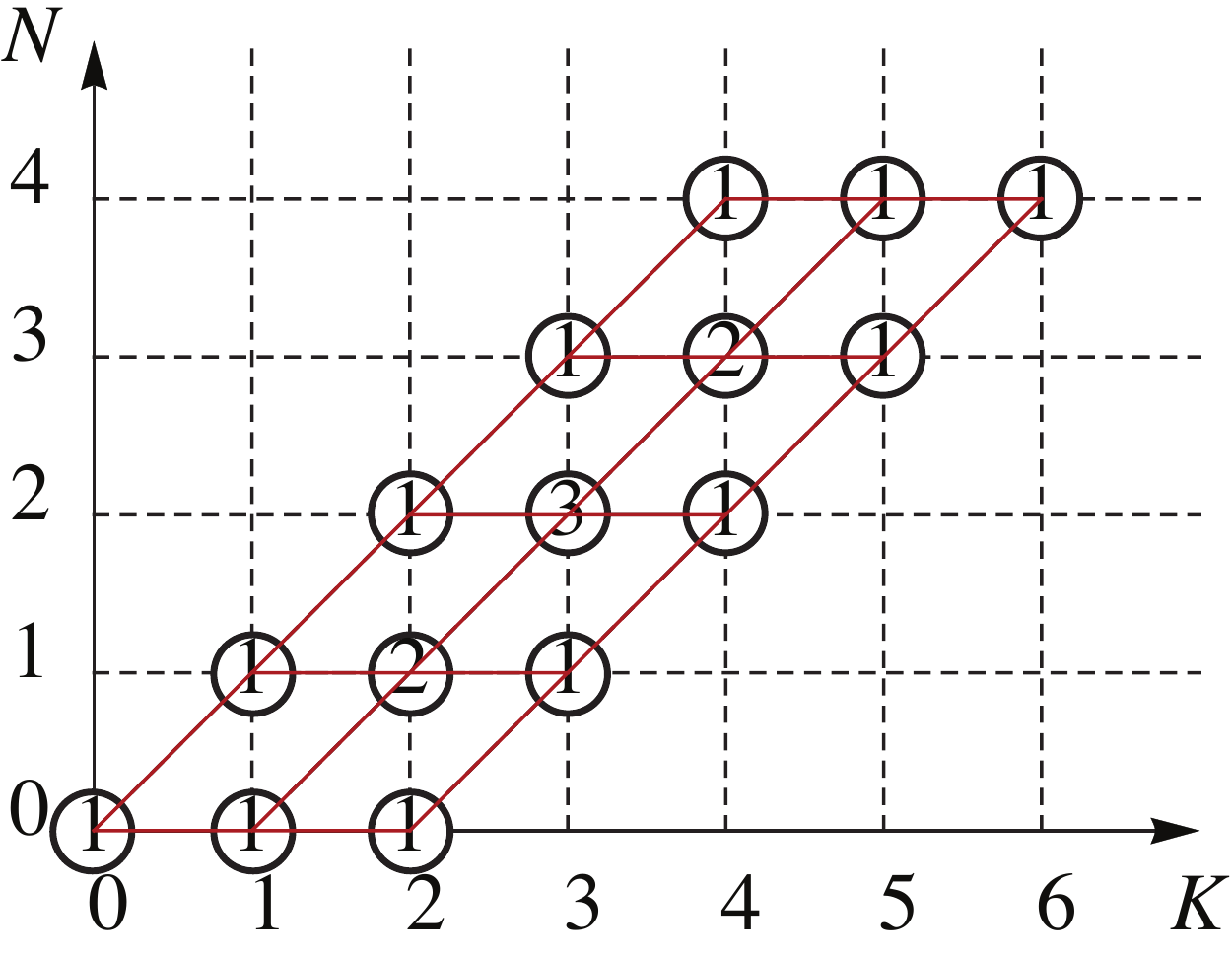}}}\\
(d) \\
\end{tabular}
\caption{Counting the bond dimensions of FS-MPS for $(K,N)=(6,4)$:
(a) dimensions for left Fock spaces $\{|L_\beta\rangle\triangleq|n_1\cdots n_k\rangle\}$;
(b) dimensions for right Fock spaces $\{|R_\gamma\rangle\triangleq|n_{k+1}\cdots n_{K}\rangle\}$;
(c) block structure of $\Psi^{L_\beta R_\gamma}$ for $k=4$;
(d) distributions of bond dimensions in each symmetry sector.}\label{fig:dimFS}
\end{figure}

To compute the rank of the matrix $\Psi^{L_\beta R_\gamma}$, it is also
instructive to employ the graphical notation. The dimension of each
symmetry sector for $\{|L_\beta\rangle\}$ can be found in Figure \ref{fig:rotate}(a)
or Figure \ref{fig:dimFS}(a). To compute the corresponding dimensions for $\{|R_\gamma\rangle\}$,
it is simple to just reverse the flows recursively in Figure \ref{fig:rotate}(a) starting from
the ending point $(K,N)$, which leads to Figure \ref{fig:dimFS}(b). Taking $k=4$ as
an example, Figure \ref{fig:dimFS}(a) shows that the left space $\{|L_\beta\rangle\}$ is composed
of three subspaces: $N_L=4$ with dimension 1, $N_L=3$ with dimension 4,
and $N_L=2$ with dimension 6, while Figure \ref{fig:dimFS}(b) shows that
the complementary right space $\{|R_\gamma\rangle\}$ is composed
of three subspaces: $N_R=4-4=0$ with dimension 1, $N_R=4-3=1$ with dimension 2,
and $N_R=4-2=2$ with dimension 1.
Due to the particle number symmetry, $\Psi^{L_\beta R_\gamma}$
has a block diagonal structure as shown in Figure \ref{fig:dimFS}(c), from which the rank can be seen to be the
sum of the ranks of smaller blocks.
In Figure \ref{fig:dimFS}(d), the ranks of smaller blocks computed by simply taking
the minimum of the dimensions of left and right subspaces are displayed.
The sum of the values on the same layer $k$ gives the theoretical bond dimension $D_k$.
In the present example, the values of $D_k$ are $\{2,4,5,4,2\}$ which can be easily read off
from Figure \ref{fig:dimFS}(d). It is notable that they are distributed symmetrically about the center of the FS-MPS chain.

\begin{figure}
\begin{tabular}{c}
{\resizebox{!}{0.22\textheight}{\includegraphics{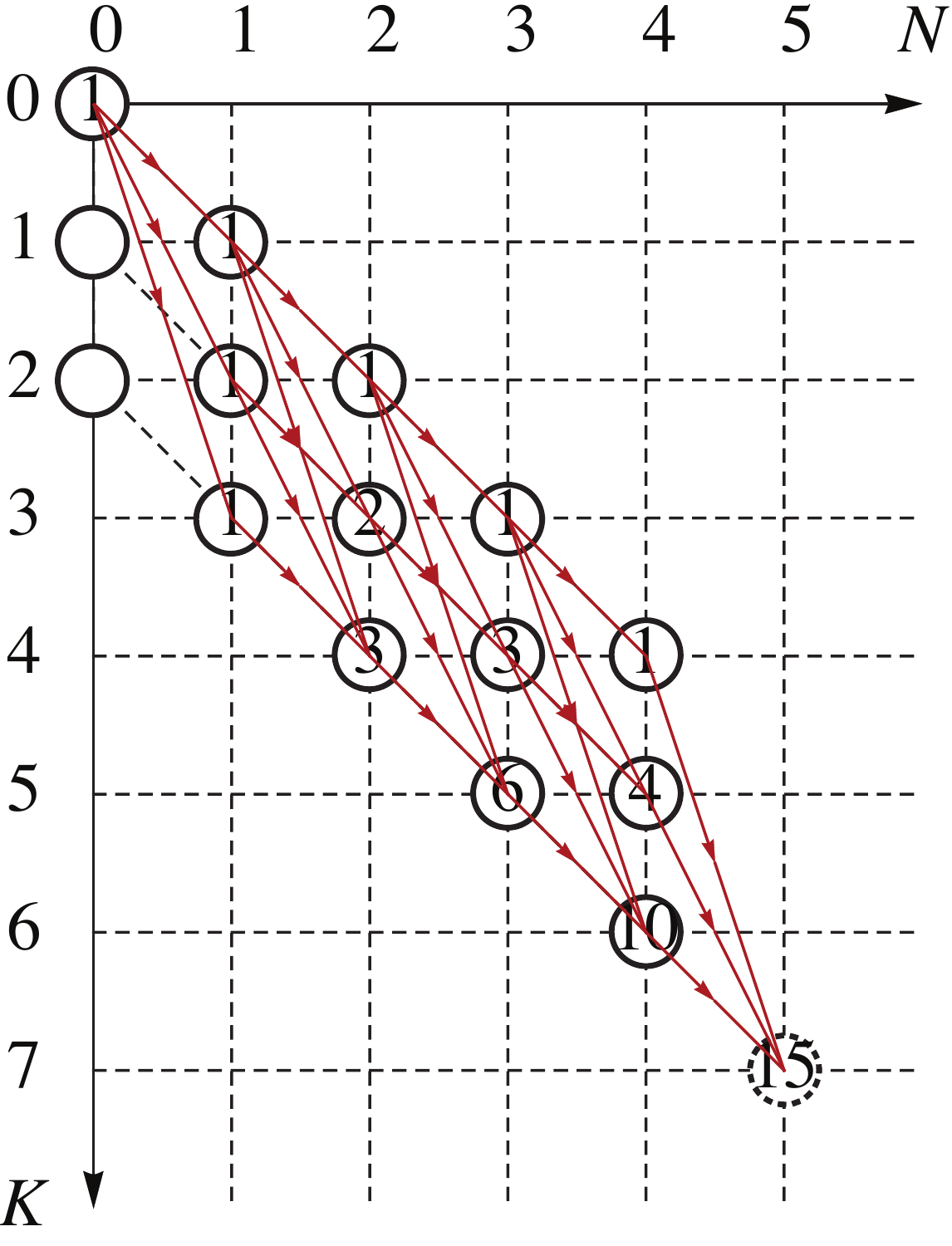}}}\\
(a) \\
{\resizebox{!}{0.22\textheight}{\includegraphics{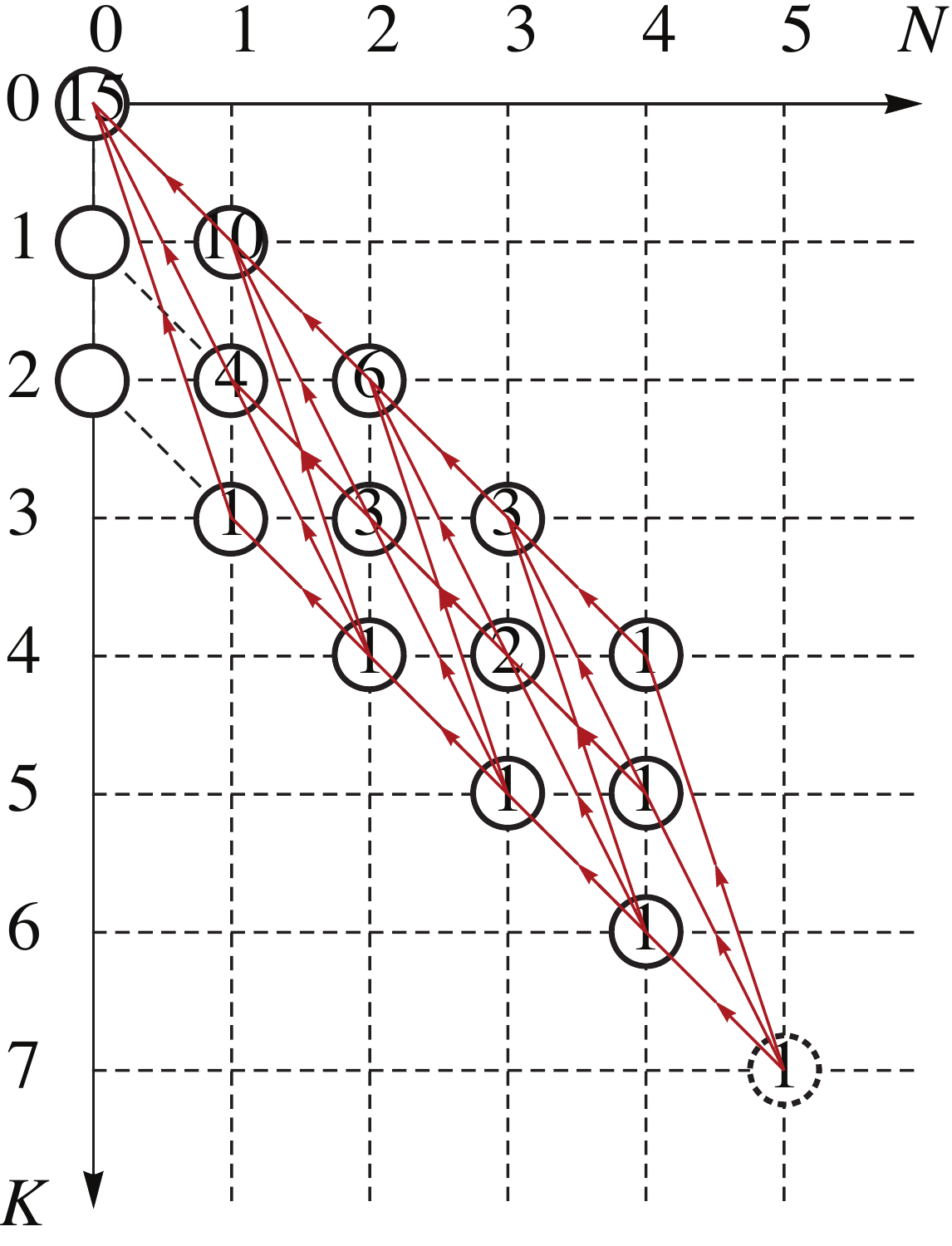}}}\\
(b) \\
{\resizebox{!}{0.22\textheight}{\includegraphics{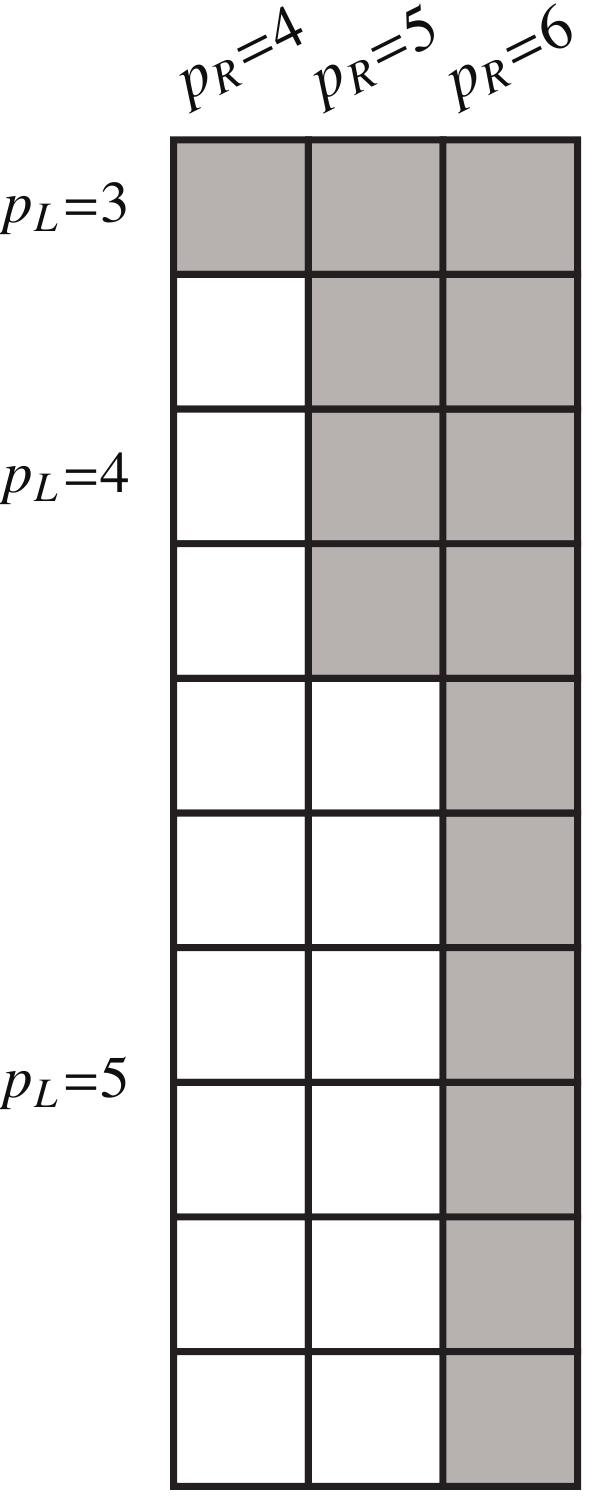}}}\\
(c) \\
{\resizebox{!}{0.22\textheight}{\includegraphics{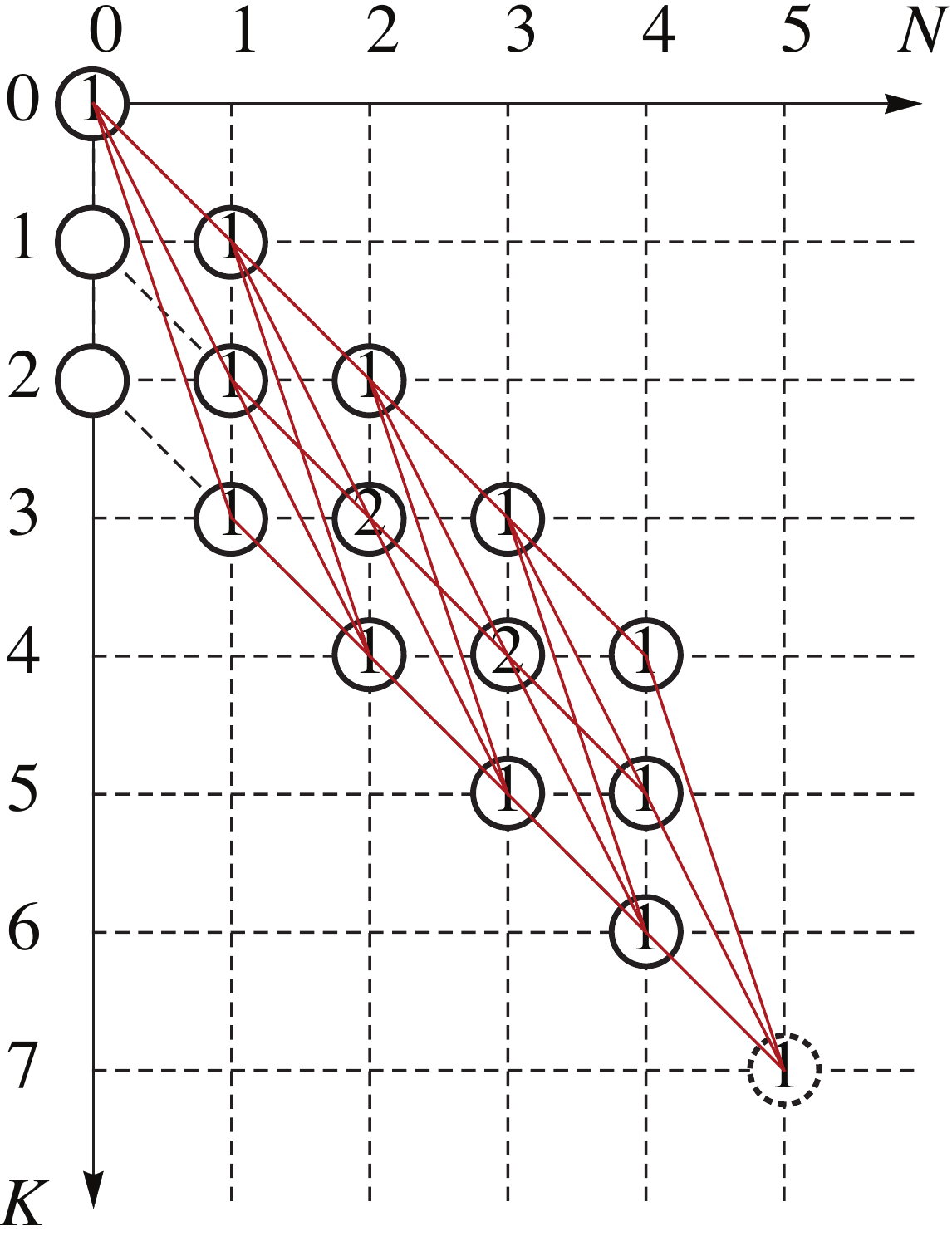}}}\\
(d) \\
\end{tabular}
\caption{Counting the bond dimensions of HS-MPS for $(K,N)=(6,4)$:
(a) dimensions for left Hilbert spaces $\{|L_\beta\rangle\triangleq|p_1\cdots p_n\rangle\}$;
(b) dimensions for right Hilbert spaces $\{|R_\gamma\rangle\triangleq|p_{n+1}\cdots p_{N}\rangle\}$;
(c) block structure of $\Psi^{L_\beta R_\gamma}$ for $n=3$;
(d) distributions of bond dimensions in each symmetry sector.}\label{fig:dimHS}
\end{figure}

The basic principle to compute the bond dimensions for HS-MPS is similar, but there are
some differences in the interpretation of the graphs. Figure \ref{fig:dimHS} displays the counterparts
of graphs in Figure \ref{fig:dimFS}. In the case of HS-MPS, the bipartition of Hilbert space
is considered, e.g., $\{|L_\beta\rangle\triangleq|p_1\cdots p_n\rangle\}$ and $\{|R_\gamma\rangle\triangleq|p_{n+1}\cdots p_{N}\rangle\}$. Taking $n=3$ as an example, Figure \ref{fig:dimHS}(a) shows the dimensions of the subspaces in $\{|L_\beta\rangle\}$
are 1, 3, and 6 for $\mathcal{P}_{3}^{k=3}$, $\mathcal{P}_{3}^{k=4}$, and
$\mathcal{P}_{3}^{k=5}$, respectively. These are the numbers of configuration strings
with the last orbital index equal to a given value $k$.
Figure \ref{fig:dimHS}(b) shows the reversed process, that is, the configuration strings
are grouped by their prefixes, and the values in circles are the numbers of configuration strings
whose first orbital index is equal to a given value $k$. Therefore, different from the FS-MPS case, although here the number of rows
of $\Psi^{L_\beta R_\gamma}$ is still given by the numbers in the layer with $n=3$ in Figure \ref{fig:dimHS}(a),
the number of columns is represented by the numbers in the next layer with $n=4$.
This shift of layers for the right space is due to the fact that the reversed process needs to be initiated
from the additional fictitious node rather than the node $(K,N)$.
Another significant difference between the FS-MPS and HS-MPS cases is that instead of a block diagonal structure for $\Psi^{L_\beta R_\gamma}$ as in Figure \ref{fig:dimFS}(c), the wavefunction $\Psi^{L_\beta R_\gamma}$ in the Hilbert-space case
has a block upper-triangular structure in the bipartitioned basis as shown in Figure \ref{fig:dimHS}(c).
This difference is again due to the fact that the prefix/suffix indices do not correspond to a physical symmetry; the final wavefunction does not transform as a single 'irrep' of the prefix/suffix. When considering renormalization from the left (also simply referred as suffix renormalization), where the recombination of configuration strings
is restricted to configuration strings with the same suffix, the bond dimension is given by the
sum of ranks for the row-wise blocks with dimensions $(1,3)$, $(3,2)$, and $(6,1)$, which gives
$1+2+1=4$ left renormalized states. On the other hand, when considering the renormalization from the right (prefix renormalization), the column-wise blocks are of dimension $(1,1)$, $(1,4)$ and $(1,10)$, which gives only $1+1+1=3$ right renormalized states.
Thus, to compute the bond dimensions for suffix renormalization,
the dimensions of the row-wise blocks can still be read off from the same layer
in Figures 3(a) and 3(b), and their ranks are summarized in Figure \ref{fig:dimHS}(d).
Again, the sum of the values in the same layer gives a symmetric sequence $\{3,4,4,3\}$.
However, there are only three virtual indices for $N=4$, and the additional fourth value
appears only due to the introduction of the additional node $(K+1,N+1)$.
Thus, the fourth value in the sequence
does not represent a real bond dimension in the suffix renormalization.
This leaves the theoretical bond dimensions $\{3,4,4\}$
for left renormalization in the case $(K,N)=(6,4)$. In contrast, when performing the prefix renormalization
from the right, the bond dimensions are given by the sequence $\{4,4,3\}$ by removing the first element
in the symmetric sequence. This asymmetric feature is quite different from the FS-MPS case.

The block upper-triangular structure and the structure of renormalization
with the prefix/suffix constraints
have several important consequences:

First, unlike in the Fock space case where applying successive SVD's to the tensor \eqref{FSrep} automatically produces an FS-MPS with particle number symmetry, the SVD for the Hilbert-space tensor \eqref{HSrep} will not lead to an HS-MPS that obeys
a prefix/suffix constraint. Rather, it will lead to a more compressed representation
because in this case the number of left states is just the rank of $\Psi^{L_\beta R_\gamma}$, which is three in case of
Figure \ref{fig:dimHS}(c), rather than 4. However, as we will show in Sec. \ref{sec:varopt},
although combining suffixes can produce a more highly
compressed representation,
it leads to difficulties in computing matrix elements among the renormalized states.
In contrast, matrix elements for the renormalized states with the prefix/suffix constraints can be factorized into products of smaller pieces, which can be computed efficiently in a recursive way
similar to the FS-MPS case. This is because the renormalized states with prefix/suffix constraints can be viewed as renormalized
states in Fock space, where the prefix/suffix labels the Fock subspace.
For instance, a suffix (left) renormalized state $|l_n\rangle$ with a number of
electrons $n$ and last orbital index $p$ can also be viewed as a
special Fock-space renormalized state $|l_k\rangle$ defined in the Fock subspace
$\mathcal{F}_{k=p}$, whose number of electrons is $n$, and where the last orbital $k$ is occupied.
In addition, maintaining the constraint leads to theoretical bond dimensions given by sums of ranks of smaller blocks, which is formally similar to the Fock-space case. As will be shown numerically in Sec. \ref{sec:numdecomp}, the redundancy introduced by the constraints is
in general not large. An interesting example to consider is a set of
noninteracting
systems. For simplicity, we examine the case where each individual subsystem
is an identical closed-shell molecule having $n$ electrons and described by $k$ spin-orbitals.
In this case, the exact wavefunction is just a product of wavefunctions of individual
systems either in the Fock-space representation or in the Hilbert-space representation
without suffix constraints, which means that the bond dimension between the
different subsystems is simply one. However, with the prefix/suffix constraint, a bond dimension of $k-n+1$ is needed to fully represent the left renormalized states
at the boundary of two subsystems. As long as the maximally allowed bond dimension
is larger than this value, then the HS-MPS wavefunction is
size extensive, even
when there are truncations within each subsystem. For example, as will be shown in Sec. \ref{sec:restriction},
the HS-MPS is flexible enough to represent products of truncated CI wavefunctions.
In general, the size extensivity of HS-MPS is ensured by using local orbitals and a proper
ordering that groups orbitals belonging to the same subsystem together, which is the same requirement as for size-extensive FS-MPS.

Second, the different bond dimensions for left and right renormalizations implies that
when performing one-site DMRG-like optimizations for HS-MPS with a given maximally allowed
bond dimension $D$, in general there are truncations during the renormalizations, and this means that the variational energies
do not necessarily decay monotonically from site to site during the sweep optimization.
However, this is a general feature of MPS whenever a 'symmetry'
is imposed on the renormalized states, as is the case here with prefix/suffix constraints. For instance,
a similar situation
happens in the spin-adapted DMRG for non-singlet states\cite{sharma_spin-adapted_2012}.
Consider the following state as an example,
\begin{eqnarray}
|\Psi_{S=1}\rangle\rangle=
\left(|l_{S=1}\rangle\rangle+|l_{S=0}\rangle\rangle\right)\times|r_{S=1}\rangle\rangle,
\end{eqnarray}
where the double bracket represents the whole spin multiplet including
all spin components\cite{rowe1975tensor} and
the product $\times$ represents the product with tensor couplings.
In a renormalization from the left, $|\Psi_{S=1}\rangle\rangle$ will lead to two spin-adapted left renormalized states
($|l_{S=1}\rangle\rangle$ and $|l_{S=0}\rangle\rangle$), whereas
it will produce only one spin-adapted right renormalized state $|r_{S=1}\rangle\rangle$
in a renormalization from right. This is analogous to the situation in Figure \ref{fig:dimHS}(c).

\subsection{Exact decomposition of a wavefunction into HS-MPS}\label{sec:decomp}
As discussed in the last section, given a wavefunction
in the tensor representation \eqref{HSrep}, a direct application
of successive SVD's does not lead to the Hilbert-space MPS \eqref{HSMPS}.
The SVD procedure needs to be modified to incorporate the prefix/suffix constraint.
For simplicity, we will only consider the case of decomposing a wavefunction into
left canonical form with the suffix constraint. The basic idea to implement the
suffix constraint with the SVD is to treat the suffix of the configuration strings as a 'symmetry' index
as mentioned in Sec. \ref{sec:formulation}.
Then, in the direct product of two Hilbert spaces to create a larger Hilbert space,
these indices can be used to set up a projection onto the target space whose basis
functions obey the ordering $p_1<p_2<\cdots<p_N$. As a simple illustration,
we consider the product of the spaces $V_1=\{|1\rangle,|2\rangle\}$
and $V_2=\{|1\rangle,|2\rangle,|3\rangle\}$ whose 'symmetry' indices are just the corresponding
orbital indices. The direct product of these two spaces
leads to the two-electron configuration space
\begin{eqnarray}
V_1\otimes V_2&=&\spn\{|1\rangle,|2\rangle\}\otimes \spn\{|1\rangle,|2\rangle,|3\rangle\}\nonumber\\
&=&
\spn\{|11\rangle,|12\rangle,|13\rangle,|21\rangle,|22\rangle,|23\rangle\},
\end{eqnarray}
Only the basis functions
$\{|12\rangle,|13\rangle,|23\rangle\}$ satisfy the requirement $p_1<p_2$ needed
to represent configuration strings for fermions.
The first one $|12\rangle$ belongs to the 'symmetry' sector with suffix equal to 2, and
the last two $|13\rangle$ and $|23\rangle$ belong to the same 'symmetry' sector with suffix equal to 3.
This information can then be reused in the construction of three electron states,
if these two-electron states are further coupled with a set of
one-electron states $\{|p_i\rangle\}$. The orbital index 'symmetry'
can be used during the successive SVD procedure \eqref{SVD}
to keep track of the combinations that produce configuration strings
satisfying $p_1<p_2<\cdots<p_N$ when performing direct products of two spaces in
the chain \eqref{FSChain}. Once the product of 'symmetry' indices is carried out,
the SVD can be performed for individual row-wise blocks as
in Figure \ref{fig:dimHS}(c) to obtain the left renormalized states.
Equivalently, such an SVD procedure can also be replaced by the diagonalization of
a pseudo-density matrix constructed by taking the diagonal 'symmetry' blocks of the
true density matrix $\rho^L_{\beta\beta'}\triangleq \sum_{R_\gamma}
\Psi^{L_\beta R_\gamma}\Psi^{L_{\beta'}R_\gamma*}$. Both procedures can be used in
the decomposition of a given wavefunction into an HS-MPS,
or in the decimation step in the DMRG optimization
of an HS-MPS, as will be discussed in the next section.
For the decimation step, the pseudo-density matrix is slightly more general, as
it can be extended to treat multiple states within a state-averaged approach.

\subsection{Restricted manifolds of HS-MPS}\label{sec:restriction}
\begin{figure}
\begin{tabular}{cc}
{\resizebox{!}{0.2\textheight}{\includegraphics{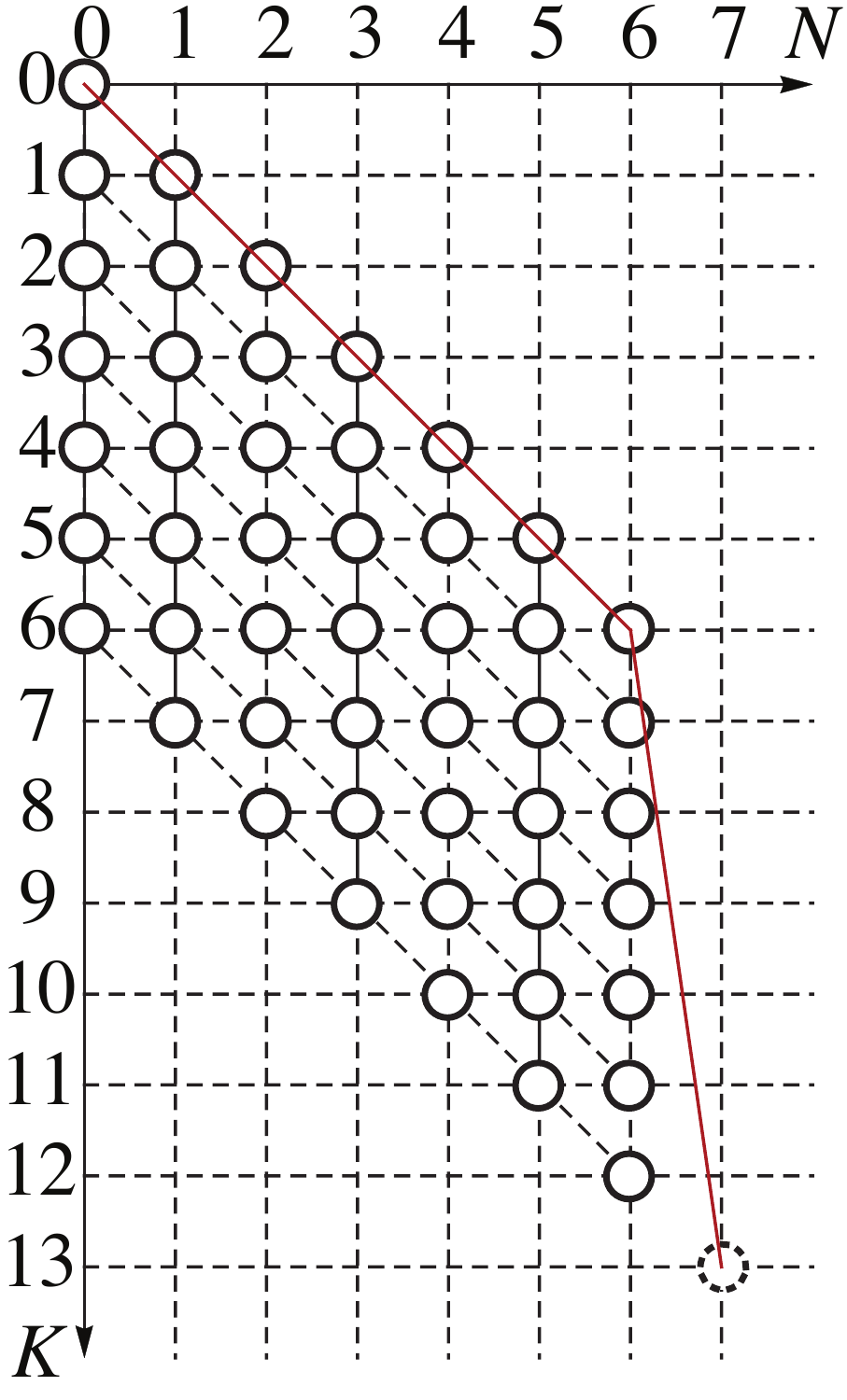}}} &
{\resizebox{!}{0.2\textheight}{\includegraphics{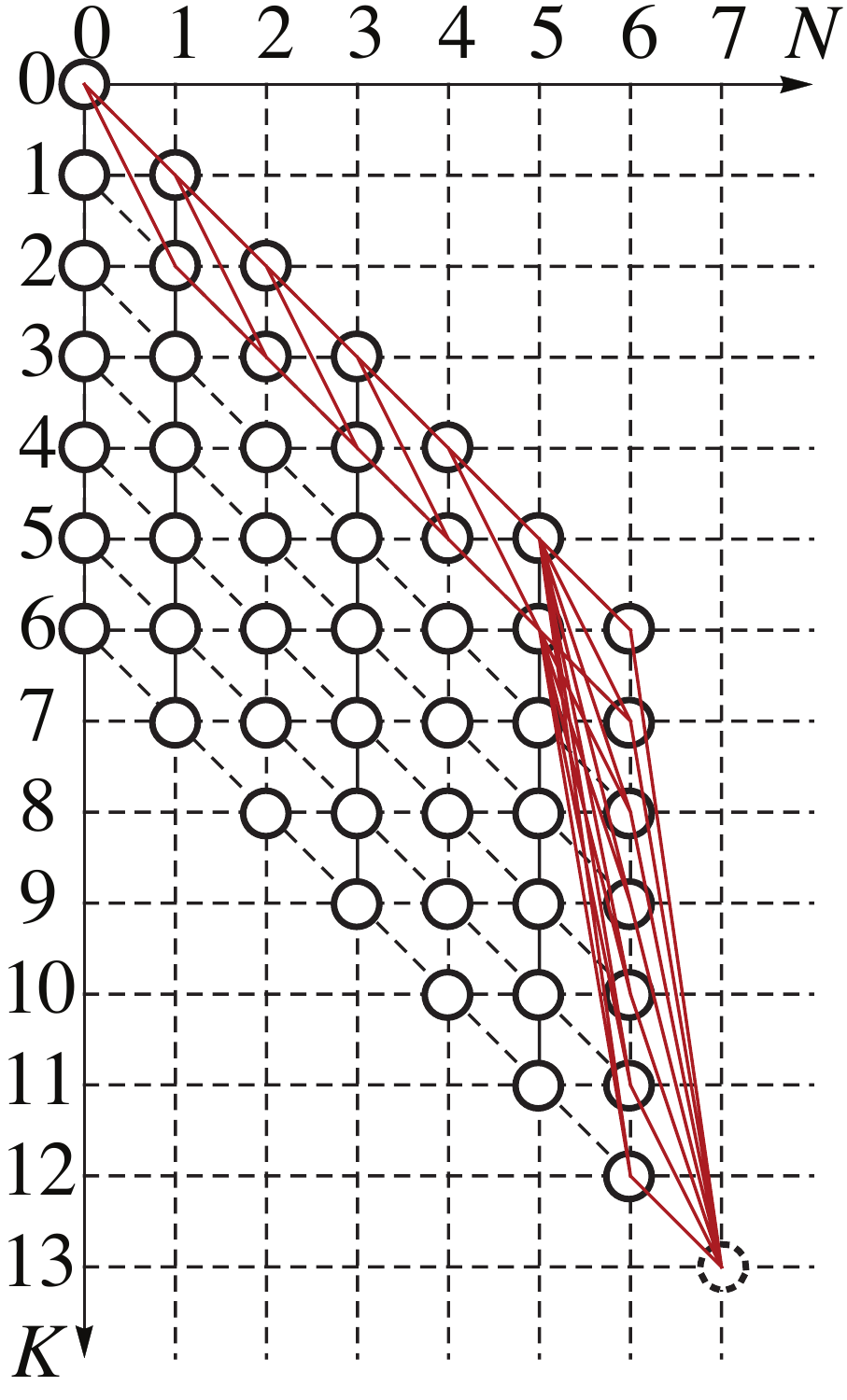}}} \\
(a) HF & (b) CIS \\
{\resizebox{!}{0.2\textheight}{\includegraphics{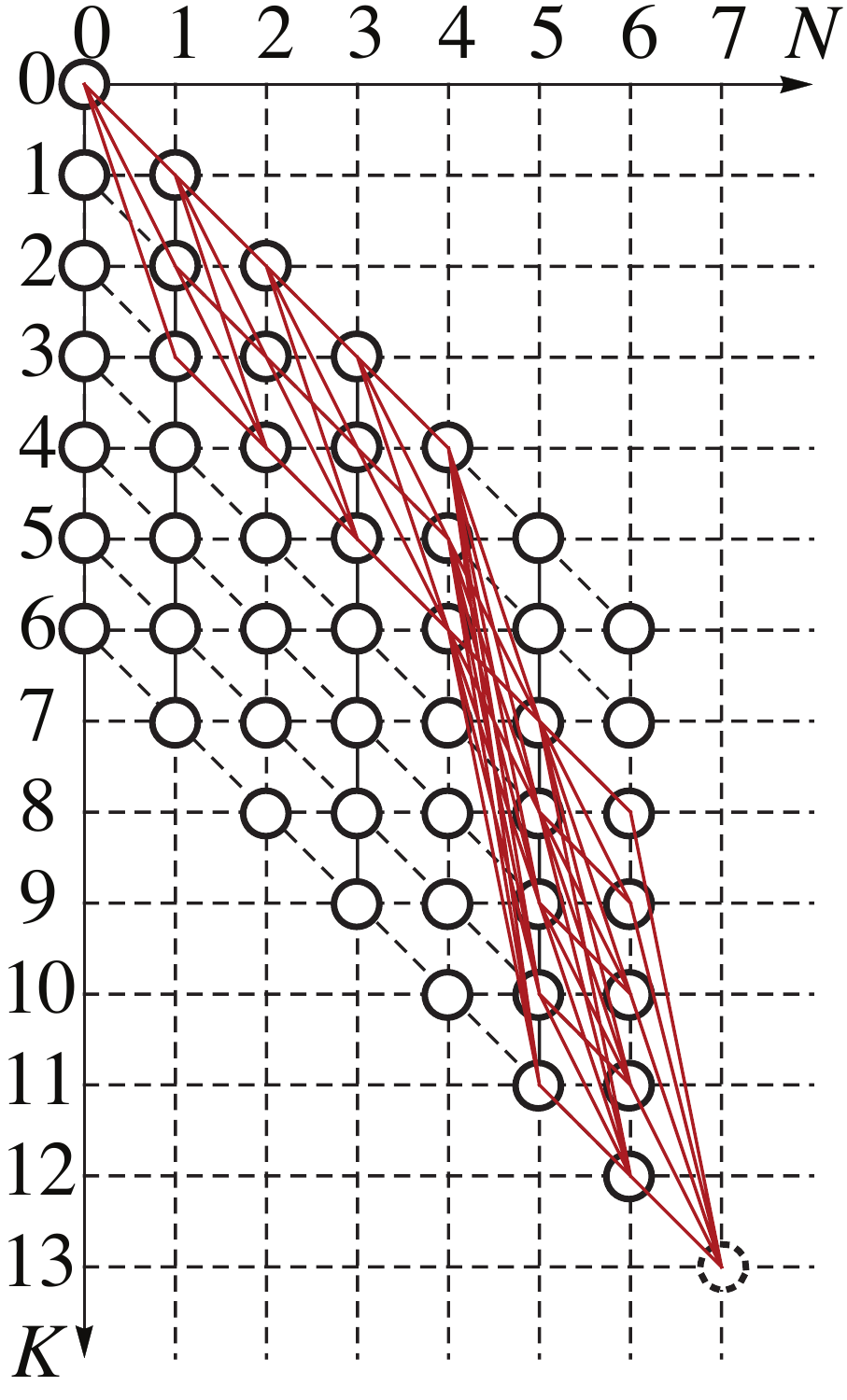}}} &
{\resizebox{!}{0.2\textheight}{\includegraphics{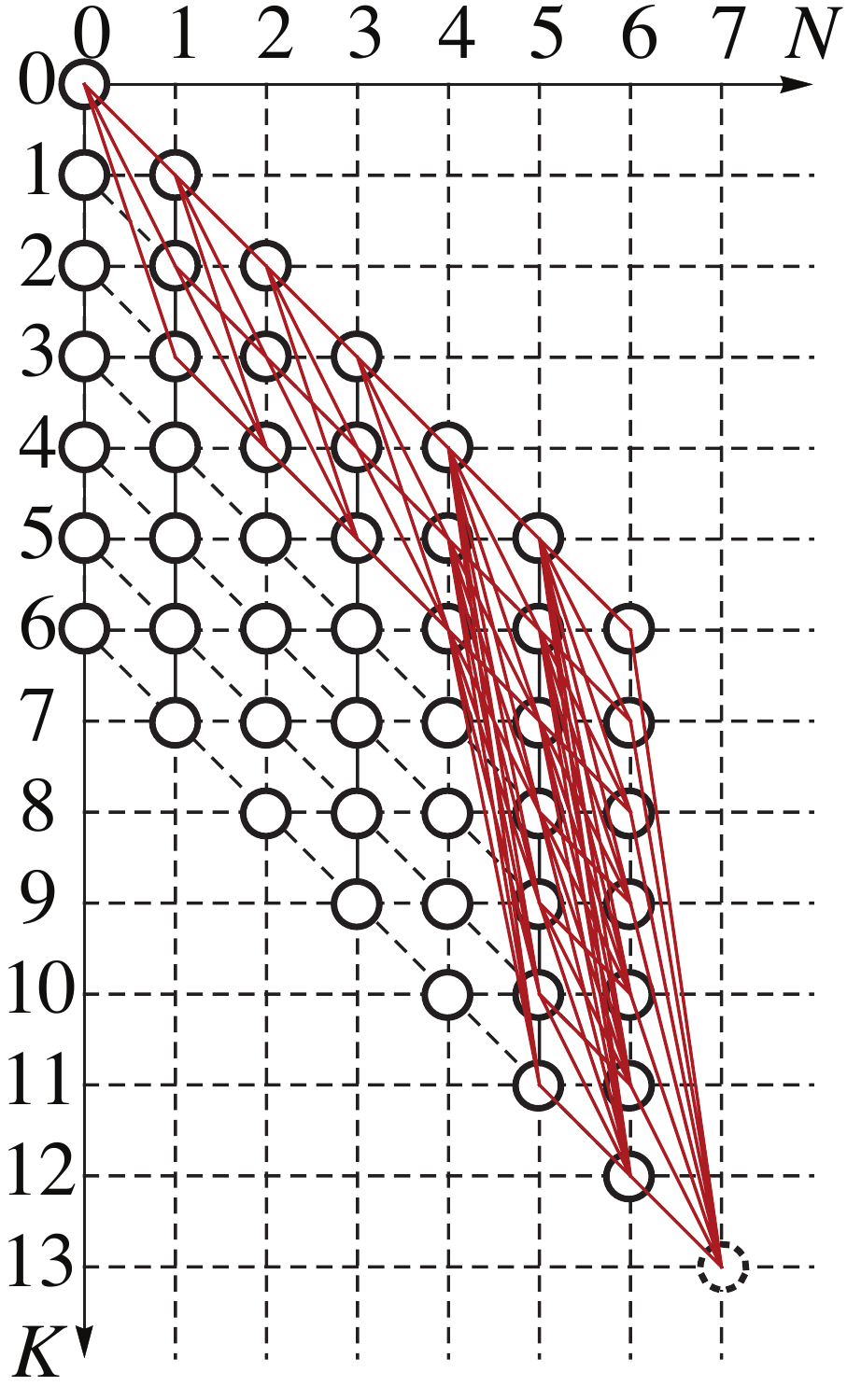}}} \\
(c) Doubles & (d) CISD \\
{\resizebox{!}{0.2\textheight}{\includegraphics{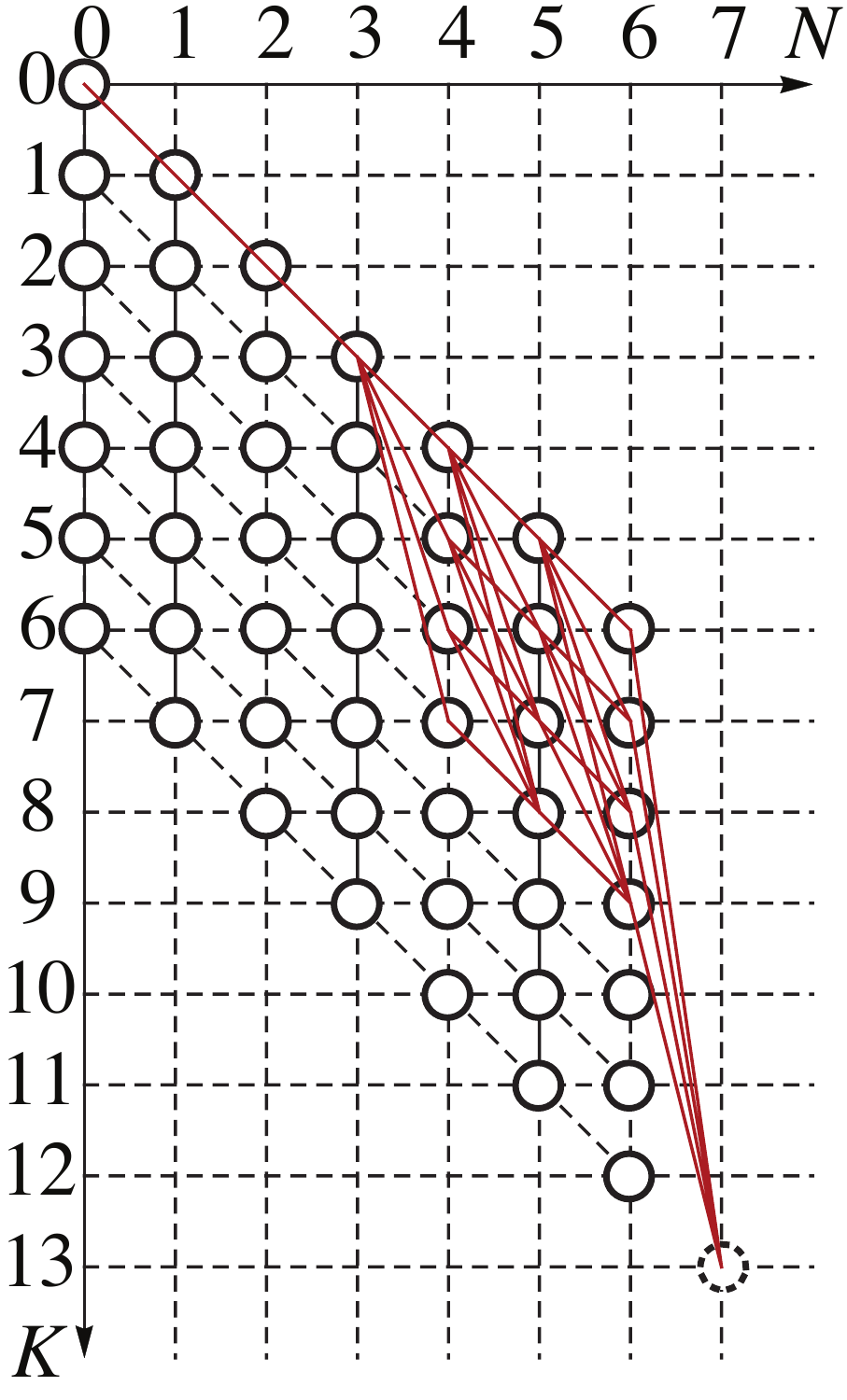}}} &
{\resizebox{!}{0.2\textheight}{\includegraphics{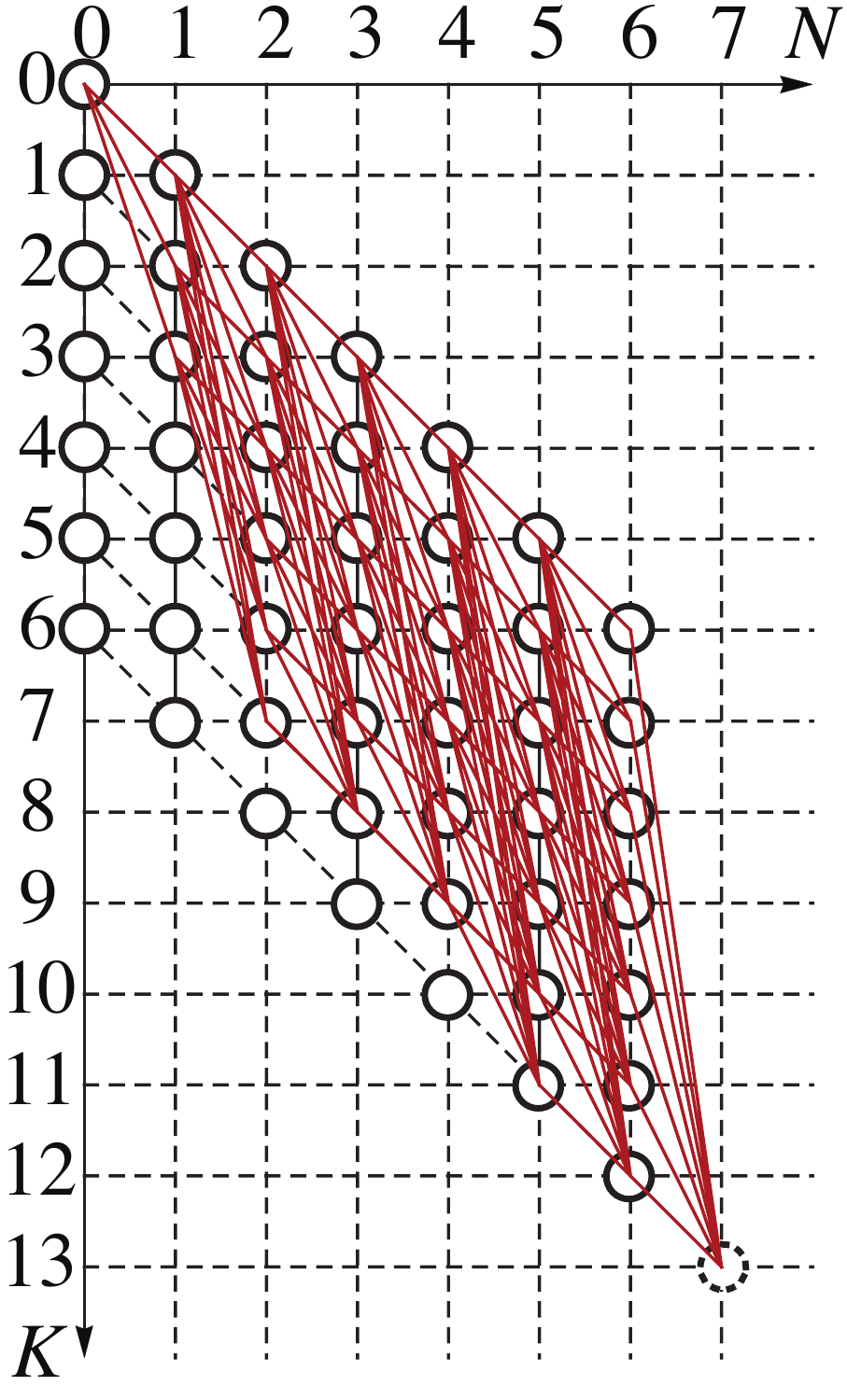}}} \\
(e) CAS & (f) MRCISD \\
{\resizebox{!}{0.2\textheight}{\includegraphics{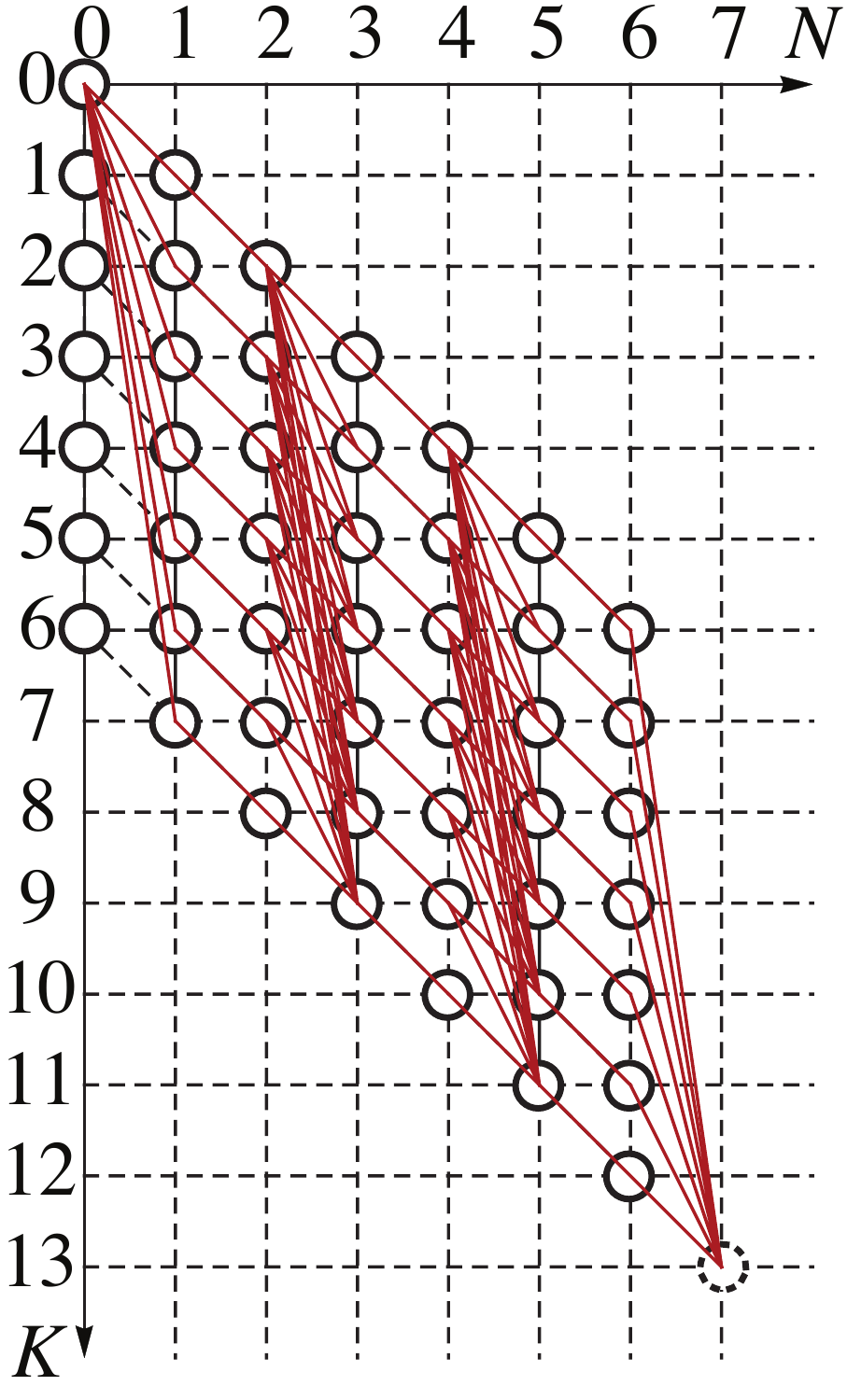}}} &
{\resizebox{!}{0.2\textheight}{\includegraphics{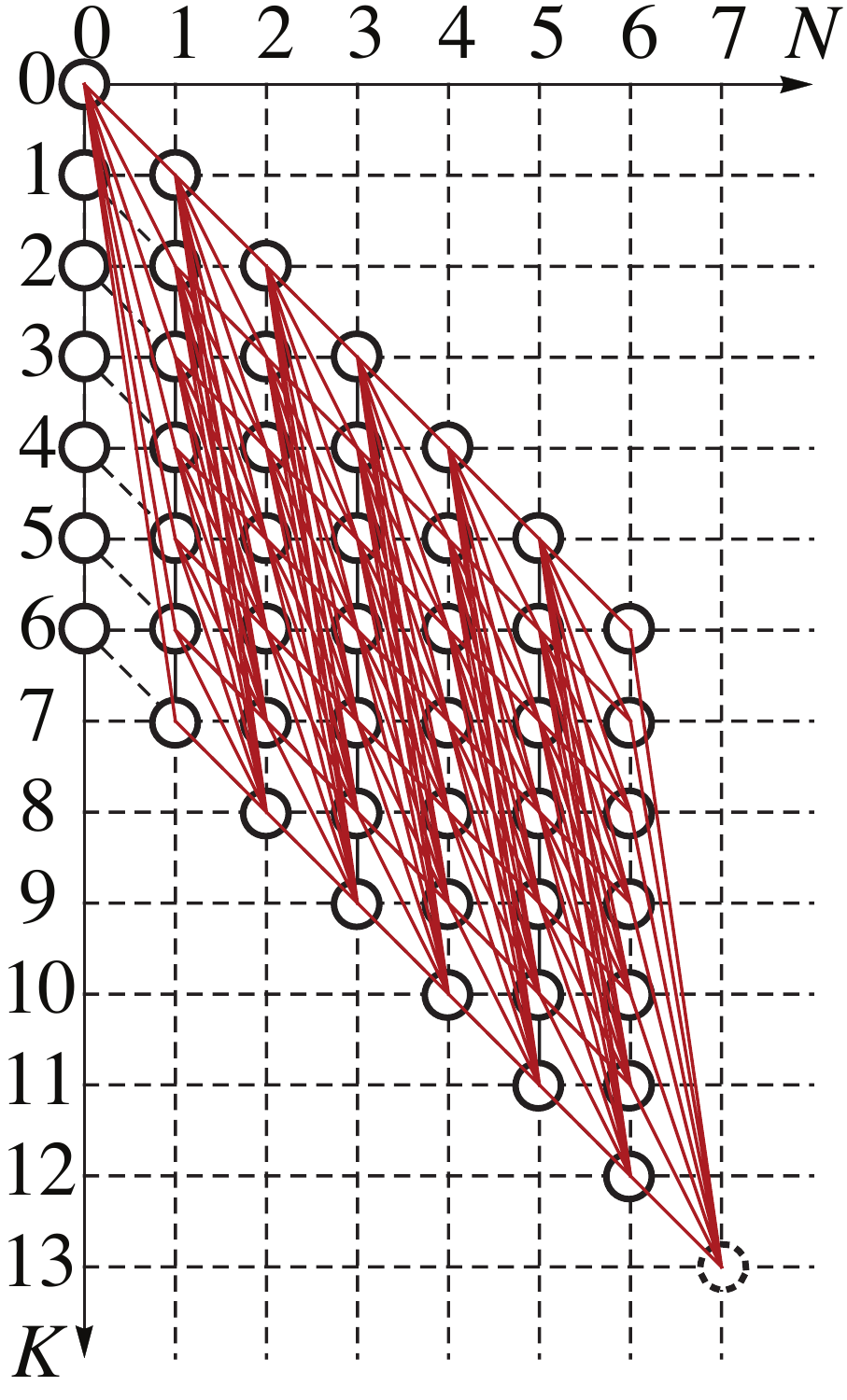}}} \\
(g) DOCI & (h) FCI \\
\end{tabular}
\caption{Graphical representations of various configuration spaces for $(K,N)=(12,6)$:
(a) Hartree-Fock determinant; (b) CI singles (CIS); (c) doubles; (d) CI singles and doubles (CISD); (e)
complete active space (CAS) with three electrons in six active spin-orbitals CAS(3e,6s); (f) multi-reference CISD (MRCISD)
based on this complete active space; (g) doubly occupied CI (DOCI); (h) the FCI space.}\label{fig:ci}
\end{figure}

A primary goal of developing the HS-MPS is to use it as a variational ansatz in the many-electron correlation problem.
For instance, the computation of the ground state can be recast into the minimization problem,
\begin{eqnarray}
E_0=\min_{\Psi\in\mathcal{M}}\frac{\langle\Psi|H|\Psi\rangle}{\langle\Psi|\Psi\rangle},\label{GSopt}
\end{eqnarray}
where $\mathcal{M}$ represents the manifold of HS-MPS.
Although the above discussions have been for the full configuration space,
it is rather straightforward to generalize the analysis to truncated configuration spaces.
Notably, this can be simply achieved via the graphical representation.
In Figure \ref{fig:ci}, various configuration spaces for the case $(K,N)=(12,6)$ are displayed.
These include the Hartree-Fock determinant, CI singles (CIS), doubles,
CI singles and doubles (CISD), complete active space
with three electrons in six active spin-orbitals CAS(3e,6s),
multi-reference CISD (MRCISD) based on this complete active space,
doubly occupied CI (DOCI)\cite{Weinhold1967}, and the FCI space. In these various cases,
the $\alpha$ and $\beta$ spin orbitals sharing the same spatial part are placed together,
and the ordering of the spatial orbitals in the CAS-based methods is: doubly occupied, active, and virtual
spin orbitals. Note that the order of the orbitals within each category does
not change the shapes of the graphs. According to Figure \ref{fig:ci}, various configuration
graphs differ in the accessible nodes and the links
among nodes. For instance, the nodes and links for CISD
are just the union of those for CIS and doubles, respectively.
The DOCI has the same accessible nodes as FCI, but the
possible links are restricted such that only the configurations
with doubly occupied spatial orbitals are present.
When applying the interpretation
of renormalization to these graphs, the accessible nodes define the
accessible values for the physical indices $p_i$ in $A^{p_i}[i]$ of HS-MPS.
The restrictions on links can be implemented with the help of the
orbital index 'symmetry' discussed in Sec. \ref{sec:decomp},
which means that in the construction of a tensor product space, the constraints on
possible couplings are used in addition to the ordering constraint.
In such a way, the HS-MPS can represent all the spaces shown in Figure \ref{fig:ci}.
Moreover, if a fixed bond dimension is used for all the virtual
indices as is usually employed in DMRG calculations, then low-rank tensor approximations to
these CI models naturally emerge from the HS-MPS. This is true for both
HS-MPS[p] and HS-MPS[h]. Note that in principle the FS-MPS can
also be used to represent these models, but this requires
introducing additional 'symmetry' labels that label states by their
number of particles and holes within the occupied, active, and virtual spaces separately.

\begin{figure}
\begin{tabular}{c}
{\resizebox{!}{0.15\textheight}{\includegraphics{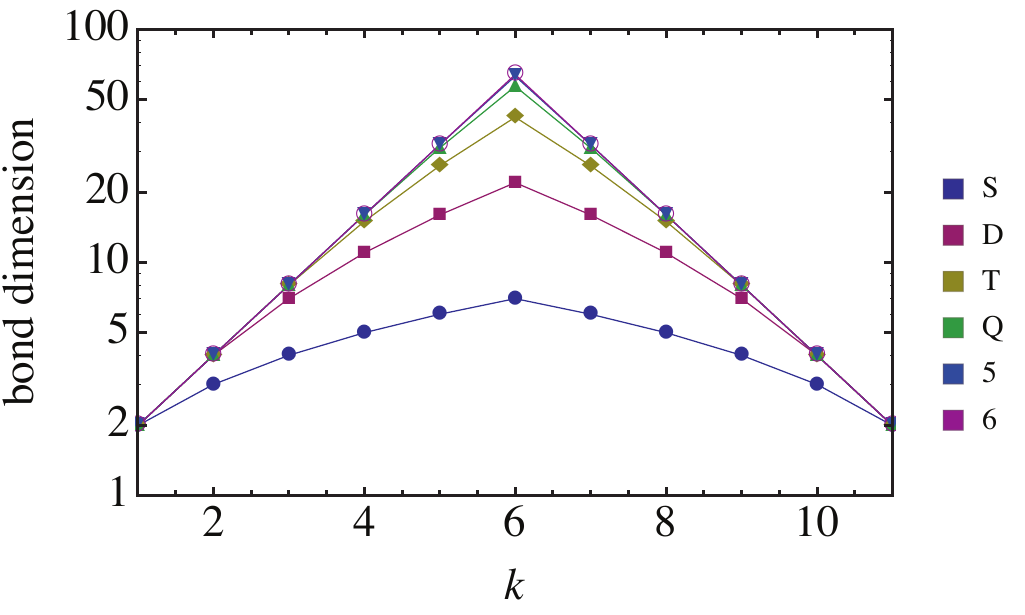}}} \\
(a) $D_k$ in FS-MPS  \\
{\resizebox{!}{0.15\textheight}{\includegraphics{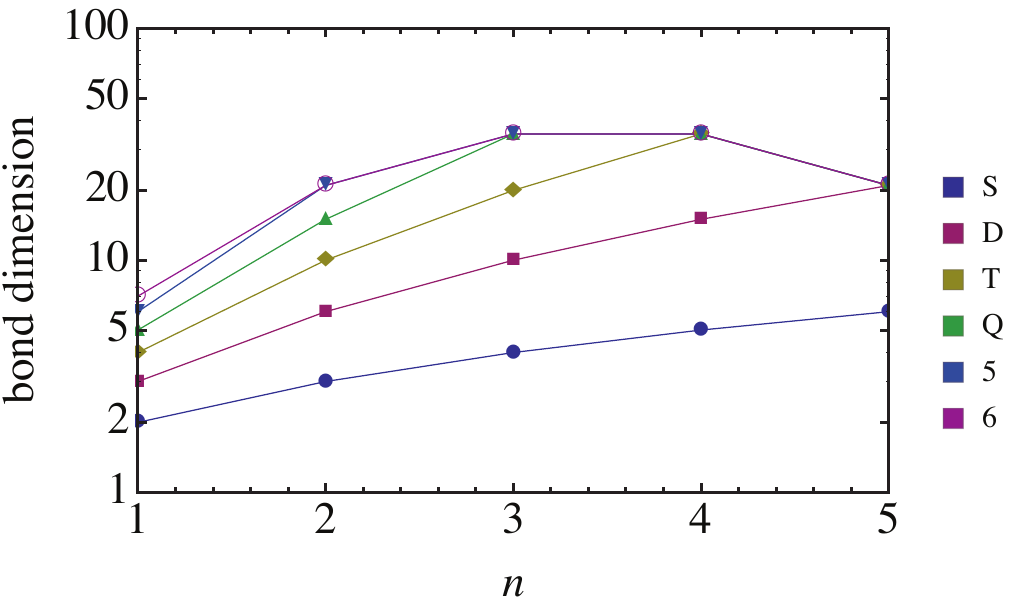}}} \\
(b) $D_n$ in HS-MPS \\
{\resizebox{!}{0.15\textheight}{\includegraphics{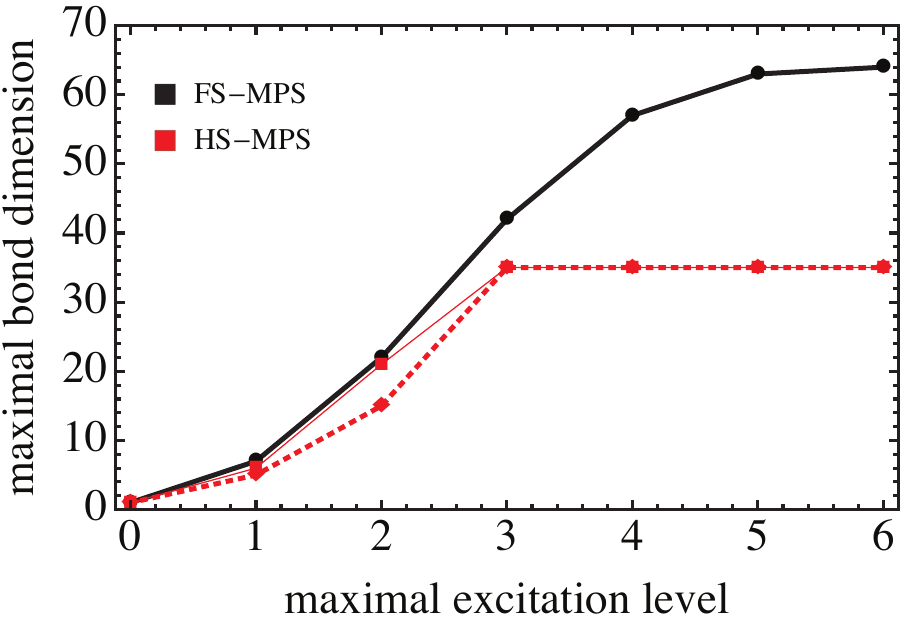}}} \\
(c) $D_{max}$ vs. CI level
\end{tabular}
\caption{Distributions of bond dimensions of FS/HS-MPS for different
CI levels in the half-filled case $(K,N)=(12,6)$.}\label{fig:bd1}
\end{figure}

The method for counting bond dimensions in Sec. \ref{sec:dims} also
applies to the HS-MPS for truncated CI models. In particular, the comparison between maximal bond dimensions for
truncated single-reference CI models helps to the better understandings of different performances of FS-MPS and HS-MPS.
We first consider the half-filled case ($K=2N$), where the bond dimensions are
the same for HS-MPS[p] and HS-MPS[h]. In Figures \ref{fig:bd1}(a) and \ref{fig:bd1}(b), the distributions
of bond dimensions of FS-MPS and HS-MPS for various CI models are illustrated for $(K,N)=(12,6)$,
respectively. For the FS-MPS case shown in Figure \ref{fig:bd1}(a), the distributions are all symmetric with the maximum
always located in the middle regardless of the excitation level. This
is the reason why in DMRG sweeps the minimal
energy is usually obtained in the middle\cite{zgid_obtaining_2008}, because this corresponds to optimizing the site
with the largest number of variational renormalized degrees of freedoms.
As shown in Figure \ref{fig:bd1}(b), the distributions for HS-MPS are drastically different,
as the maximal bond dimension is located at the right boundary for low excitation levels,
and gradually moves to the middle as the maximal excitation level increases.
Generally, for the half-filled case the maximal bond dimension of FS-MPS is found to be $2^N$, while that
of HS-MPS is found to be $2C_{N}^{(N-1)/2}$ for odd $N$ and
$C_{N+1}^{N/2}$ for even $N$, respectively. Clearly, all of these scale factorially with $N$.

Figure \ref{fig:bd1}(c) displays the increase of maximal bond dimension versus
the increase of the maximal excitation level,
which shows that the maximal bond dimension of the HS-MPS
is generally smaller than that of the FS-MPS at any excitation level at half-filling.
In particular, it becomes saturated after the maximal excitation
level exceeds $N/2$. This phenomenon can be explained by the plot shown in Figure \ref{fig:bd1}(b).
As long as the maximal excitation level is greater than $N/2$, only the bond dimensions for
those sites near the left boundary ($n<N/2$) are increased.
These observations lead to a very practical strategy to maximize
the accuracy and computational efficiency of HS-MPS:
We can merge the last two sites of HS-MPS into a larger two-site tensor $A^{p_{N-1}p_{N}}[N-1,N]$ and treat it exactly in DMRG sweep optimizations, such that in the
CISD case the wavefunctions can be represented exactly with
a bond dimension equal to the number of electron pairs $N(N-1)/2$. This should be compared with the FS-MPS case,
where the necessary bond dimension to recover the CISD limit scales
as $\mathcal{O}(K)$.
The effect of this strategy is shown in
the red dashed line in Figure \ref{fig:bd1}(c), which shows that the bond
dimensions to represent CIS and CISD are reduced, while the
computational cost is not greatly increased.

\begin{figure}
\begin{tabular}{c}
{\resizebox{!}{0.15\textheight}{\includegraphics{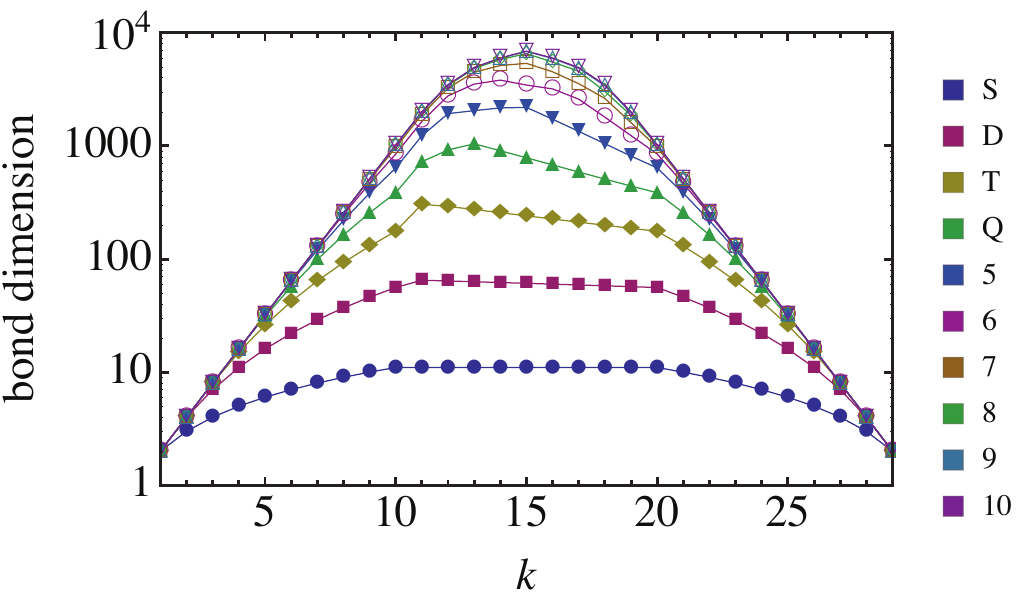}}} \\
(a) $D_k$ in FS-MPS \\
{\resizebox{!}{0.15\textheight}{\includegraphics{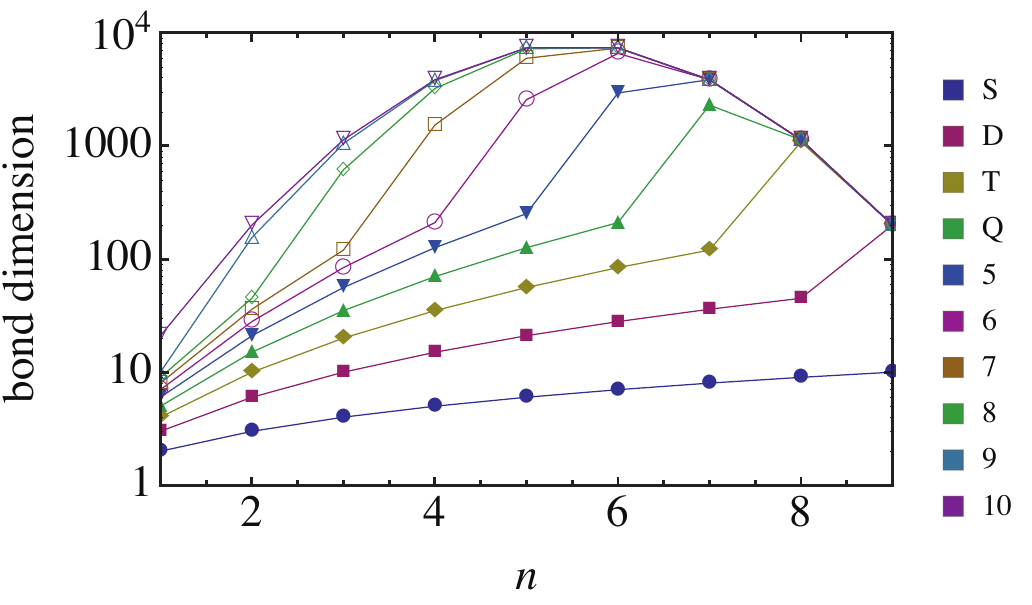}}} \\
(b) $D_n$ in HS-MPS[p] \\
{\resizebox{!}{0.15\textheight}{\includegraphics{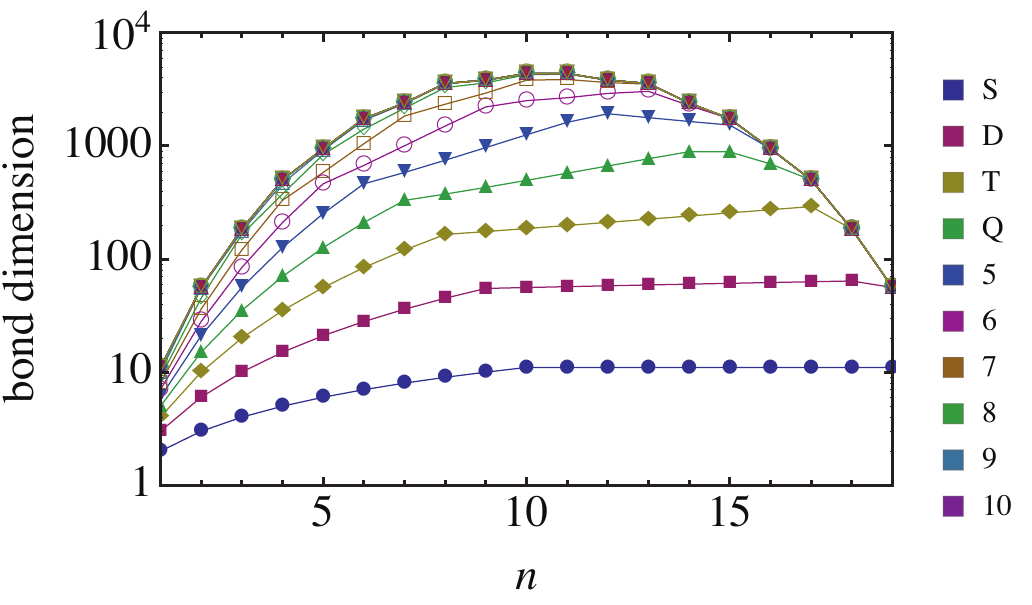}}} \\
(c) $D_n$ in HS-MPS[h] \\
{\resizebox{!}{0.15\textheight}{\includegraphics{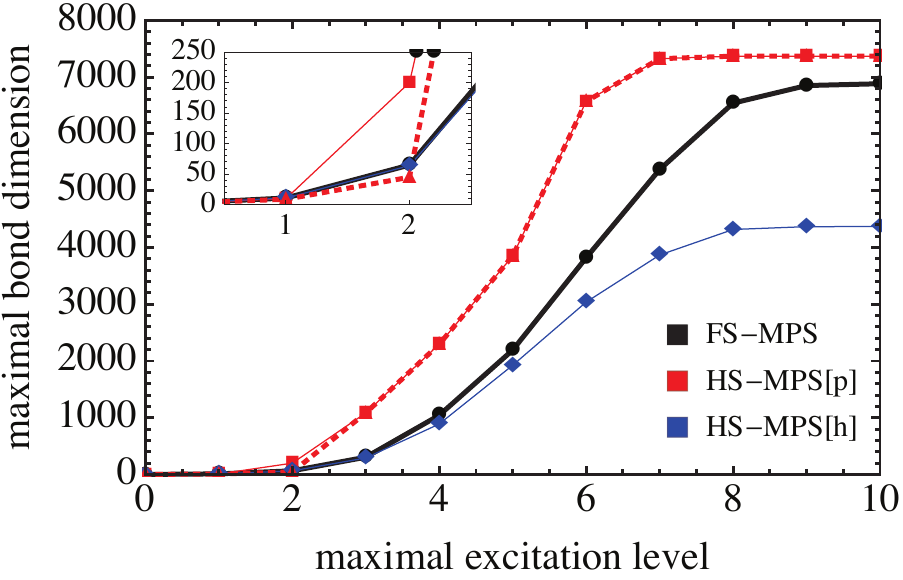}}} \\
(d) $D_{max}$ vs. CI level \\
\end{tabular}
\caption{Distributions of bond dimensions of FS/HS-MPS for different
CI levels in the non-half-filled case $(K,N)=(30,10)$.}\label{fig:bd2}
\end{figure}

For  cases away from half-filling, the HS-MPS[p] and
HS-MPS[h] have different structures. Figure \ref{fig:bd2} shows the distributions
of bond dimensions for FS-MPS, HS-MPS[p] and HS-MPS[h] in the case of $(K,N)=(30,10)$.
The bond dimensions of FS-MPS for truncated CI models now
become less symmetric as shown in Figure \ref{fig:bd2}(a) due to the
fact that the configuration graphs for truncated CI are not symmetric
away from the half-filling. In the low-filling limit, the maximal bond dimensions
at a given CI level follow the ordering:
HS-MPS[p] $>$ FS-MPS $>$ HS-MPS[h], suggesting that using the
HS-MPS[h] is more effective to represent the CI spaces. For HS-MPS[p], if as mentioned above the last two sites are treated
together, then it will be more efficient in recovering the CISD limit (see inset in Figure \ref{fig:bd2}(d)).
However, when higher excitations are targeted, HS-MPS[p]
becomes less effective. From these two examples with different fillings, we conclude that
the relative strengths of the two MPS representations (FS-MPS and HS-MPS)
in general depend on the values of $(K,N)$, and also on the actual distribution
of configuration coefficients in the CI spaces.

\subsection{Evaluation of matrix elements}\label{sec:matrixElements}
Having defined various manifolds of HS-MPS, now we turn to the evaluations
of the matrix elements $\langle\Psi|H|\Psi\rangle$ and $\langle\Psi|\Psi\rangle$ in Eq. \eqref{GSopt}.
This is achieved by generalizing the complementary operator approach used in the Fock-space DMRG\cite{Xiang1996,white_ab_1999,chan_highly_2002}.
The simplicity of derivations in FS-MPS case is due to the factorization
for the overlap between two basis vectors
\begin{eqnarray}
\langle n_1'n_2'\cdots n_K'| n_1n_2\cdots n_K\rangle
=\delta_{n_1'n_1}\delta_{n_2'n_2}\cdots
\delta_{n_K'n_K}.\label{overlap1}
\end{eqnarray}
By direct applying this fundamental property, it is
easy to show that for renormalized states defined by
\begin{eqnarray}
|l_k\rangle&=&\sum_{n_1\cdots n_k}
|n_1\cdots n_k\rangle L^{n_1\cdots n_k},\\
|r_k\rangle&=&\sum_{n_{k+1}\cdots n_K}
|n_{k+1}\cdots n_K\rangle R^{n_{k+1}\cdots n_K},
\end{eqnarray}
the following factorization relation holds for
their overlap
\begin{eqnarray}
\langle l'_k r_k'|l_k r_k\rangle=\langle l'_k|l_k\rangle
\langle r'_k|r_k\rangle,\label{factorization1}
\end{eqnarray}
which breaks the computation of overlaps into products
of smaller pieces that can be computed recursively for FS-MPS.
In the Hilbert-space case, for two ordered orbital strings $p_1'p_2'\cdots p_N'$ and $p_1p_2\cdots p_N$,
a similar factorization still holds
\begin{eqnarray}
\langle p_1'p_2'\cdots p_N'|p_1p_2\cdots p_N\rangle
=\delta_{p_1'p_1}\delta_{p_2'p_2}\cdots \delta_{p_N'p_N}.\label{overlap2}
\end{eqnarray}
However, in general for renormalized states defined by
\begin{eqnarray}
|l_i\rangle&=&\sum_{p_1<\cdots< p_i}
|p_1\cdots p_i\rangle L^{p_1\cdots p_i},\\
|r_i\rangle&=&\sum_{p_{i+1}<\cdots <p_N}
|p_{i+1}\cdots p_N\rangle R^{p_{i+1}\cdots p_N},
\end{eqnarray}
an analogous relation to Eq. \eqref{factorization1} does not hold.
A counterexample can be simply given as follows:
\begin{eqnarray}
|l_i'r_i'\rangle&\triangleq&(|12\rangle+|56\rangle)\wedge(|34\rangle+|78\rangle),\nonumber\\
|l_ir_i\rangle&\triangleq&(|12\rangle+|34\rangle)\wedge(|56\rangle+|78\rangle),\nonumber\\
\langle l_i'r_i'|l_ir_i\rangle&=&\langle 1278|1278\rangle
+\langle 5634|3456\rangle\nonumber\\
&\ne& \langle 12|12\rangle\langle 78|78\rangle=\langle l_i'|l_i\rangle\langle r_i'|r_i\rangle,\label{Contradict}
\end{eqnarray}
where $\wedge$ indicates the wedge product of two states in the Hilbert-space case as dictated by the antisymmetry.
The violation is due to the exchange of two configuration substrings $|5634\rangle=|3456\rangle$
in $|l_i'r_i'\rangle$; such operations are obviously forbidden in Fock space by definition.
Fortunately, if $|l_i\rangle$ and $|r_i\rangle$ are suffix and prefix renormalized states, respectively,
and their combination satisfies the ordering requirements ($p_{i}'<p_{i+1}'$ and $p_{i}<p_{i+1}$),
then it can be shown that the simple factorization relation holds
\begin{eqnarray}
\langle l'_i r_i'|l_i r_i\rangle=\langle l'_i|l_i\rangle
\langle r'_i|r_i\rangle,\label{factorization2}
\end{eqnarray}
by directly applying Eq. \eqref{overlap2}. These conditions simply exclude the situation shown in Eq. \eqref{Contradict}.
The fundamental relation \eqref{factorization2} will be extensively
used in the following derivation of matrix elements.

To illustrate how the matrix elements of the Hamiltonian can be evaluated
in HS-MPS case, the one-electron Hamiltonian is taken as an example.
Given the bra state $\langle l_i' r_i'|$ and the ket state $|l_i r_i\rangle$,
their 'symmetry' indices define two partitions for the orbital indices, which
can be denoted by $L'R'$ and $LR$, respectively. The matrix element over $H_1$
can thus be separated into a sum of four parts,
\begin{eqnarray}
\langle l_i' r_i'|H_1|l_i r_i\rangle
=
\sum_{pq\in (L'+R')(L+R)}
h_{pq}\langle l_i' r_i'|a_p^\dagger a_q|l_i r_i\rangle.\label{H1eT4}
\end{eqnarray}
The summations over $pq\in\{L'L,R'R\}$ can be evaluated by using Eq. \eqref{factorization2} directly, e.g.,
\begin{eqnarray}
&&\sum_{pq\in L'L}
h_{pq}\langle l_i' r_i'|a_p^\dagger a_q|l_i r_i\rangle\nonumber\\
&=&
\sum_{pq\in L'L} h_{pq}\langle (a_p l_i') r_i'| (a_ql_i) r_i\rangle\nonumber\\
&=&
\sum_{pq\in L'L} h_{pq}\langle (a_p l_i')| (a_ql_i)\rangle\langle  r_i'| r_i\rangle\nonumber\\
&=&
\langle l_i'|H_1|l_i\rangle\langle  r_i'| r_i\rangle,\label{H1sum1}
\end{eqnarray}
whereas the summations over $pq\in\{L'R,R'L\}$
require the expansion of the left and right renormalized states
in order to apply Eq. \eqref{factorization2}, e.g.,
\begin{eqnarray}
&&\sum_{pq\in L'R}
h_{pq}\langle l_i' r_i'|a_p^\dagger a_q|l_i r_i\rangle\nonumber\\
&=&
\sum_{pq\in L'R} h_{pq}(-1)^i\langle (a_p l_i') r_i'|l_i (a_qr_i)\rangle\nonumber\\
&=&
\sum_{pq\in L'R}\sum_{p'_{i+1}r'_{i+1}}\sum_{l_{i-1}p_{i}} h_{pq}(-1)^i A^{p'_{i+1}*}_{r'_{i}r'_{i+1}}[i+1] \nonumber\\
&&\times\langle (a_p l_i') p'_{i+1}r'_{i+1}|l_{i-1}p_{i}(a_qr_i)\rangle A^{p_{i}}_{l_{i-1}l_{i}}[i]  \nonumber\\
&=&
\sum_{pq\in L'R}\sum_{p'_{i+1}r'_{i+1}}\sum_{l_{i-1}p_{i}} h_{pq}(-1)^i  A^{p'_{i+1}*}_{r'_{i}r'_{i+1}}[i+1]\nonumber\\
&&\times\langle (a_p l_i')|l_{i-1}\rangle
\langle p'_{i+1}|p_{i}\rangle\langle r'_{i+1}|(a_qr_i)\rangle A^{p_{i}}_{l_{i-1}l_{i}}[i]\nonumber\\
&=&
(-1)^i\sum_{p}\sum_{x}
\left(\sum_{l_{i-1}} \langle l_i'|a_p^\dagger|l_{i-1}\rangle A^{x}_{l_{i-1}l_{i}}[i] \right)\nonumber\\
&&\times\left(\sum_{r'_{i+1}}A^{x*}_{r'_{i}r'_{i+1}}[i+1]\langle r'_{i+1}|S_p|r_i\rangle\right),\label{H1sum2}
\end{eqnarray}
where the one-electron complementary operator $S_p$ is defined by $S_p\triangleq\sum_q h_{pq}a_q$.
Note that in both Eqs. \eqref{H1sum1} and \eqref{H1sum2},
the summation restrictions on $p$ and $q$ have been eliminated in the final expressions, as
they are now implicitly imposed by the nonzero conditions of the matrix elements such as
$\langle l_i'|a_p^\dagger|l_{i-1}\rangle$. Thus, Eq. \eqref{H1eT4} can be finally written as
\begin{eqnarray}
\langle l_i' r_i'|H_1|l_i r_i\rangle
&=&
\langle l_i'|H_1|l_i\rangle\langle r_i'|r_i\rangle
+
\langle l_i'|l_i\rangle\langle r_i'|H_1|r_i\rangle\nonumber\\
&+&\;(-1)^i\sum_{p}\sum_{x}
\left[
\left(
\sum_{l_{i-1}}\langle l_i'|a_p^\dagger|l_{i-1}\rangle
A^{x}_{l_{i-1}l_{i}}[i]\right)\right.\nonumber\\
&&\times\left.\left(
\sum_{r'_{i+1}}A^{x*}_{r'_i r'_{i+1}}[i+1]\langle r_{i+1}'|S_p|r_{i}\rangle
\right)+c.c.\right].
\end{eqnarray}
In the same way, the expression
for the total Hamiltonian \eqref{Hnp} can be found to be
\begin{widetext}
\begin{eqnarray}
\langle l_i' r_i'|H|l_i r_i\rangle
&=&
\langle l_i'|H|l_i\rangle\times\langle r_i'|r_i\rangle
+
\langle l_i'|l_i\rangle\times
\langle r_i'|H|r_i\rangle
+(-1)\sum_{pr}
\langle l_i'|a_p^\dagger a_r|l_i\rangle
\times
\langle
r_i'|Q_{pr}|r_i\rangle\nonumber\\
&+&
\left[(-1)^{i}\sum_{p}\sum_x
\langle l_i'|a_p^\dagger|l_{i-1}\rangle
A^x_{l_{i-1}l_i}
\times
\left(
A^{x*}_{r_{i}'r_{i+1}'}\langle r_{i+1}'|\frac{1}{2}S_p+S^R_p|r_i\rangle
-\sum_q
(A^{q*}A^{x*})_{r_{i}'r_{i+2}'}
\langle r_{i+2}'|P_{pq}|r_i\rangle
\right)\right.\nonumber\\
&&+\;
(-1)^{i}
\sum_q\sum_{x}
\left(
A_{l_{i-1}'l_{i}'}^{x*}
\langle l_{i-1}'|\frac{1}{2}S_q-S_q^L|l_i\rangle
-(-1)^{i-1}\sum_p
(A^{x*}A^{p*})_{l_{i-2}'l_{i}'}\langle l_{i-2}'|P_{pq}|l_i\rangle
\right)\times \langle r_i'|a_q^\dagger|r_{i+1}\rangle A^{x}_{r_i r_{i+1}}\nonumber\\
&&+\;\left.
\sum_{(pq)}\sum_{xy}\langle l_i'|a_p^\dagger a_q^\dagger|l_{i-2}\rangle (A^{x}A^{y})_{l_{i-2}l_{i}}
\times (A^{x*}A^{y*})_{r_{i}'r_{i+2}'}
\langle r_{i+2}'|P_{pq}|r_i\rangle
+c.c.
\right],\label{H2e}
\end{eqnarray}
\end{widetext}
where we have omitted the explicit summation notation for $l_{i-1}$, $r_{i+1}'$, $l_{i-2}$, etc.,
as well as the site index $k$ in $A[k]$ for brevity, and only the summations over
orbital indices are explicitly retained. The additional complementary operators in Eq. \eqref{H2e} are defined by
\begin{eqnarray}
Q_{pr}&=&\sum_{qs}v_{pqrs}a_q^\dagger a_s,\\
P_{pq}&=&\sum_{rs}v_{pqrs}a_r a_s,\\
S_{p}^R&=&\sum_{qrs}v_{pqrs}a_q^\dagger a_r a_s,\\
S_{q}^L&=&\sum_{prs}v_{pqrs}a_p^\dagger a_r a_s,
\end{eqnarray}
with $v_{pqrs}$ related to the antisymmetrized two-electron integral
$\langle pq||rs\rangle$ by
\begin{eqnarray}
v_{pqrs}
=\left\{
\begin{array}{cc}
-\langle pq||rs\rangle,& p<q,\;r<s\\
0,& otherwise
\end{array}\right..
\end{eqnarray}
In summary, the Hamiltonian matrix \eqref{H2e} can be factorized into a sum of products just
as in the Fock-space case. Albeit with a very complicated form, the physical meaning of Eq. \eqref{H2e}
is quite clear: the first two terms are 'local' terms,
the third term describes density-density (Coulomb and exchange)
interaction, and the last three terms describe interactions due to
either one-electron or two-electron charge transfers between
left and right renormalized states. They all have counterparts in the FS-MPS case\cite{chan_highly_2002}.
However, the difference is that here the expression \eqref{H2e} explicitly involves the site tensors $A[i-1]$, $A[i]$, $A[i+1]$, and
$A[i+2]$, which makes the computation more complicated than the Fock-space case, as
discussed in the next section.

\subsection{Variational optimization by DMRG algorithm}\label{sec:varopt}
To perform the optimization in Eq. \eqref{GSopt}, the DMRG algorithm can be generalized to HS-MPS.
The basic idea of DMRG is to optimize the site tensors one-by-one, that is,
using a sweep algorithm that optimizes the sites from left to right, and then right to left, until
the energy difference between two sweeps converges to below a predefined threshold.
For simplicity, we will only discuss the so-called one-site algorithm, and the extension to a two-site algorithm
is straightforward. In the one-site algorithm, the tensor $A^{p_i}_{l_{i-1}l_i}[i]$, or in other words
the Hilbert-space counterpart of the mapping Eq. \eqref{Ak}, is
optimized while keeping all other sites fixed.
According to \eqref{GSopt}, this amounts to solving a quadratic optimization problem for
$A[i]$ under the left canonical constraint
\begin{eqnarray}
\sum_{l_{i-1}p_i}A^{p_i*}_{l_{i-1}l_i}[i]A^{p_i}_{l_{i-1}l_i'}[i]=\delta_{l_{i}l_i'}.\label{leftCanon}
\end{eqnarray}
Direct gradient-based optimizations are possible for such kind of problems\cite{manopt}, but in DMRG this optimization subproblem
is solved elegantly via three steps: blocking, solving a CI problem, and decimation.

\begin{figure}
\begin{tabular}{c}
{\resizebox{!}{0.22\textheight}{\includegraphics{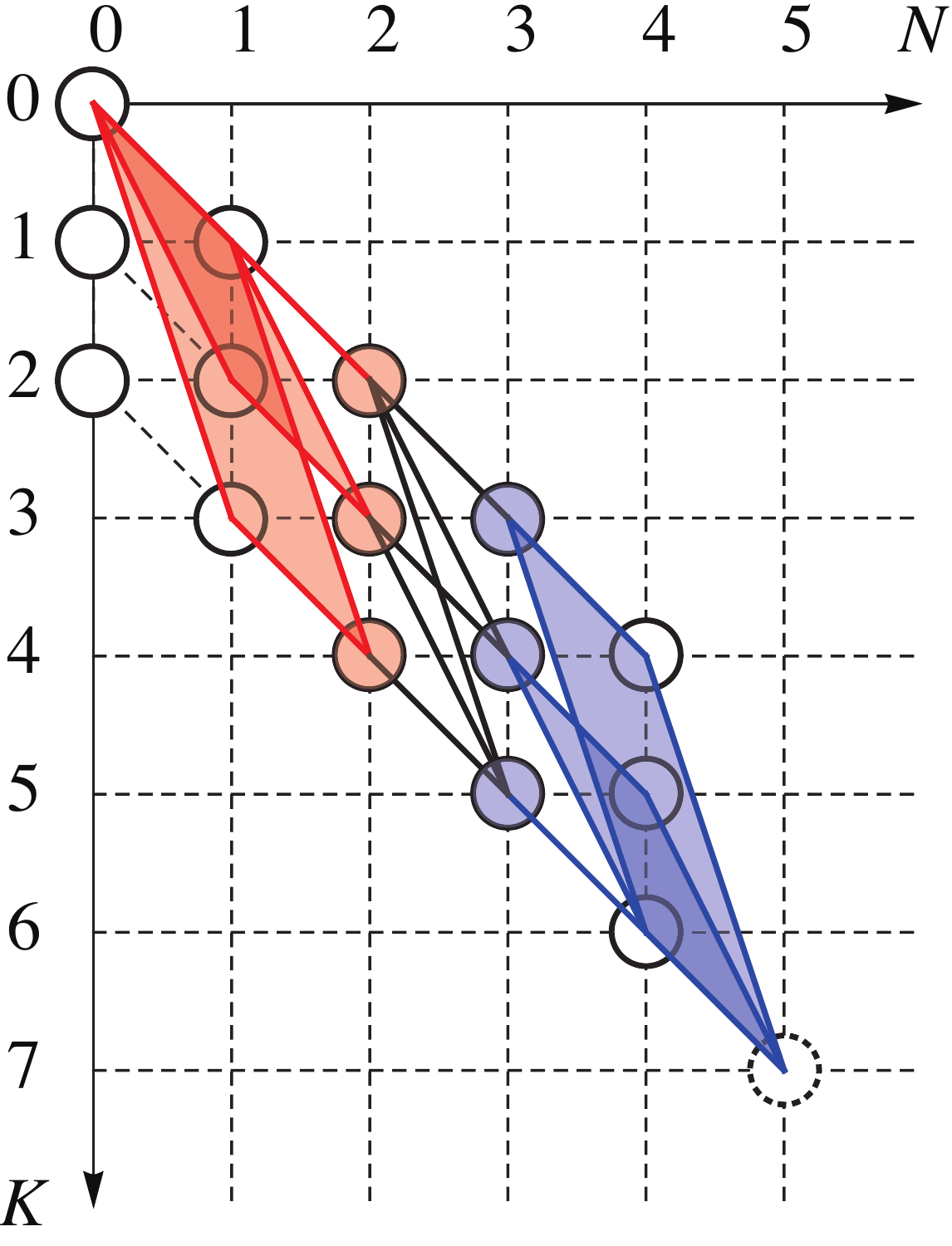}}}\\
\end{tabular}
\caption{Contracted configuration space spanned by
$\{|l_{i-1}p_i r_{i}\rangle\}$ for $(K,N)=(6,4)$ and $i=3$.
The red circles on the layer $N=2$ represent the subspaces spanned by $\{|l_{i-1}\rangle\}$.
The blue circle on the layer $N=3$ with $K=p_i$ represents the subspace spanned by
the direct product of the orbital $p_i$ with all $\{|r_{i}\rangle\}$.
All the orbital index 'symmetry'-allowed couplings between $\{|l_{i-1}\rangle\}$ and $\{|p_i\rangle\}$ to form a superblock space $\{|L_i\rangle\triangleq|l_{i-1}p_{i}\rangle\}$ are represented by the paths (black lines) between the layers $N=2$ and $N=3$.}\label{fig:contractedci}
\end{figure}

The central CI part amounts to solving a CI problem in the space spanned by the configurations $\{|l_{i-1}p_i r_{i}\rangle\}$,
\begin{eqnarray}
\sum_{l_{i-1}p_i r_{i}}\langle l'_{i-1}p'_i r'_{i}|H|l_{i-1}p_i r_{i}\rangle
\Psi^{l_{i-1}p_{i}r_{i}}=E
\Psi^{l'_{i-1}p'_{i}r'_{i}},\label{CIsub}
\end{eqnarray}
where $|l_{i-1}\rangle$ and $|r_{i}\rangle$ are the left (suffix) and right (prefix) renormalized states with $i-1$ and $N-i$ electrons, respectively. Graphically, this space is nothing but
a contracted CI space as illustrated in Figure \ref{fig:contractedci} for $(K,N)=(6,4)$ and $i=3$,
where the colored regions represent the left (red) and right (blue) renormalized states.
To solve this problem, in the blocking step the proper superblock space $\{|L_i\rangle\triangleq|l_{i-1}p_{i}\rangle\}$
is first formed by combining $\{|l_{i-1}\rangle\}$ and $\{|p_i\rangle\}$ under the
ordering constraint by using the orbital index 'symmetry'. Then the standard eigenvalue problem \eqref{CIsub}
is solved with the Hamiltonian matrix elements computed via Eq. \eqref{H2e}.
For computational efficiency and memory savings, the matrix representations of
operators in the superblock space, e.g.,
$\langle L_{i}'|a_p^\dagger a_r|L_{i}\rangle\triangleq\langle l_{i-1}'p_i'|a_p^\dagger a_r|l_{i-1}p_i\rangle$,
are never formed explicitly. They are contracted with the trial vectors
denoted by $c_{l_{i-1}p_{i},r_{i}}$ in the direct CI algorithm, e.g.,
\begin{eqnarray}
\sigma_{l_{i-1}'p_{i}',r_{i}}^{(pr)}&\triangleq&
\sum_{l_{i-1}p_{i}}\langle l_{i-1}'p_i'|a_p^\dagger a_r|l_{i-1}p_i\rangle c_{l_{i-1}p_{i},r_{i}}\label{bk1}\\
&=&
\sum_{l_{i-1}}\langle l_{i-1}'|a_p^\dagger a_r|l_{i-1}\rangle
c_{l_{i-1}p_{i}',r_{i}}
-\delta_{pp_i'}c_{l_{i-1}'r,r_{i}}\nonumber\\
&+&\;(-1)^{i-1}
[\delta_{pp_i'}
\sum_{l_{i-1}p_i}\langle l_{i-1}'|a_r|l_{i-1}p_{i}\rangle c_{l_{i-1}p_{i},r_{i}}\nonumber\\
&&+\sum_{l_{i-1}}\langle l_{i-1}'p_{i}'|a_p^\dagger|l_{i-1}\rangle
c_{l_{i-1}r,r_{i}}].\label{bk2}
\end{eqnarray}
This reduces the cost from $\mathcal{O}(K^4D^3)$ in Eq. \eqref{bk1} to
$\mathcal{O}(K^3D^3)$ in Eq. \eqref{bk2}, assuming $K\gg N$
and the dimensions of $\{|l_{i-1}\rangle\}$ and $\{|r_{i}\rangle\}$ are both $D$.
This kind of reduction which employs the sparse structure of
$\langle l_{i-1}'p_i'|a_p^\dagger a_r|l_{i-1}p_i\rangle$ is used extensively
in the present algorithm, such that the computational cost for the matrix-vector product
in the CI problem in the one-site algorithm scales as $\mathcal{O}(K^{3}D^3)$.
After the wavefunction $\Psi^{l_{i-1}p_{i}r_{i}}$ has been obtained by solving the CI problem,
the algorithms in Sec. \ref{sec:decomp} can be employed to obtain
$A^{p_i}_{l_{i-1}l_i}[i]$. In this decimation step, the pseudo-density matrix approach is used,
which allows to perform state-average calculations for both ground and excited states and
also to add noise to avoid getting stuck in local minimum during the optimizations.
With the newly obtained tensor $A^{p_i}_{l_{i-1}l_i}[i]$, the renormalized operators defined in the
contracted space $\{|l_{i}\rangle\}$ are also computed directly
without constructing the superblock operators. The most time consuming
part in this step comes from the renormalization of complementary operators, e.g.,
\begin{eqnarray}
\langle l_{i}'|Q_{qs}|l_{i}\rangle
&\triangleq&
\sum_{l_{i-1}'p_{i}',l_{i-1}p_{i}}
A^{p_i'*}_{l_{i-1}'l_{i}'}
\langle l_{i-1}'p_i'|Q_{qs}|l_{i}p_{i}\rangle
A^{p_i}_{l_{i-1}l_{i}}\label{rm1}\\
&=&
\sum_{l_{i-1}p_i}(\sum_{l_{i-1}'}A^{p_i*}_{l_{i-1}'l_{i}'}
\langle l_{i-1}'|Q_{qs}|l_{i-1}\rangle)
A^{p_i}_{l_{i-1}l_{i}}\nonumber\\
&-&
\sum_{l_{i-1}'p_{i}}(\sum_{p_i'}A^{p_i'*}_{l_{i-1}'l_{i}'}v_{p_{i}'qp_{i}s})A^{p_i}_{l_{i-1}'l_{i}}\nonumber\\
&+&
(-1)^{i-1}
[
\sum_{l_{i-1}p_i}(\sum_{p}\langle l_{i}'|a_p^\dagger|l_{i-1}\rangle v_{pqp_is})A^{p_i}_{l_{i-1}l_{i}}\nonumber\\
&&+
\sum_{l_{i-1}'p_i'}A^{p_i'*}_{l_{i-1}'l_{i}'}
(\sum_{r}v_{p_i'qrs}\langle l_{i-1}'|a_r|l_{i}\rangle)],\label{rm2}
\end{eqnarray}
where the cost has been reduced from $\mathcal{O}(K^4D^3+K^3D^3)$ in Eq. \eqref{rm1}
to $\mathcal{O}(K^3D^3+K^4D^2)$ in Eq. \eqref{rm2}.

In summary, since there are $N$ sites that need to be optimized, the computational cost of one sweep in the present one-site algorithm
scales as $\mathcal{O}\left(N(K^3D^3+K^4D^2)\right)$. In comparison, the corresponding cost for Fock-space DMRG
scales $\mathcal{O}(K^3D^3+K^4D^2)$, which is cheaper by a factor of $N$.
This is mainly due to the fact that the dimension of the physical index in HS-MPS is
larger than that in FS-MPS: The former scales as $\mathcal{O}(K)$ while the latter is just a constant 2
in spin-orbital basis. This makes the two-site algorithm for HS-MPS even
more expensive, since the dimension of the CI subspace spanned by
$\{|l_{i-1}p_{i}p_{i+1}r_{i+1}\rangle\}$ becomes $\mathcal{O}(K^2D^2)$,
while in the Fock-space case the corresponding CI dimension is $\mathcal{O}(D^2)$
for the space spanned by $\{|l_{k-1}n_{k}n_{k+1}r_{k+1}\rangle\}$.
Therefore, in the following calculations the one-site algorithm is employed for HS-MPS.

The use of Abelian symmetries ($S_z$ spin projection and $D_{2h}$ point group symmetry) is rather
straightforward. There are two ways to implement $S_z$ symmetry. One is to simply separate the $\alpha$
and $\beta$ orbitals into two parts, such that the total wavefunction
can be represented by a rectangular matrix $\Psi^{I_{\alpha}J_{\beta}}$,
with $I_\alpha$ and $J_\beta$ are $\alpha$
and $\beta$ strings. Then we can represent both the $\alpha$ and
$\beta$ string spaces by HS-MPS, that is to use two configuration graphs for different spins,
and finally connect them by an virtual index $\gamma$, viz.,
$\Psi^{I_{\alpha}J_{\beta}}=\sum_{\gamma}s_{\gamma}\Psi^{I_{\alpha}}_\gamma
\Psi^{J_{\beta}}_\gamma$ in the SVD form. This works for FCI, but is not
quite suitable for the truncated CI models shown in Figure \ref{fig:ci}.
The other more general way to is to use the orbital ordering
in Figure \ref{fig:ci}, and label the renormalized states by a tuple
containing the orbital index 'symmetry', $S_z$ value, and the irreducible
representation of the point group, etc.,  similar to the usage of Abelian symmetries in the FS-MPS case\cite{chan_highly_2002}.
The only difference is that the particle number symmetry need not to be considered in HS-MPS,
since it has already been taken care of by the Hilbert-space formulation.
With this symmetry information, the symmetry-allowed couplings for two
spaces in the direct product procedure can be determined, such that
the CI space $\{|l_{i-1}p_i r_i\rangle\}$ in the optimization can be symmetry-adapted.
In the renormalization step, as the pseudo-density matrix is always totally symmetric
in the case of Abelian symmetry, it is block diagonal in the superblock space and compatible
with the prefix/suffix constraint. Thus, the resulting new renormalized
states still carry well defined symmetry properties. The adaptation to
non-Abelian symmetry will be discussed in Sec. \ref{sec:generalization},
although we have not implemented in this work.

\subsection{Further generalizations}\label{sec:generalization}
Several generalizations of the HS-MPS and the prefix/suffix renormalization
can be envisaged.

\begin{figure}
\begin{tabular}{c}
{\resizebox{!}{0.18\textheight}{\includegraphics{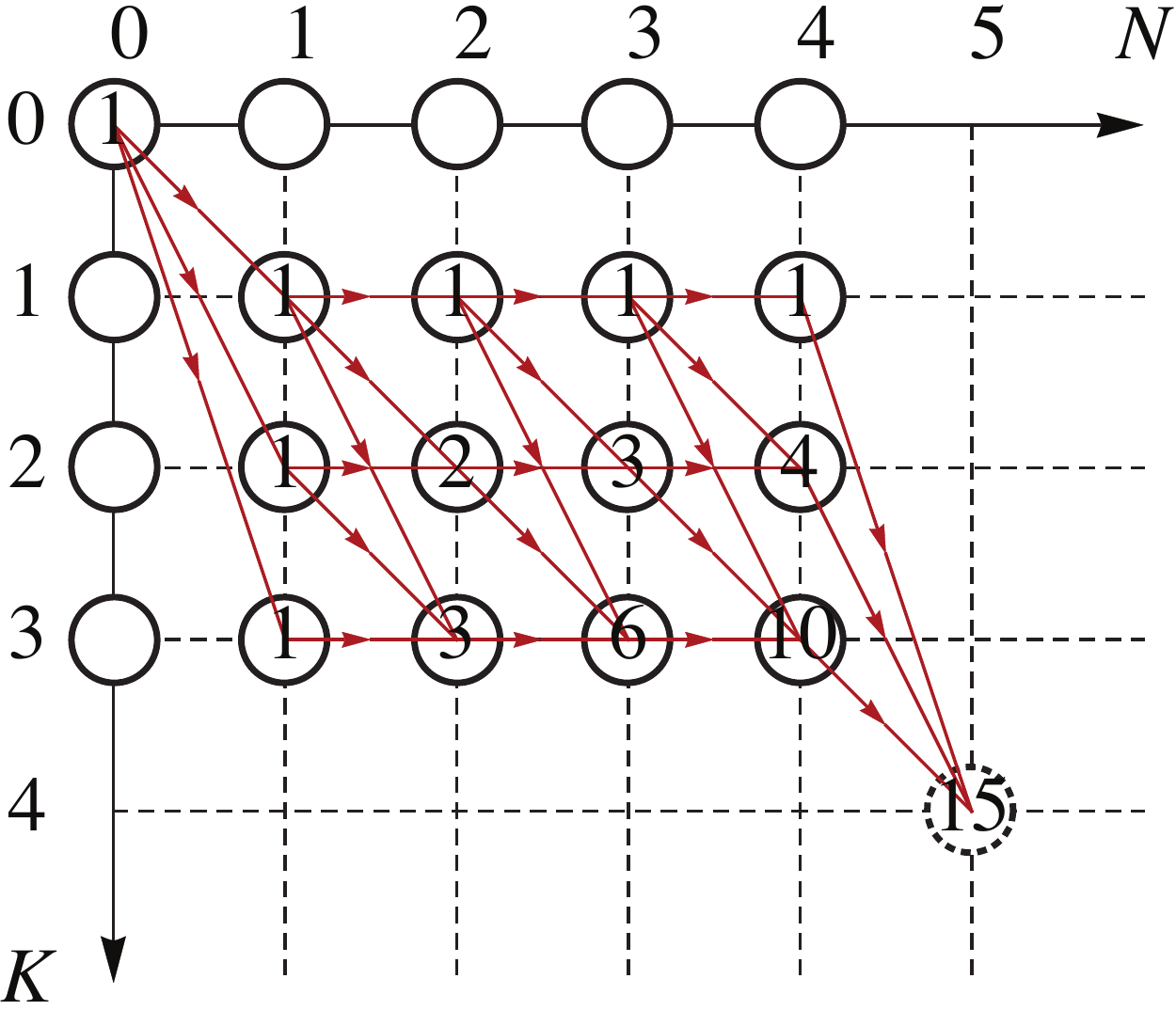}}} \\
\end{tabular}
\caption{Configuration graph for bosons with $(K,N)=(3,4)$. The equivalence to Figure \ref{fig:rotate}(c)
or Figure \ref{fig:dimHS}(a) can be seen simply via reshaping.}\label{fig:boson}
\end{figure}

First, HS-MPS can be extended to represent  fully symmetric tensors
by representing their independent part, viz., $p_1\le p_2\le \cdots \le p_N$,
similar to the antisymmetric case in Eq. \eqref{HSrep}. This can be used to
solve bosonic problems with fixed number of particles.
Working with this constraint is as simple as for the fermionic case.
Figure \ref{fig:boson} displays the corresponding configuration graph
for the bosonic case with $(K,N)=(3,4)$, and the HS-MPS is still constructed
by adapting the accessible range of physical indices and
the coupling rules. In fact, this graph is equivalent to
Figure \ref{fig:rotate}(c) or Figure \ref{fig:dimHS}(a) for fermions with $(K,N)=(6,4)$ by simply
reshaping the graph. In general, there is a one-to-one map
between configurations of $N$ fermions in $K$ orbitals
and configurations of $N$ bosons in $K'=K-N+1$ orbitals,
both of which lead to the same space dimension $C_{K}^N$.
Table \ref{tab:map} illustrates this mapping for $(K,N)=(6,4)$, where the
$i$-th orbital index $p_i$ in determinant strings for fermions
is mapped to the $i$-th orbital index $p_i-i+1$ in permanent strings
for bosons ($i\in\{1,\cdots,N\}$), which is equivalent to the reshaping of the configuration
graphs. This correspondence is in fact used
in the current implementation for fermions to make the
storage of site tensors more compact by employing an
effective physical dimension $K'=K-N+1$ instead of $K$
for the physical index $p_i$.

\begin{table}
\caption{The correspondences among
Fock-space occupation strings for fermions,
Slater determinants, and permanents for bosons.
The mapping from determinant strings to permanent strings is
to change $p_i$ to $p_i-i+1$ ($i\in\{1,\cdots,N\}$),
which corresponds to the reshaping of
configuration graphs.}
\begin{tabular}{cccc}
\hline\hline
No. & Occupation & Determinant & Permanent \\
\hline
1 &  111100 & 1234 &  1111 \\
2 &  111010 & 1235 &  1112 \\
3 &  111001 & 1236 &  1113 \\
4 &  110110 & 1245 &  1122 \\
5 &  110101 & 1246 &  1123 \\
6 &  110011 & 1256 &  1133 \\
7 &  101110 & 1345 &  1222 \\
8 &  101101 & 1346 &  1223 \\
9 &  101011 & 1356 &  1233 \\
10 & 100111 & 1456 &  1333 \\
11 & 011110 & 2345 &  2222 \\
12 & 011101 & 2346 &  2223 \\
13 & 011011 & 2356 &  2233 \\
14 & 010111 & 2456 &  2333 \\
15 & 001111 & 3456 &  3333 \\
\hline\hline
\end{tabular}\label{tab:map}
\end{table}

Next, we propose a sketch of spin adaptation. The advantage of using spin-adapted renormalized states in MPS is the reduction of
the bond dimension as exemplified in Sec. \ref{sec:numvar} as well as the increase of
efficiency in the computation of matrix elements by using the Wigner-Eckart theorem.
The basic idea of spin
adaptation in MPS is to recursively use tensor couplings in the direct product of
two spin complete subspaces, which means they are invariant
under the action of $S^2$ defined in the corresponding underlying Fock spaces,
such that the renormalized states can be chosen as eigenfunctions of $S^2$.
Here, we assume the same spatial orbital is used for both spins, and consequently the number of
spin orbitals is even, i.e., $K=2k$ with $k$ being the number of spatial orbitals.
To ensure that every subspace encountered is spin complete, in FS-MPS two adjacent spin-orbital sites are merged into a single
site such that the local physical space becomes spin complete, and which can be decomposed into
a direct sum of two subspaces: $S=0$ with the
vacuum and doubly occupied states, and $S=1/2$ with singly occupied states.
All the intermediate states formed from the tensor coupling
of these sites for spatial orbitals can then be made into eigenfunctions of $S^2$.
This is essentially similar to the construction of a canonical basis, viz., the Gelfand states
adapted to the subgroup chain $U(1)\subset U(2)\cdots \subset U(k-1)\subset U(k)$,
in the unitary group approach (UGA)\cite{paldus1974group}.

\begin{figure}
\begin{tabular}{c}
{\resizebox{!}{0.2\textheight}{\includegraphics{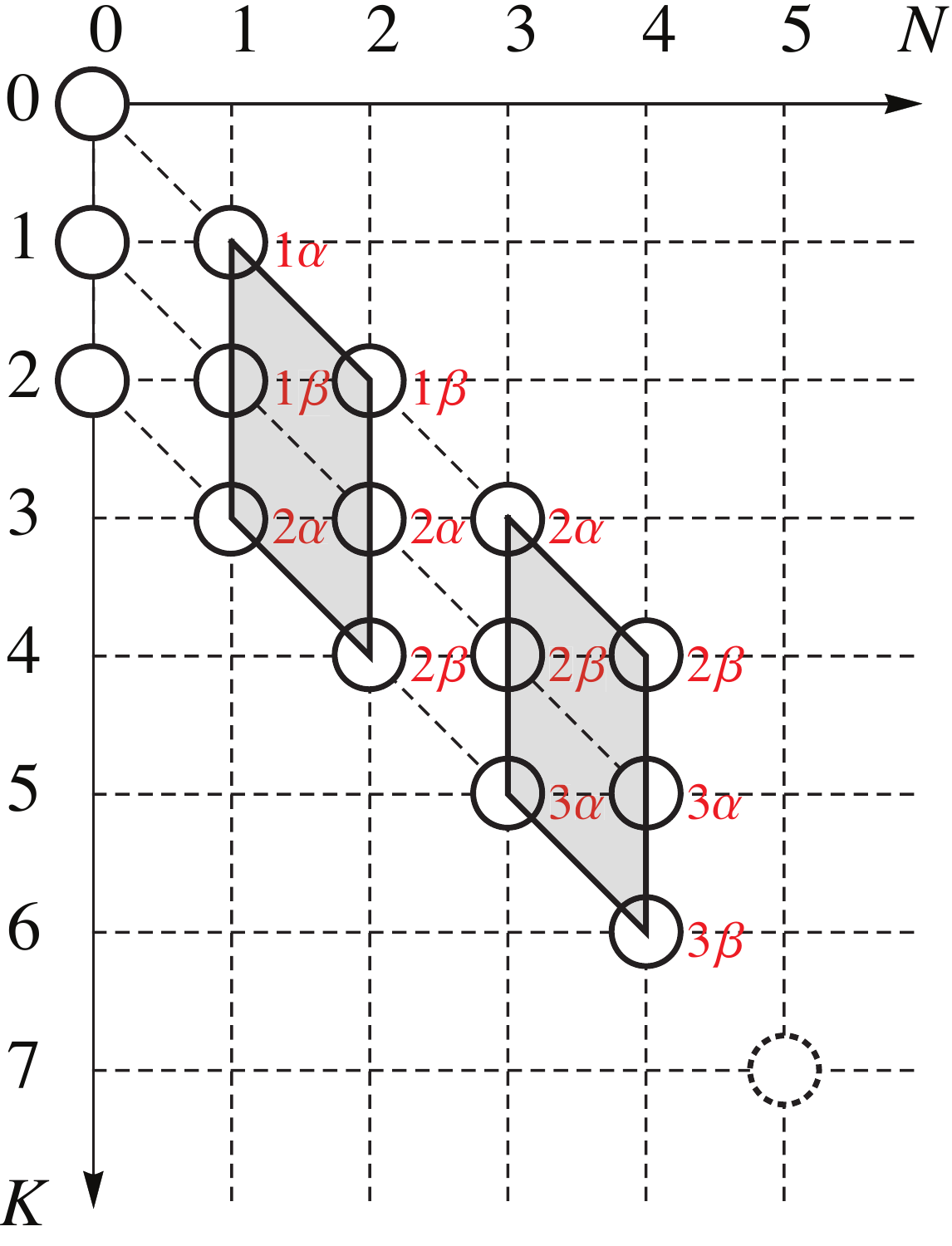}}}\\
(a) \\
{\resizebox{!}{0.2\textheight}{\includegraphics{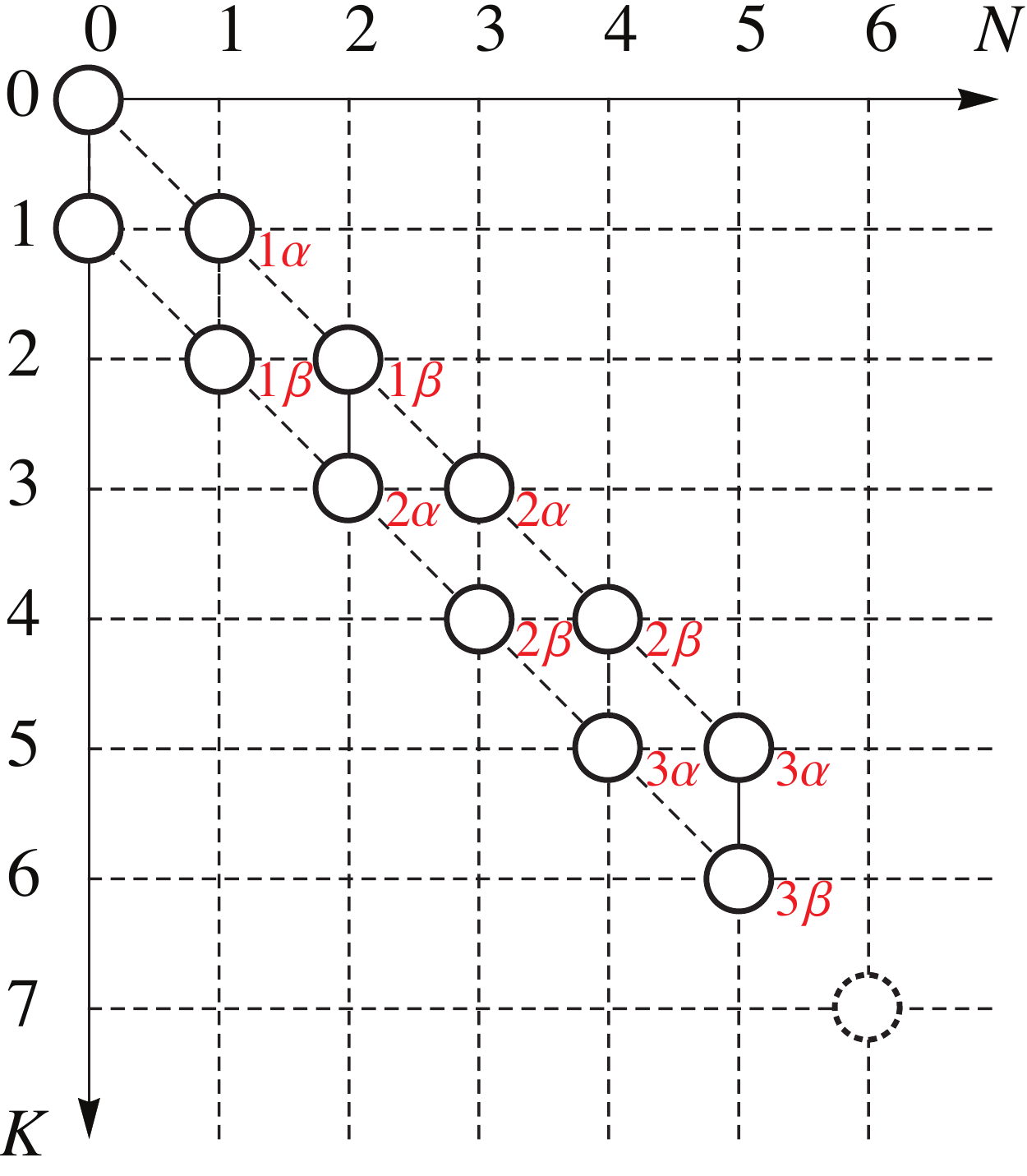}}}\\
(b) \\
{\resizebox{!}{0.24\textheight}{\includegraphics{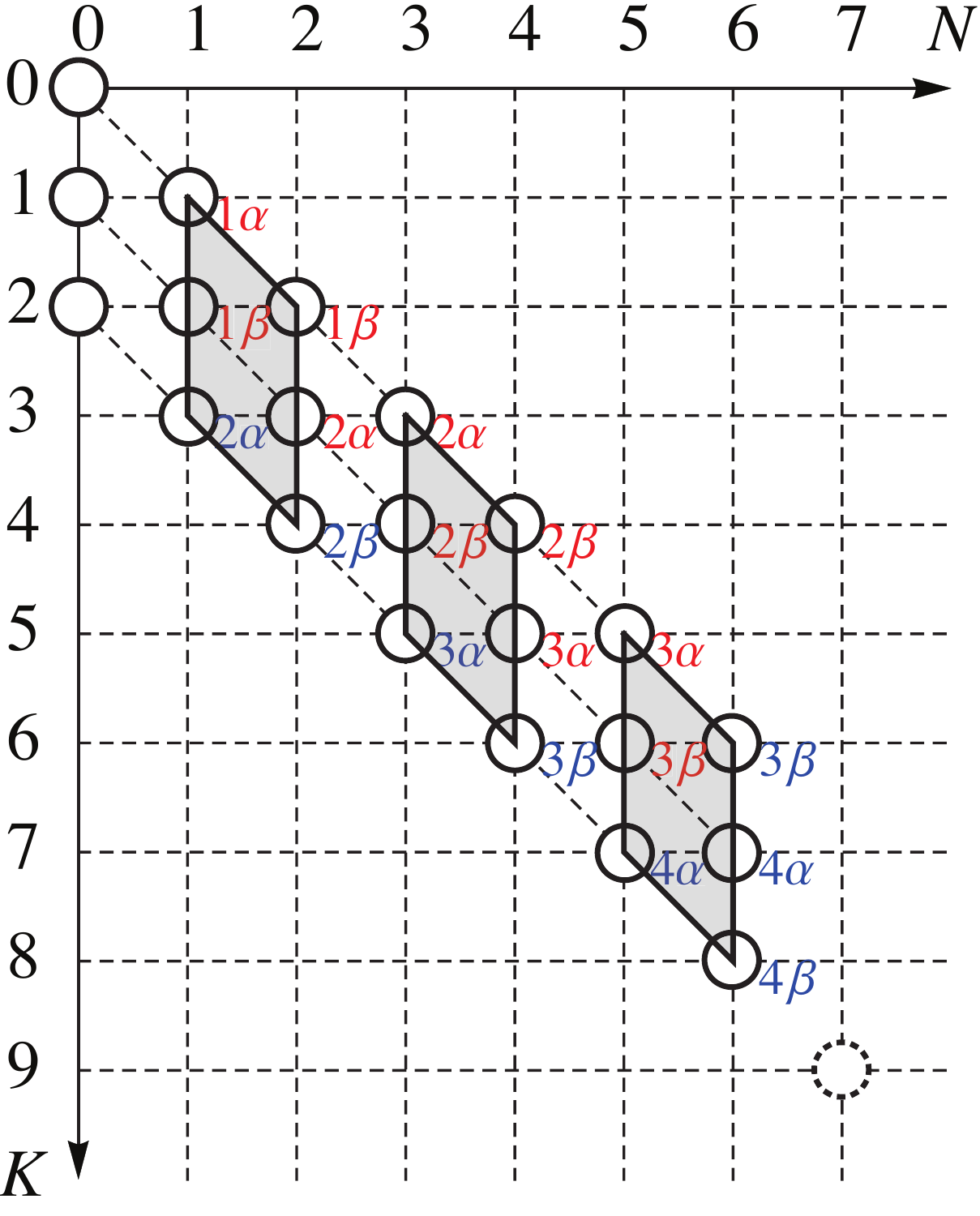}}}\\
(c) \\
\end{tabular}
\caption{On the spin adaptation of HS-MPS for even- and odd-electron systems:
(a) $(K,N)=(6,4)$; (b) $(K,N)=(6,5)$; (c) $(K,N)=(6,5)$ combined with
a fictitious electron in a fictitious spatial orbital such that $(K,N)=(8,6)$ for the entire system.}\label{fig:spin}
\end{figure}

The spin adaptation of HS-MPS is nontrivial, because it is not always natural to obtain
spin complete subspaces. For even-electron systems $N=2n$ as shown in Figure \ref{fig:spin}(a)
for $(K,N)=(6,4)$, the adjacent two sites can be merged into a single site, which
forms a spin-complete space for two electrons in $(K-N+2)/2=k-n+1$ spatial
orbitals as indicated by the shaded region.
The basis for the local physical space corresponding to
the  new merged physical index can be simply chosen as the set of spin-adapted pairs.
Supposing $i<j<k<l$ represent the indices for the spatial orbitals, it is seen
that there are two kinds of elementary pairs: the singlet pair formed by one doubly occupied
orbital denoted by $[i^2]_S$ and the singlet or triplet pairs formed by two singly
occupied orbitals denoted by $[ij]_{S,T}$. These pairs can be further coupled to form four-electron states, six-electron states, etc., up to $N=2n$ electron states. There are five possible ways to couple two electron pairs,
\begin{eqnarray}
&\protect [i^2]_S \times [j^2]_S,\;\; [i^2]_S \times [jk]_{S,T},\nonumber\\
&\protect [ij]_{S,T} \times [k^2]_{S},\;\; [ij]_{S,T} \times [kl]_{S,T},\;\; [ij]_{S} \times [jk]_{S,T}.\label{paircoupling}
\end{eqnarray}
The last one deserves special explanation. Unlike the case $[ij]_{S,T} \times [kl]_{S,T}$, the coupling $[ij]_{S,T}\times [jk]_{S,T}$ can only result in four states as the $j$-th spatial orbital becomes double occupied, viz.,
\begin{eqnarray}
\protect[ij]_S\times [jk]_S &=& [ij^2k]_S,\nonumber\\
\protect[ij]_S\times [jk]_T &=& [ij]_T\times [jk]_S=[ij^2k]_T,\nonumber\\
\protect[ij]_T\times [jk]_T &=& [ij^2k]_S\oplus [ij^2k]_T.
\end{eqnarray}
This shows that if all of the couplings are allowed, then there will be a linear dependence
among the superblock states in DMRG calculations. Although the
resulting CI problem is a generalized eigenvalue problem and can in principle also be solved, for numerical reasons
it is usually advantageous to work with MPS without such redundancies.
Thus, for simplicity one could just choose the path $[ij]_{S} \times [jk]_{S,T}$ as
the only allowed coupling, as in \eqref{paircoupling}. This type of geminal spin-coupling scheme in HS-MPS is in the same spirit as
the Serber type spin functions for even number of electrons\cite{pauncz}, where
the spins of two electrons are coupled first and then the spin pairs are coupled sequentially.
The DOCI shown in Figure \ref{fig:ci} can be viewed as a special case of
the spin-adapted HS-MPS, where only the $[i^2]_S$ type pairs appear in the
construction of the final singlet wavefunction.

The case for odd-electron systems for $(K,N) = (6,5)$ is depicted in Figure \ref{fig:spin}(b). The same strategy
does not work, e.g., the configuration space for the first two electrons
spanned by $\{|1_\alpha1_\beta\rangle, |1_\alpha 2_\alpha\rangle, |1_\beta 2_\alpha\rangle\}$
is not spin complete due to the lack of $|1_\beta 2_\beta\rangle$.
On the other hand, combining the first three electrons forms a spin-complete space,
but the configuration space for the remaining two does not.
To solve this problem, a fictitious electron in a fictitious spatial orbital can
be added. This makes the number of electrons even and the pairwise spin adaptation
scheme can be applied. For the case shown in Figure \ref{fig:spin}(b), this approach
leads to Figure \ref{fig:spin}(c), which is quite similar to Figure \ref{fig:spin}(a).
To ensure the correct number of electrons for the physical system, the local space for the last pair must be
restricted to the type $[ij]_{S,T}$ where $j$ is the fictitious spatial orbital.
Consider the example in Figure \ref{fig:spin}(c), obtained by applying the coupling scheme in Eq. \eqref{paircoupling}.
It is seen that for the first two electrons
the only pairs that contribute to the final states are $[1^2]_S$ and $[12]_{S}$, the pairs for the next two electrons
are $[2^2]_S$ and $[23]_{S}$, while the pairs for the last two electron are $[34]_{S,T}$.
There are only three possible couplings, viz., $[1^2]_S\times [2^2]_S \times [34]_{S,T}$,
$[1^2]_S\times [23]_S \times [34]_{S,T}$, and
$[12]_S\times [23]_S \times [34]_{S,T}$, such that
the dimension of the configuration space is $4\times 3=12$.
It is twice as large as that of the physical system ($C_{6}^{5}=6$),
as it should be due to the introduction of the fictitious electron and orbital. The use of
the fictitious system essentially modifies the structure of the HS-MPS as indicated in Figure \ref{fig:spin}.
Finally, we mention that the idea of using an extra-electron system
has been used before
e.g., in spin-orbit configuration interaction\cite{SOCI} to make the Hamiltonian matrix real-symmetric for odd-electron systems.

\subsection{Connection to other theories}\label{sec:connection}
Before proceeding to numerical studies, we provide several special
cases to better understand the HS-MPS ansatz, and to establish
connections with other methods. First, we give an analytic construction of HS-MPS for the simplest
nontrivial case $(K,N)=(4,3)$. To this end, the FCI wavefunction \eqref{FCI}
can be viewed as a multivariate polynomial of orbital indices $p_i$, viz.,
\begin{eqnarray}
f(\mathbf{p})&=&c_{123}p_1p_2p_3+
c_{124}p_1p_2p_4 \nonumber\\
&+&c_{134}p_1p_3p_4
+c_{234}p_2p_3p_4,
\end{eqnarray}
and the factorization leads to
\begin{eqnarray}
f(\mathbf{p})&=&p_1(
p_2(c_{123}p_3+
c_{124}p_4)+p_3(c_{134}p_4))
+p_2(c_{234}p_3p_4)\nonumber\\
&=&
\left[\begin{array}{c}
p_1 \\
p_2 \\
\end{array}\right]^T
\left[\begin{array}{cc}
p_2 & c_{124}p_3\\
0 & c_{234}p_3 \\
\end{array}\right]
\left[\begin{array}{c}
c_{123}p_3+c_{124}p_4\\
p_4 \\
\end{array}\right].\label{poly}
\end{eqnarray}
From this expression, the nonzero elements of the HS-MPS \eqref{HSMPS} can be
read off, and the bond dimensions can be seen to be $\{2,2\}$.
Thus, the HS-MPS is related to the factorized form of multivariate polynomials,
which are known as multivariate Horner schemes for efficient computations of the value of
polynomials. It is a generalization of the well known Horner scheme\cite{knuth2} for computing
the value of a univariate polynomial
\begin{eqnarray}
f(x)=\sum_{i=0}^{n}a_i x^i=a_0+x(a_1+x(\cdots (a_{n-1}+x a_n)\cdots)).
\end{eqnarray}
Both the HS-MPS and Horner scheme share the same recursive structure.

Next, we examine an interesting class of states that have the HS-MPS
as natural representations. We refer to them as 'path-weighted' states
because their wavefunction coefficients are products of
weights for each segment of the path from $(0,0)$ to $(K+1,N+1)$
on the configuration graph for HS-MPS. For example, the nonzero
CI coefficient $\Psi^{(p_1p_2p_3)}$ for $N=3$ can be written as
\begin{eqnarray}
\Psi^{(p_1p_2p_3)}=a_{0,p_1}b_{p_1,p_2}c_{p_2,p_3}d_{p_3,K+1},\;\;
d_{p_3,K+1}\equiv1,
\end{eqnarray}
where $a,b,c,d$ are the weights associated with the path
$0\rightarrow p_1\rightarrow p_2\rightarrow p_3\rightarrow K+1$
for the orbital index on the configuration graph. This kind of state
can be represented exactly by an HS-MPS with bond dimension equal to the
dimension of the physical index $K-N+1$. Taking $K=5$ as example,
the wavefunction tensor is expressed as
\begin{eqnarray}
\Psi^{p_1p_2p_3}=A^{p_1}[1]A^{p_2}[2](A[3]A[4])^{p_3}.\label{ANN}
\end{eqnarray}
In the path-weighted state, the site tensors can then be written as
\begin{eqnarray}
A[1]&=&
\left[
\begin{array}{c}
a_{0,1} \\
a_{0,2} \\
a_{0,3} \\
\end{array}\right]^T,
\quad
A[2]=
\left[
\begin{array}{ccc}
b_{1,2} & b_{1,3} & b_{1,4} \\
0 & b_{2,3} & b_{2,4} \\
0 & 0 & b_{3,4} \\
\end{array}\right],\nonumber\\
\quad
A[3]&=&
\left[
\begin{array}{ccc}
c_{2,3} & c_{2,4} & c_{2,5} \\
0 & c_{3,4} & c_{3,5} \\
0 & 0 & c_{4,5} \\
\end{array}\right],
\quad
A[4]=
\left[
\begin{array}{c}
1\\
1\\
1\\
\end{array}\right],\label{PW}
\end{eqnarray}
where the auxiliary site $A[4]$ is introduced just to represent
the paths to the virtual site $(K+1,N+1)$. Because of the special
structure of the path-weighted state, the tensors for the sites 2 and 3 in
Eq. \eqref{PW} are written in a compact form, and their full (rank-3) forms can be recovered as
\begin{eqnarray}
A^{p_2}_{\alpha_1\alpha_2}[2]
=
\left[\begin{array}{ccc}
\left[\begin{array}{ccc}
b_{1,2} & 0 & 0\\
0 & 0 & 0\\
0 & 0 & 0\\
\end{array}\right], &
\left[\begin{array}{ccc}
0 & b_{1,3} & 0\\
0 & b_{2,3} & 0\\
0 & 0 & 0\\
\end{array}\right], &
\left[\begin{array}{ccc}
0 & 0 & b_{1,4}\\
0 & 0 & b_{2,4}\\
0 & 0 & b_{3,4}\\
\end{array}\right] \\
\end{array}\right].\label{FullForm}
\end{eqnarray}
The coefficients $b_{p_i,p_j}$ are a kind of transition amplitude,
and the left canonical condition \eqref{leftCanon}
ensures that they are normalized as
\begin{eqnarray}
&|b_{1,2}|^2=1,\; |b_{1,3}|^2+|b_{2,3}|^2=1,\nonumber\\
&|b_{1,4}|^2+|b_{2,4}|^2+|b_{3,4}|^2=1.
\end{eqnarray}
Furthermore, if the transition amplitudes are site independent and also independent of the source, viz., $(a,b,c)_{ij}=w_{j}$, then the wavefunction can be regarded as a kind of 'orbital-weighted' state whose coefficients only depend on the occupied orbitals
\begin{eqnarray}
\Psi^{(p_1p_2p_3)}=w_{p_1}w_{p_2}w_{p_3}.
\end{eqnarray}
In the extreme case $w_{p}=1$, the state becomes a maximally entangled state,
\begin{eqnarray}
|\Psi\rangle=\sum_{p_1<p_2<\cdots <p_N}|p_1p_2\cdots p_N\rangle.
\end{eqnarray}

From this analysis, we also observe that there are similarities between both FS-MPS and HS-MPS and artificial neural networks (ANNs).
They all share the same recursive structure. The virtual indices in MPS can be regarded as the hidden layers in ANNs,
the renormalized states can be viewed as neurons, and the site tensors can be seen as the
synaptic weights. The physical indices can be interpreted as an additional set of inputs for each layer.
Specifically, Figure \ref{fig:dimFS} can be viewed as an ANN with $K-1$ hidden layers, which accepts
the vacuum as the input into the origin and outputs the final wavefunction at the site $(K,N)$ via
the propagation through the $K-1$ hidden layers by the recursive chain \eqref{FSChain}.
The example in Eq. \eqref{ANN} then just corresponds to the single neuron
per each orbital index 'symmetry' case. The graphically contracted function (GCF) method of
Shepard et al.\cite{shepard2005general,shepard2006hamiltonian,shepard2014multifacet1,shepard2014multifacet2,gidofalvi2014wave} also shares a similar structure, where the contraction is based on the Shavitt graph\cite{paldus1974group,shavitt1977,shavitt1978matrix}
for the spin-adapted configuration space. A single GCF\cite{shepard2005general} is in fact a ``path-weighted'' state,
where an arc weight is associated with each path between two nodes in the Shavitt graph.
In the latter multifacet generalization\cite{shepard2014multifacet1}, the number of intermediate states per symmetry sector
is allowed to be greater than one.
Consequently, the multifacet GCF wavefunction becomes a special spin-adapted
Fock-space MPS, where the physical indices are just the step vector in GUGA,
the number of renormalized states per symmetry is controlled, and
the wavefunction is in either left or right canonical form and optimized by
a gradient-based algorithm instead of the DMRG optimization in mixed canonical form.
The further connection of spin-adapted FS-MPS with the Shavitt graph will
be made clear in the next section.

\section{Illustrative calculations}\label{sec:results}
The DMRG algorithm for HS-MPS has been implemented for
generic second quantized Hamiltonians of the form \eqref{Hnp}.
The symmetries of spin projection $S_z$ and point groups ($D_{2h}$ and its subgroups) have also been implemented.
In all of the following calculations, the restricted (open-shell)
Hartree-Fock orbitals and transformed molecular integrals
are generated by the \textsc{Pyscf} package\cite{PYSCF}.
The ordering of orbitals in FS-MPS and HS-MPS[p] is chosen
in the order of increasing orbital energies, while
in HS-MPS[h] such an ordering is reversed to place the
orbitals of higher energies in the leftmost
sites. The  site tensors of the HS-MPS
are initialized from the HS-MPS for CISD, and then
the physical indices and bond dimensions are
gradually increased to the desired values for more accurate calculations.
The use of local orbitals and reordering of orbitals are not examined
in the present paper. For a fair comparison with the HS-MPS results, the Fock-space DMRG calculations
were performed with the \textsc{Block} code\cite{chan_highly_2002,chan_algorithm_2004,sharma_spin-adapted_2012}  using only $S_z$ symmetry,
unless otherwise stated.

\subsection{Decomposition of many-electron wavefunctions into HS-MPS}\label{sec:numdecomp}
\begin{figure}
\begin{tabular}{c}
{\resizebox{!}{0.2\textheight}{\includegraphics{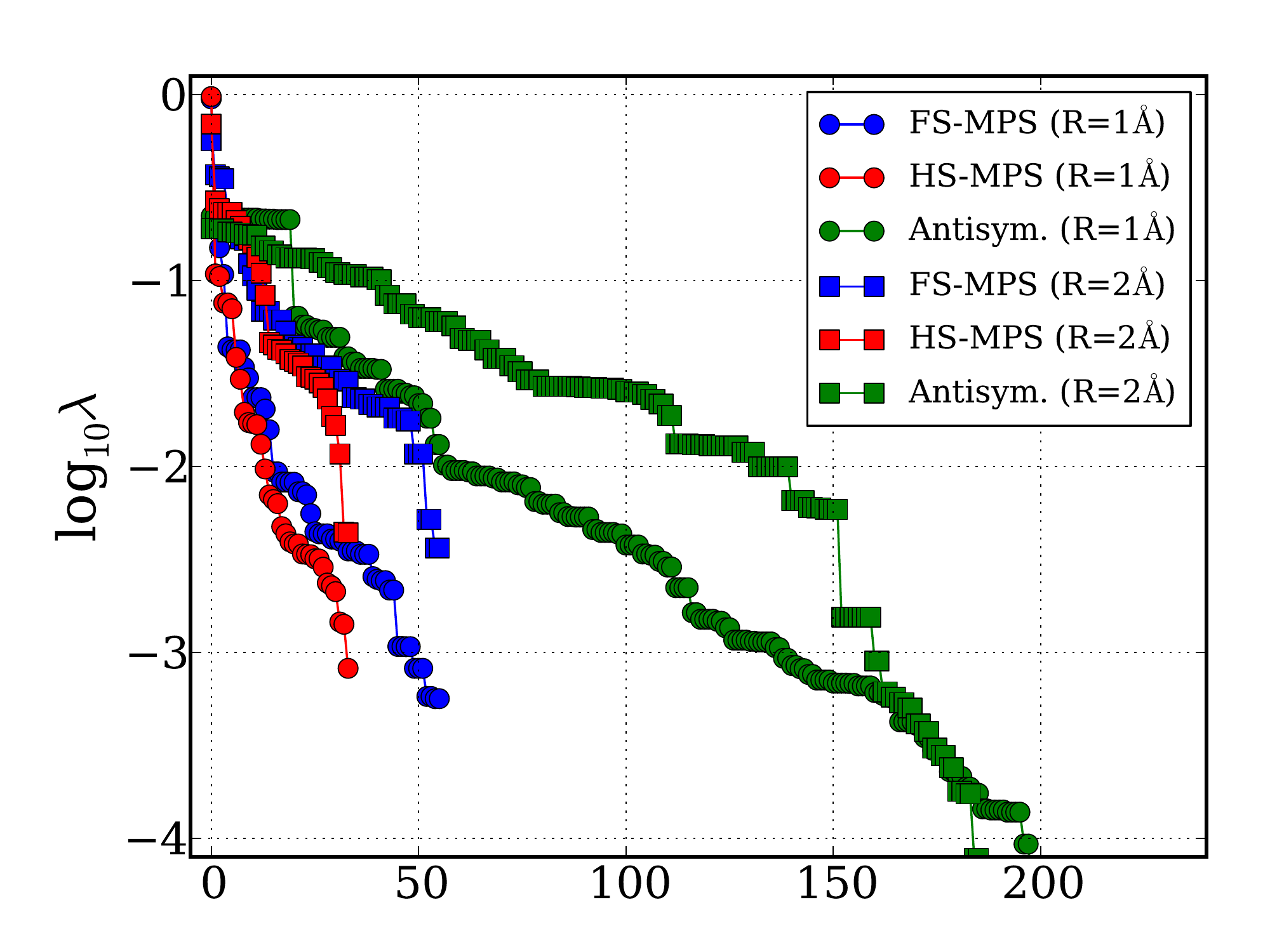}}}\\
(a) \\
{\resizebox{!}{0.2\textheight}{\includegraphics{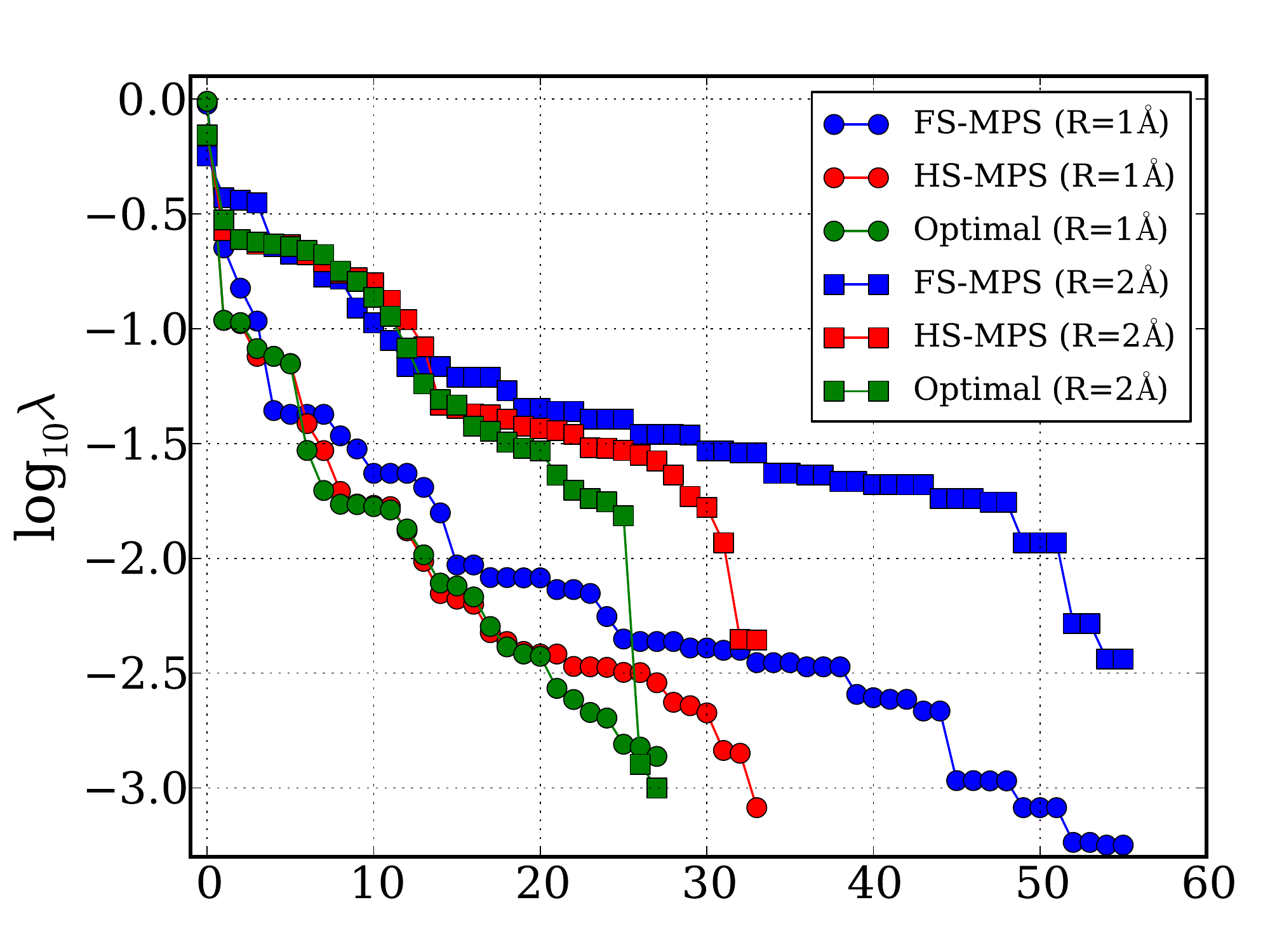}}}\\
(b) \\
\end{tabular}
\caption{The decay of Schmidt values at the middle of various MPS for H$_6$ with $R$=1{\AA} and $R$=2{\AA}.}\label{fig:h6s}
\end{figure}

\begin{figure}
\begin{tabular}{c}
{\resizebox{!}{0.2\textheight}{\includegraphics{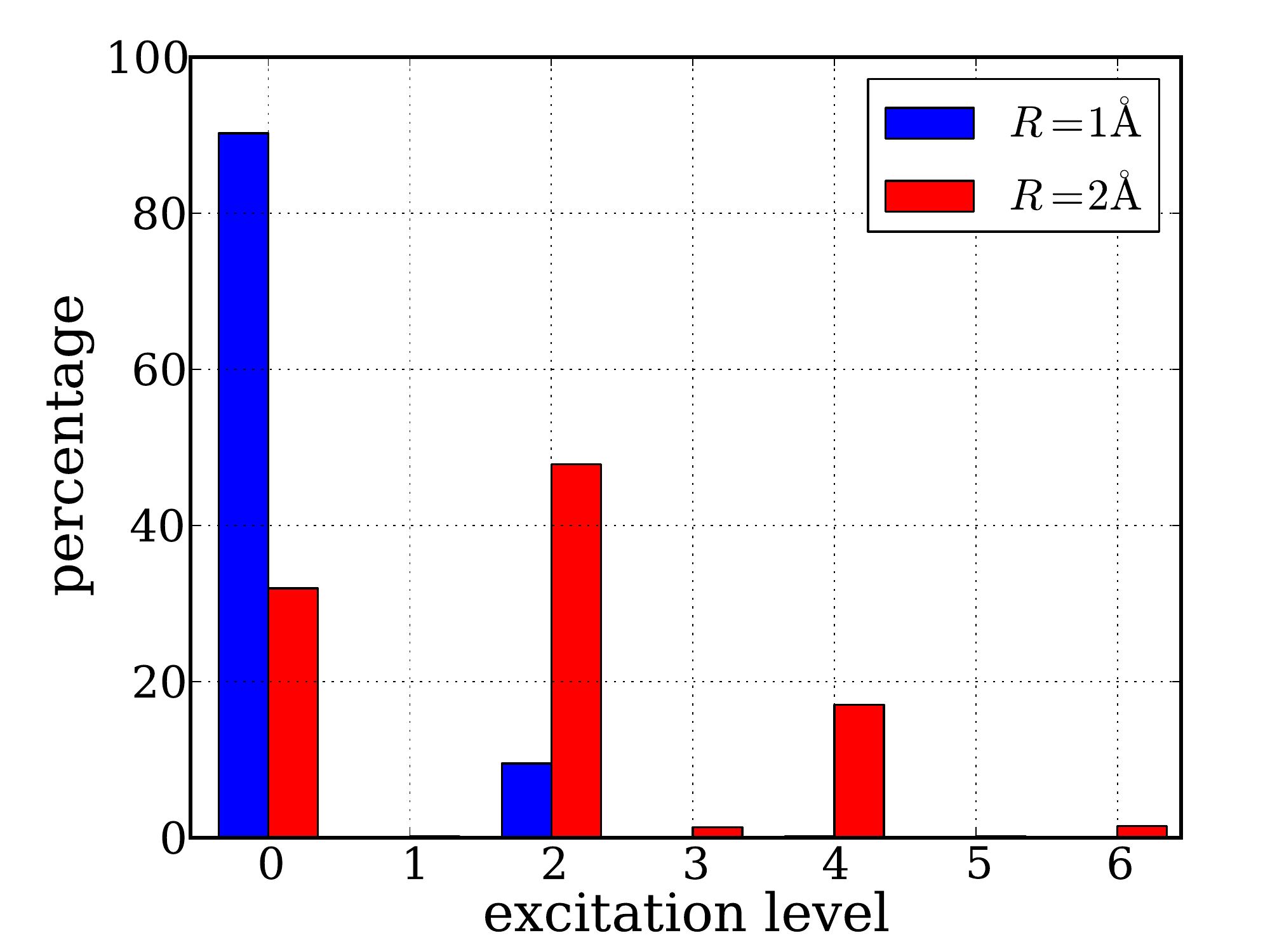}}}\\
(a) Population of configurations \\
{\resizebox{!}{0.2\textheight}{\includegraphics{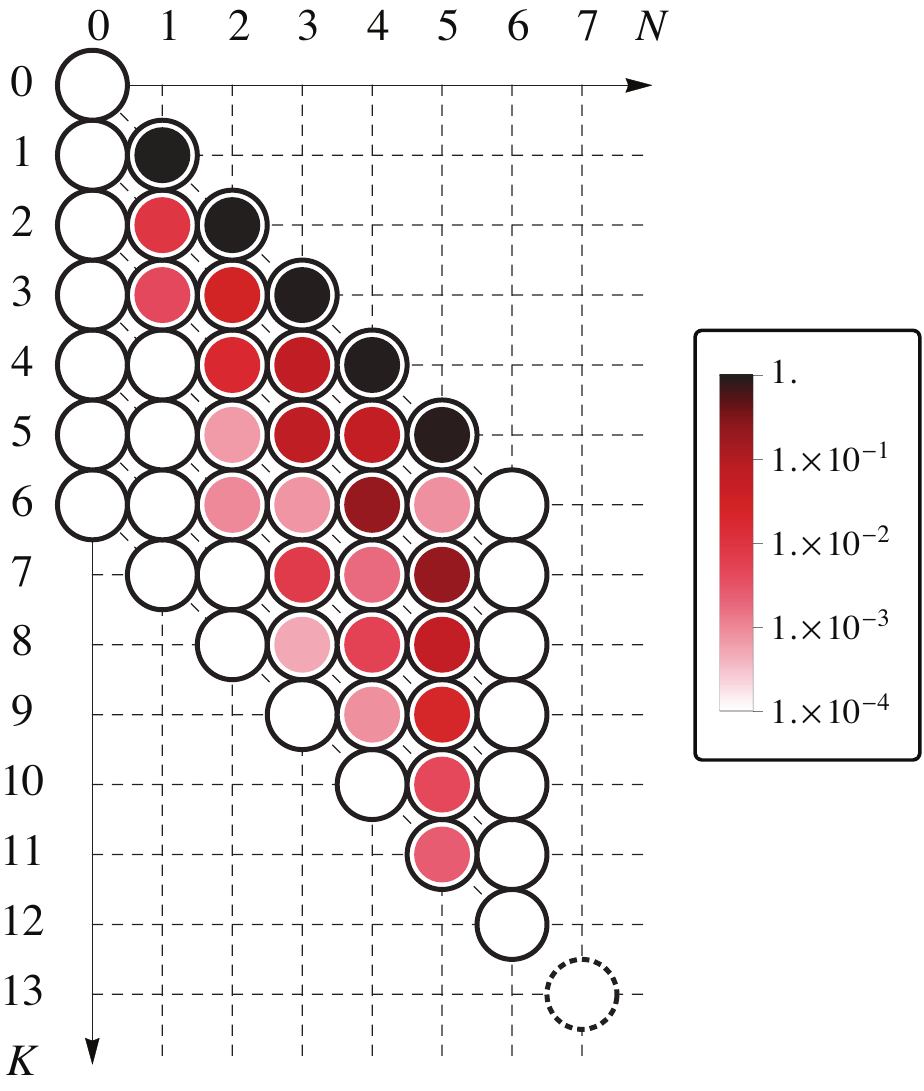}}}\\
(b) Weights of symmetry sectors for $R$=1{\AA}. \\
{\resizebox{!}{0.2\textheight}{\includegraphics{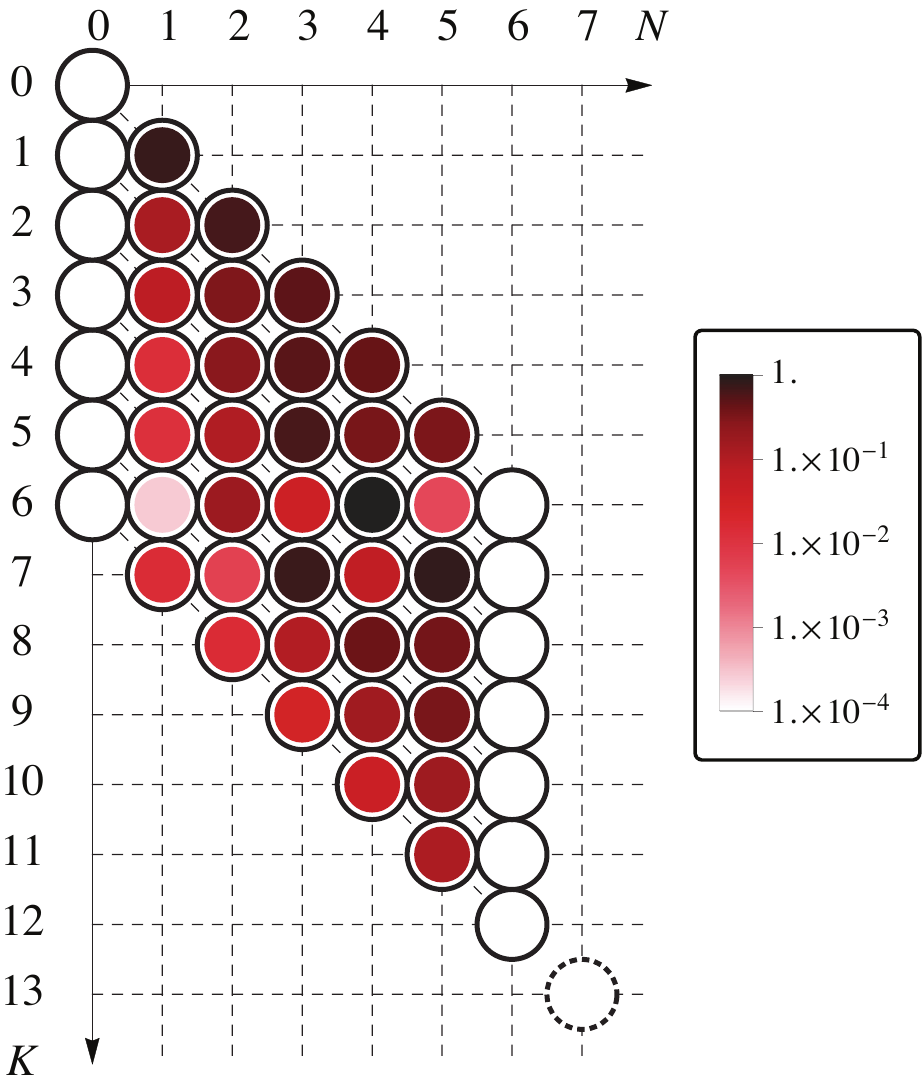}}}\\
(c) Weights of symmetry sectors for $R$=2{\AA}. \\
\end{tabular}
\caption{Comparison of the FCI wavefunctions and the corresponding HS-MPS for H$_6$ with $R$=1{\AA} and $R$=2{\AA}.}\label{fig:h6pop}
\end{figure}

Before discussing the variational calculations, we examine the decomposition of a given FCI
wavefunction into HS-MPS using the method presented in Sec. \ref{sec:decomp}.
Due to the large memory cost to store  the FCI wavefunction tensor,
such decompositions
cannot been carried out for large $K$ and $N$.
The simple linear molecule H$_6$ with bond distance
$R$=1{\AA} and $R$=2{\AA} is studied using FCI with the STO-3G basis,
which corresponds to the case $(K,N)=(12,6)$ shown in
Figure \ref{fig:bd1}. The theoretical bond dimensions
for HS-MPS are $\{7, 21, 35, 35, 21\}$, while for FS-MPS they are
$\{2, 4, 8, 16, 32, 64, 32, 16, 8, 4, 2\}$. Figure \ref{fig:h6s}
shows the decay of Schmidt values at the middle of various MPS.
As demonstrated in Figure \ref{fig:h6s}(a), even in this
simple molecule the singular values decay extremely slowly
for the decomposition of the antisymmetric Hilbert-space tensor
\eqref{Antisym}. Figure \ref{fig:h6s}(b)
shows that the decay rates are comparable for both FS-MPS
and HS-MPS. Further, compared with the optimal
decomposition without any constraint (see green points in Figure \ref{fig:h6s}(b)),
the prefix/suffix constraints in HS-MPS do not introduce
too much redundancy. In particular,
the rates of decay almost coincide for the first twenty
singular values. Clearly, increasing the bond distance makes
the system more strongly correlated such that the decay of Schmidt values becomes much more slow.
The percentages of configurations with different excitation levels
are shown in Figure \ref{fig:h6pop}(a), which illustrates the
dramatic increase of contributions from higher excitations.
The weights of sectors with different
orbital index 'symmetry' are plotted in Figure \ref{fig:h6pop}(a)
for $R$={1\AA} and Figure \ref{fig:h6pop}(b) for $R$=2{\AA}, respectively.
The increase of renormalized states with large populations is
clearly revealed when the bond distance is increased.

\subsection{Variational calculations of ground states}\label{sec:numvar}

\begin{figure}
\begin{tabular}{c}
{\resizebox{!}{0.2\textheight}{\includegraphics{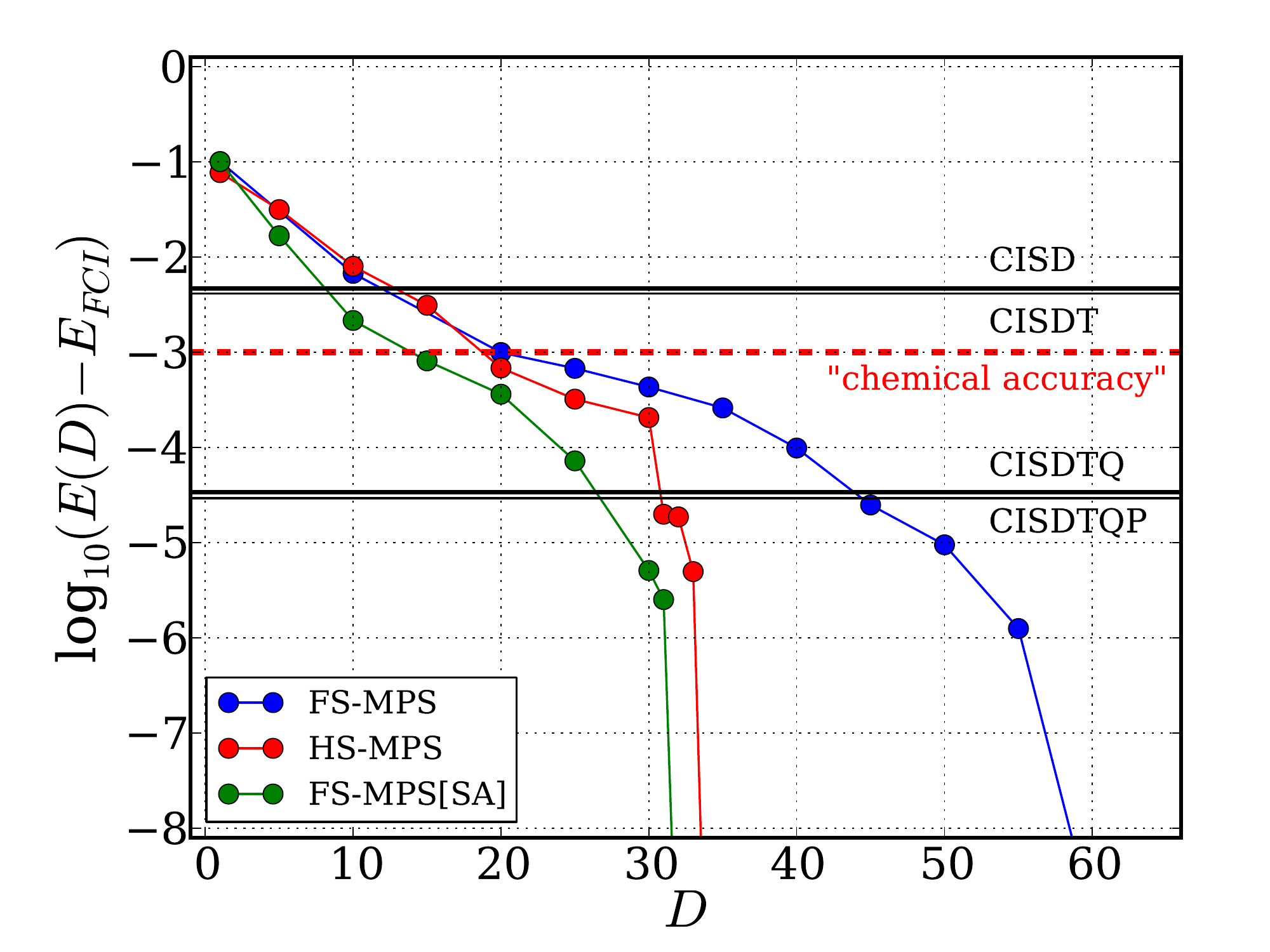}}}\\
(a) $R$=1{\AA} \\
{\resizebox{!}{0.2\textheight}{\includegraphics{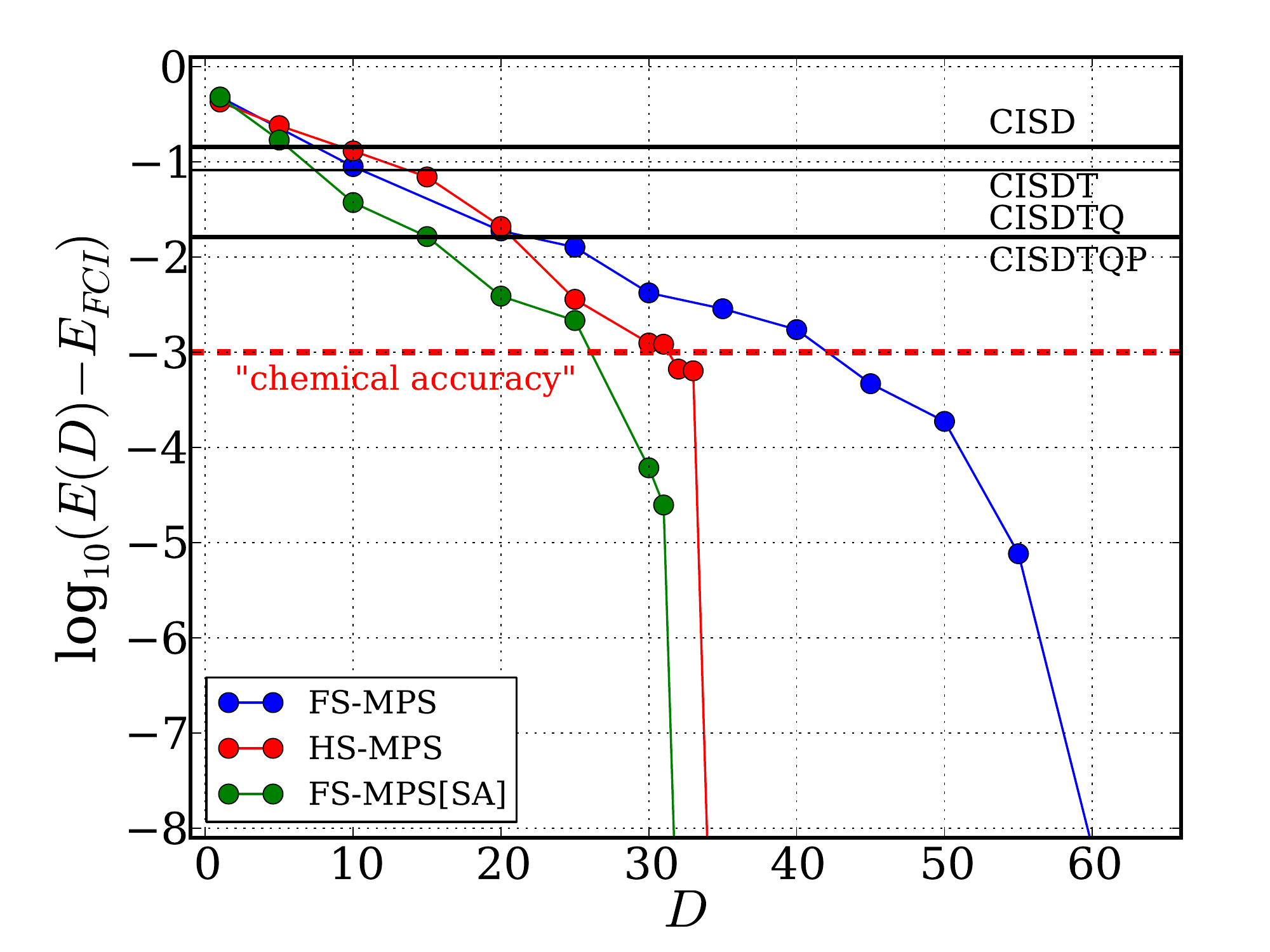}}}\\
(b) $R$=2{\AA} \\
\end{tabular}
\caption{Errors of variational energies as a function of $D$ for H$_6$ with $R$=1{\AA} in the STO-3G basis.}\label{fig:h6e}
\end{figure}

\begin{figure}
\begin{tabular}{c}
{\resizebox{0.5\textwidth}{!}{\includegraphics{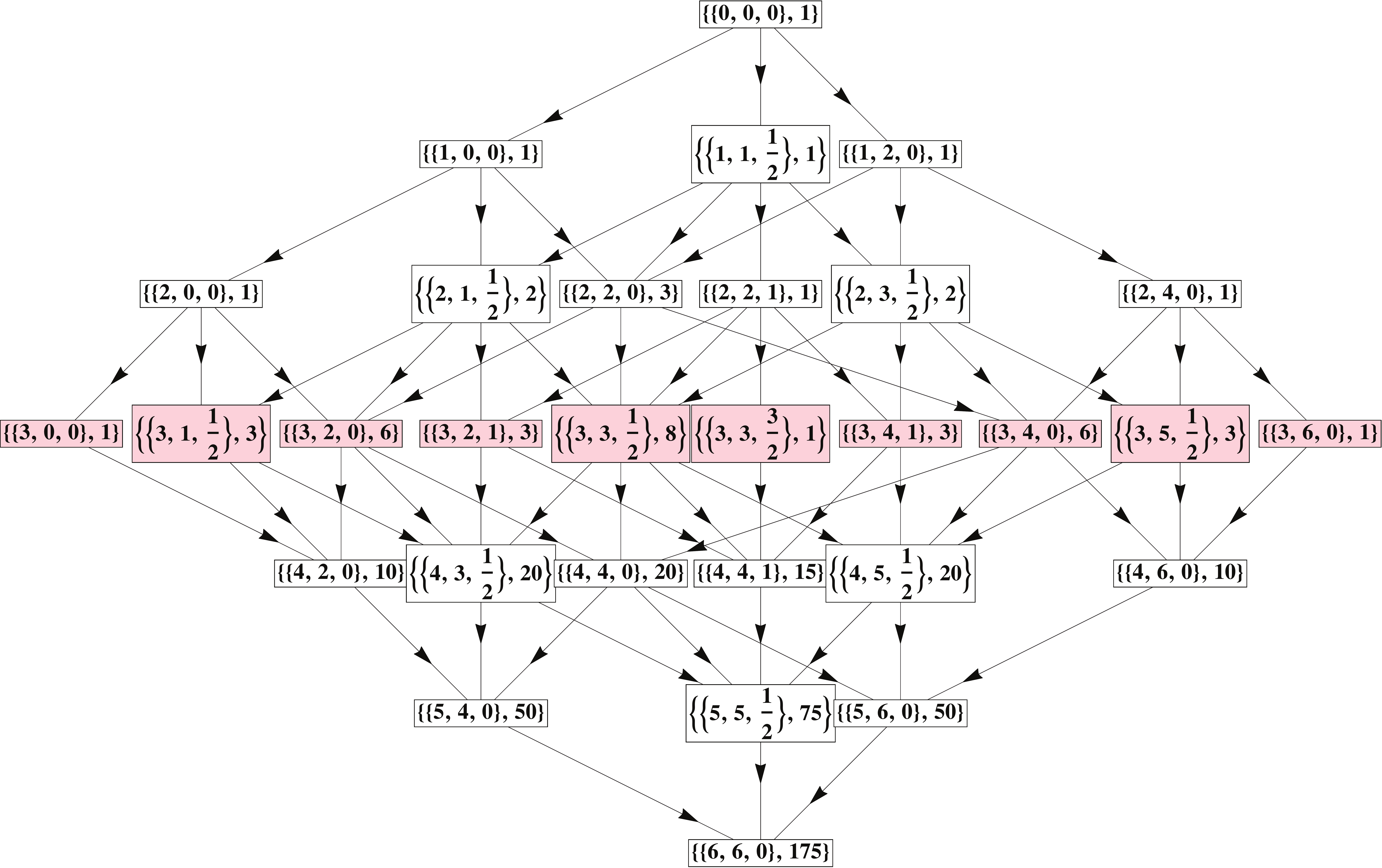}}}\\
\end{tabular}
\caption{Configuration graph in terms of spin-adapted intermediate states.
It is the same as the Shavitt graph although plotted differently. The numbers in each block $\{\{K,N,S\},f_{K,N,S}\}$ are the number of spatial orbitals $K$, the number of electrons $N$, total spin $S$, and the dimension $f_{K,N,S}$.
The red row corresponds to the largest bond dimension.}\label{fig:h6sa}
\end{figure}

Now we turn to the variational calculations via DMRG.
Figure \ref{fig:h6e} shows the errors of variational energies as a
function of the maximal bond dimension $D$ for H$_6$/STO-3G employing
FS-MPS and HS-MPS. To illustrate the effect of spin adaptation,
the results obtained by spin-adapted FS-MPS are also shown.
The red dotted lines represent the so-called 'chemical accuracy' (1$mE_h$),
while the black lines represent the errors of various truncated CI.
It is seen that HS-MPS converges faster than FS-MPS for this molecule.
This can be traced back to Figure \ref{fig:bd1}(c), which shows
that at the half filling the required bond dimension of HS-MPS for a given
excitation level is generally smaller than that of FS-MPS. The spin adapted FS-MPS converges even faster, because the spin adaptation
reduces the number of renormalized states by using the whole spin
multiplet as a single state. This is more evident by plotting the
spin-adapted counterpart of the configuration graph, see
Figure \ref{fig:h6sa} for the full configuration graph obtained in a similar way as
Figure \ref{fig:dimHS}(a). The numbers in each block $\{\{K,N,S\},f_{K,N,S}\}$ are the number of spatial orbitals $K$, the number of electrons $N$, total spin $S$, and the dimension $f_{K,N,S}$ for this space computed via the Paldus formula\cite{paldus1974group}.
The physical dimension of spin-adapted FS-MPS is four, indicating four possible couplings between intermediate nodes depending on the changes
of particle number $N$ and total spin $S$, viz.,
$(\Delta N,\Delta S)\in\{(0,0),(1,1/2),(1,-1/2),
(2,0)\}$. Figure \ref{fig:h6sa} is equivalent to the Shavitt graph
in GUGA although plotted in a slightly different way, and the four types of couplings correspond to the
so-called step vector\cite{shavitt1977}. The graph for the right renormalization as a counterpart of Figure \ref{fig:dimHS}(b)
can be derived by reversing the flows from bottom to top in Figure \ref{fig:h6sa}.
In this way, the maximal bond dimension can be found at the middle
of the hydrogen chain. By summing the dimensions of the
subspaces for $K=3$ (blocks in red color), the maximum bond dimension
is found to be 35. This explains the decay pattern of the spin-adapted FS-MPS observed
in Figure \ref{fig:h6e}.

\begin{figure}
\begin{tabular}{c}
{\resizebox{!}{0.22\textheight}{\includegraphics{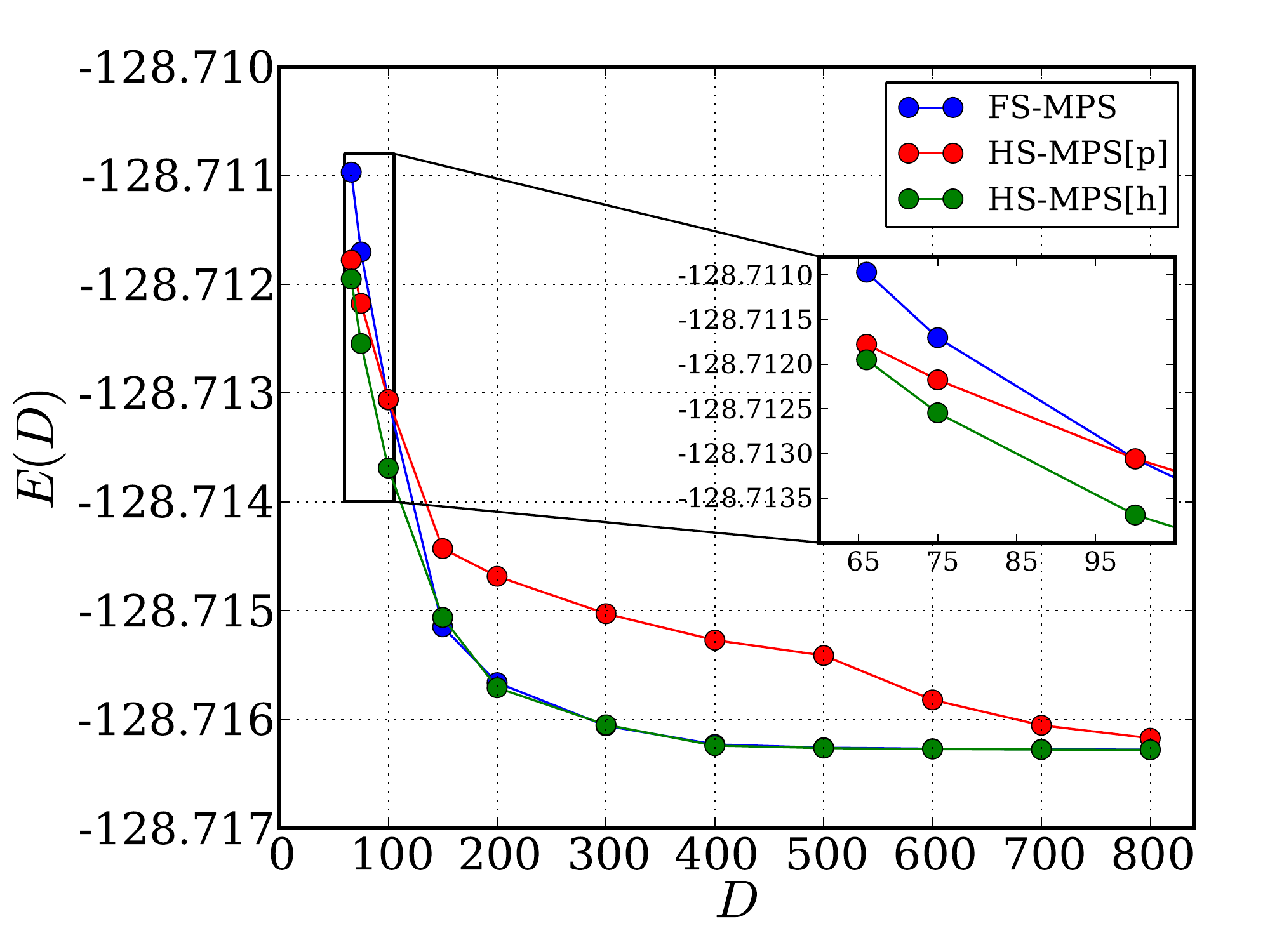}}}\\
(a) \\
{\resizebox{!}{0.22\textheight}{\includegraphics{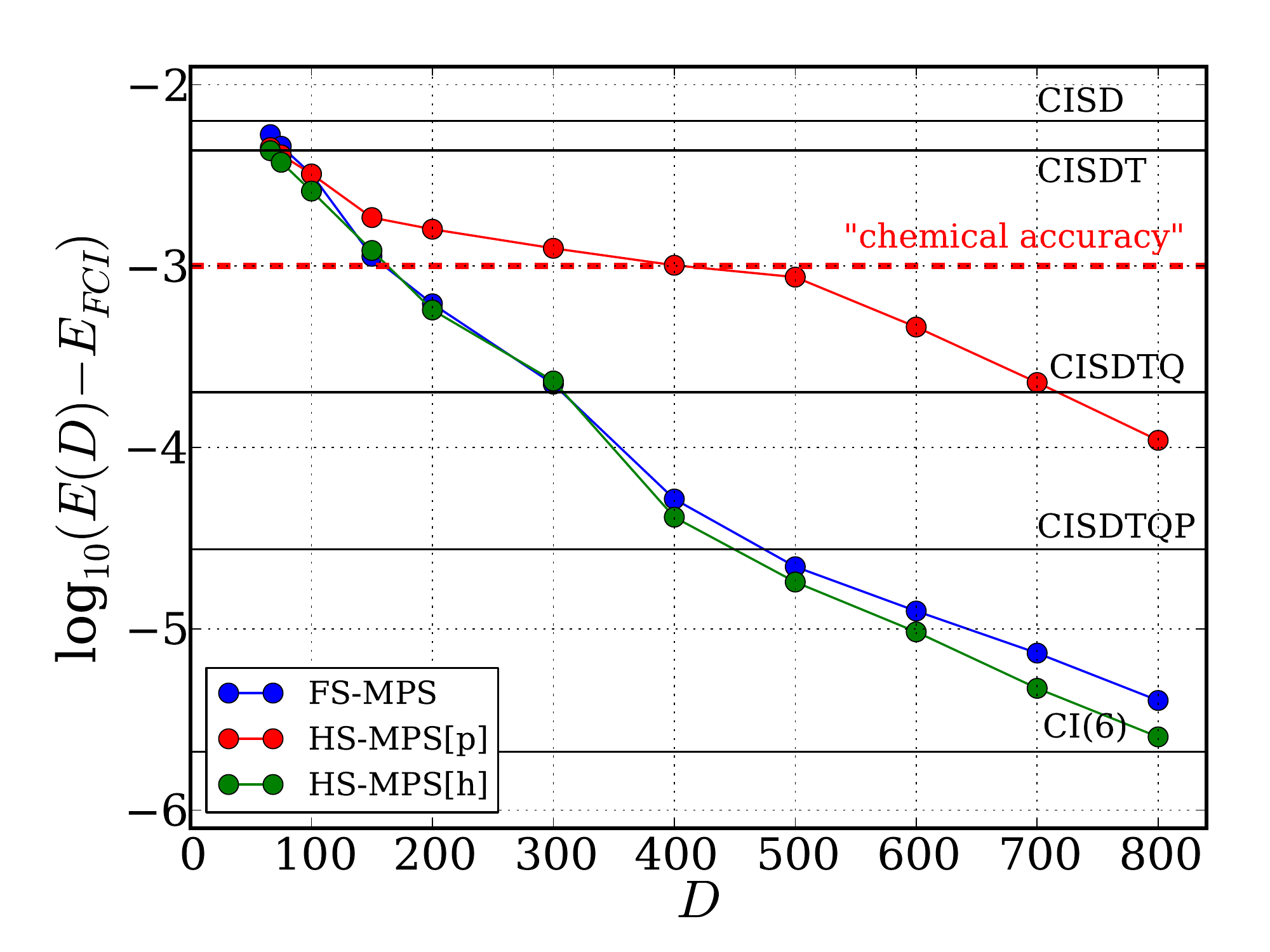}}}\\
(b) \\
\end{tabular}
\caption{Errors of variational energies as a function of $D$
for Ne in the DZP basis.}\label{fig:ne6e}
\end{figure}

For the case away from half filling, we examine the neon atom with Dunning's
DZP basis\cite{DunningDZ,DunningChapter}, which corresponds to the case $(K,N)=(30,10)$ shown
in Figure \ref{fig:bd2}. In this case, the energies computed
by FS-MPS, HS-MPS[p], and HS-MPS[h] at a given bond dimension
are generally different as shown in Figure \ref{fig:ne6e}.
In particular, HS-MPS[h] converges slightly faster than FS-MPS, while HS-MPS[p] converges more slowly than the other two curves.
Such behaviors can be anticipated from Figure \ref{fig:bd2}(d), where it is shown
that for a given CI level the maximal bond dimension required by HS-MPS[h]
is indeed the smallest. More interestingly, according to Figure \ref{fig:ne6e}(a),
the energies obtained by HS-MPS[p] are lower than those obtained by
FS-MPS for $D$ smaller than 100. This also correlates well
with the inset in Figure \eqref{fig:bd2}(d), which shows
there is a crossing point around 100 between the red dashed line
for HS-MPS[p] and the black solid line for FS-MPS. The comparison
with various forms of CI in Figure \ref{fig:ne6e} reveals the important point that to achieve
the same accuracy as a given CI model, the bond dimensions
required by FS-MPS and HS-MPS with the full set of physical indices are
significantly smaller the maximal bond dimension shown in Figure \ref{fig:bd2}
for the given maximal CI level. For instance, to achieve
CISDTQ accuracy, the bond dimensions required by FS-MPS, HS-MPS[p],
and HS-MPS[h] are about 300, 700, and 300, respectively.
All of these are much smaller than the maximal bond dimensions shown in Figure \ref{fig:bd2},
which are 1044, 2295, and 890 for FS-MPS, HS-MPS[p], and HS-MPS[h], respectively.
The fundamental reason is that the configurations
included in a truncated CI model based on excitation rank do not
contribute equally to the correlation energies. Further,
some of the excluded higher excitations in a truncated CI model may be
more important. This is particularly true in
the strong correlation regime, as will be shown
in the next example, where it can be seen that
the FS/HS-MPS sample the full set of important configurations
very efficiently.

The water molecule H$_2$O with geometry ($R_{OH}$=1{\AA}) taken from
Ref. \cite{bauschlicher1986benchmark} is employed as an example to test the convergence of HS-MPS[p]
for various truncated CI models. Table \ref{tab:h2oCI} shows
the results computed with Dunning's DZ basis\cite{DunningDZ}, which corresponds to
$(K,N)=(28,10)$. The active spaces for CASCI and MRCISD were 6-electron-in-8-spatial-orbital CAS(6e,8o) with Hartree-Fock orbitals.
It is seen that both CISDTQ and MRCISD reach 1$mE_h$ accuracy
in this case, although they cover only a small portion of the FCI space.
The convergence of HS-MPS for these two models is similar to the convergence for FCI, and the 1$mE_h$ accuracy is reached at around $D=300$
in all cases (CISDTQ, MRCISD, and FCI). It is also interesting
to compare the results of HS-MPS ($D=45$) with the CISD results.
The CISD wave function can be exactly represented by an HS-MPS
with $D=45$ and a restricted set of physical indices, see the first column of
Table \ref{tab:h2oCI}. Then, it is apparently surprising to see that
with the same bond dimension $D=45$, an increase of error (21.2$mE_h$) is observed when the physical indices are not
restricted in the HS-MPS for FCI, as compared with the
CISD error (8.3$mE_h$). This, however, is because if there are truncations
  in the DMRG, which is the case for HS-MPS when using the one-site algorithm
  (although not the case for FS-MPS) then the DMRG algorithm does not strictly correspond to an
  energy minimization. In particular, the density matrix criterion for
  choosing the left and right renormalized space is no longer optimal with
  respect to the energy metric.
However, when the bond dimension becomes larger, the error will
be reduced again, e.g., the error becomes 4.2$mE_h$ when $D$ goes to
100, which is better than the CISD energy.

It is also interesting to examine a case of  static correlation, that
arises when the water molecule is symmetrically
stretched to $R_{OH}$=3{\AA}. To avoid convergence to the wrong spin state,
we used a 'biased' Hamiltonian with $\lambda=1$, viz.,
\begin{eqnarray}
H(\lambda)=H+V(\lambda),\quad V(\lambda)=\lambda S_- S_+,
\end{eqnarray}
to shift the energies of states with spin $S>0$ upwards. Alternatively,
we also performed state average calculations targeting two
states at a time without adding the bias, in the case of FCI.
This approach leads to slightly faster convergence as shown in Table \ref{tab:h2oCI}.
It is notable that while  CISDTQ itself has a large error, of about 12$mE_h$,
the HS-MPS for the various CI models display similar convergence patterns
as for the case of $R_{OH}$=1{\AA}, and thus the error of HS-MPS is
quite 'parallel' to the parent model.
The data further shows that the HS-MPS for FCI captures
many important higher excitations at a bond dimension
of around 400, due to the unrestricted physical dimension,
as compared with HS-MPS for CISDTQ with the same bond dimension
where the physical dimensions are restricted.

\begin{table}
\caption{Convergence of HS-MPS for various CI models (energy in $mE_h$) for H$_2$O with the DZ basis.}\scriptsize
\begin{tabular}{ccccccc}
\hline\hline
$D$	&	CISD	&	CISDT	&	CISDTQ	&	CASCI	&	MRCISD	&	FCI			\\
\hline
 & \multicolumn{6}{c}{$R_{OH}$=1{\AA}, $E_{FCI}=-76.156699 E_h$}\\
error$^a$	&	8.3 	&	7.1 	&	0.3 	&	90.2 	&	0.8 	&	0.0 			\\
\hline
45	&	0.0 	&	15.8 	&	21.0 	&	0.4 	&	30.0 	&	21.2 			\\
100	&	0.0 	&	0.5 	&	4.0 	&	0.0 	&	7.6 	&	4.2 			\\
150	&	0.0 	&	0.0 	&	2.3 	&	0.0 	&	1.8 	&	2.4 			\\
200	&	0.0 	&	0.0 	&	1.4 	&	0.0 	&	1.2 	&	1.5 			\\
300	&	0.0 	&	0.0 	&	1.0 	&	0.0 	&	0.6 	&	1.1 			\\
400	&	0.0 	&	0.0 	&	0.4 	&	0.0 	&	0.4 	&	0.5 			\\
500	&	0.0 	&	0.0 	&	0.1 	&	0.0 	&	0.2 	&	0.3 			\\		
 & \multicolumn{6}{c}{$R_{OH}$=3{\AA}, $E_{FCI}=-75.866193 E_h$}\\
error$^a$	&	131.9 	&	100.5 	&	12.3 	&	63.0 	&	0.5 	&	0.0 			\\
\hline
45	&	0.0	&	12.8 &	34.6 	&	8.4 	&	57.7 	&	42.8, 46.9$^b$ 	\\
100	&	0.0	&	1.6	&	11.8 	&	0.3 	&	38.9 	&	17.6, 20.7$^b$ 	\\
150	&	0.0	&	0.0	&	2.7 	&	0.0 	&	15.3 	&	5.8, 5.9$^b$ 	\\
200	&	0.0	&	0.0	&	2.0 	&	0.0 	&	5.6 	&	3.9, 2.8$^b$ 	\\
300	&	0.0	&	0.0	&	1.0 	&	0.0 	&	1.5 	&	1.9, 1.6$^b$ 	\\
400	&	0.0	&	0.0	&	0.8 	&	0.0 	&	1.1 	&	1.2, 0.8$^b$ 	\\
500	&	0.0	&	0.0	&	0.2 	&	0.0 	&	1.0 	&	1.1, 0.5$^b$ 	\\
\hline\hline
\multicolumn{7}{l}{$^a$ Errors of truncated CI relative to FCI energies.}\\
\multicolumn{7}{l}{$^b$ State-average DMRG results.}
\end{tabular}
\label{tab:h2oCI}
\end{table}

\subsection{State-average calculations of excited states}\label{sec:ex}
While it is seen for the $K>N$ case that the FS-MPS generally provides
better energies as compared with HS-MPS[p],
we can expect HS-MPS[p] to be more efficient for describing a
manifold of low-lying states.
There are several reasons to support this.
First, one  major difference between HS-MPS[p] and FS-MPS is that
HS-MPS[p] has a larger physical dimension $K-N+1$ compared to 2 in the
FS-MPS. This results in a much larger CI subproblem during the
DMRG site optimization, such that the excited states can be expected to
be better described in the local variational space of a given site.
Second, considering the extreme case where all
low-lying states arise from  excitations from the highest
occupied molecular orbital (HOMO) to the virtual orbitals,
to represent all such states in an HS-MPS[p] only the last
site needs to be changed. As in our setting the last two
sites of HS-MPS[p] are treated exactly without any truncation,
this implies that all these excited states can be described without
increasing the bond dimension because they can share
the same $(N-1)$-electron states. In the occupation
number representation used in the FS-MPS, to represent all these
excitations the bond dimension will at least increase by one over
that for a single state.

To test this expectation, state-averaged
calculations were performed for the lowest six states of H$_2$O
with the same basis set and geometries ($R_{OH}$=1{\AA})
as employed in the last section. Only the $S_z=0$ sector is targeted and states with different spatial
symmetries are all state-averaged in the construction
of the pseudo-density matrix for renormalizations.
Table \ref{tab:h2oEx} lists the computed energies with FS/HS-MPS, and for comparison the
results from single-state optimizations are also shown.
The most striking feature is that when going from single-state to multi-state
optimizations, the mean absolute errors (MAE) of the FS-MPS with relatively small bond dimensions
increase significantly by a factor about 6-7. In comparison,
the MAE of HS-MPS do not increase dramatically. Consequently,
increasing the number of states makes the MAE of the HS-MPS become
smaller than those of the FS-MPS with bond dimensions 45, 100, and
200. When the bond dimension is sufficiently large (around 400),
then again the MAE of the FS-MPS become smaller than those of the HS-MPS
as in the single-state calculations. The comparison with the CISD
results is also interesting. The CISD results indicate a large
bias towards the ground state as the orbitals are optimized
for the ground state. However, as shown in Table \ref{tab:h2oEx} HS-MPS with $D=45$
already  rectifies this bias by adjusting the structure
of the renormalized states. Consequently,
the MAE of HS-MPS with $D=45$ becomes much smaller than that of CISD
in the multi-state calculations.

\begin{table}
\caption{Errors with respect to FCI energies (in $mE_h$) for the lowest six states of H$_2$O in the DZ basis.}\scriptsize
\begin{tabular}{cccccccccccc}
\hline\hline
&  &  &\multicolumn{4}{c}{FS-MPS ($D$)}& &\multicolumn{4}{c}{HS-MPS ($D$)}\\
 \cline{4-7}\cline{9-12}
State	&	FCI/$E_h$	    & CISD &	45	&	100	&	200	&	400	&	&	45	&	100	&	200	&  400	\\
\hline
$1^1A_1$	&	-76.156699 	& 8.3  &	40.3 	&	20.0 	&	4.9 	&	1.5 	&	&	29.9 	&	20.6 	&	3.2 	&	1.3 	\\
$1^3B_2$	&	-75.876664 	& 85.9 &	49.3 	&	24.6 	&	5.2 	&	1.8 	&	&	46.9 	&	22.3 	&	3.9 	&	2.7 	\\
$1^1B_2$	&	-75.847464 	& 85.5 &	48.0 	&	24.5 	&	5.1 	&	1.7 	&	&	45.3 	&	22.3 	&	4.1 	&	2.6 	\\
$1^3A_1$	&	-75.798565 	& 85.0 &	65.2 	&	25.3 	&	4.9 	&	1.9 	&	&	55.5 	&	23.9 	&	4.8 	&	3.2 	\\
$1^3A_2$	&	-75.793979 	& 86.3 &	62.0 	&	28.0 	&	9.6 	&	2.1 	&	&	54.6 	&	22.6 	&	4.7 	&	2.7 	\\
$1^1A_2$	&	-75.774742 	& 84.9 &	60.4 	&	28.3 	&	10.5 	&	2.1 	&	&	53.6 	&	22.3 	&	4.8 	&	2.6 	\\
MAE	&		&	72.6 &  54.2 	&	25.1 	&	6.7 	&	1.8 	&	&	47.7 	&	22.3 	&	4.2 	&	2.5 	\\
\\
$1^1A_1{}^a$	&	-76.156699 	& 8.3 &	8.0 	&	3.5 	&	1.1 	&	0.1 	&	&	21.2 	&	4.2 	&	1.6 	&	0.5	\\
\hline\hline
\multicolumn{11}{l}{$^a$ Single-state DMRG results.}
\end{tabular}
\label{tab:h2oEx}
\end{table}

The open-shell molecule BeH ($R$=1.3426{\AA}) is used as another
example. The lowest five $\Sigma^+$ states and four $\Pi$ states
are computed with FS-MPS and HS-MPS. In this case, both the spin projection $S_z$ and
point group symmetry (Abelian subgroup $C_{2v}$) are used.
The results computed with a ROHF reference and the aug-cc-pVDZ basis\cite{avdz1,avdz2}
are shown in Table \ref{tab:behEx}. In this case, the theoretical bond dimensions
of HS-MPS are $\{60, 1610, 9920, 1610\}$ for $(K,N)=(64,5)$. However,
it is seen that milli-Hartree accuracy can be achieved with a very small bond dimension such as 100.
The HS-MPS is more accurate than the FS-MPS with the same bond dimensions $D$ of 10, 50, and 100. This remains the case until
a very high accuracy of less than 0.1$mE_h$ is targeted. In the latter regime,
FS-MPS becomes more advantageous as the situation ($K\gg N$) is similar to
that shown in Figure \ref{fig:bd2}.

\begin{table}
\caption{Errors with respect to FCI energies (in $mE_h$) for the lowest five $\Sigma^+$ states
and four $\Pi$ states of BeH in the aug-cc-pVDZ basis.}\scriptsize
\begin{tabular}{cccccccccccc}
\hline\hline
&  &  &\multicolumn{4}{c}{FS-MPS ($D$)}& &\multicolumn{4}{c}{HS-MPS ($D$)}\\
 \cline{4-7}\cline{9-12}
State	&	FCI/$E_h$	& CISD &	10	&	50	&	100	& 200 	& &	10	&	50	&  100	& 200 \\	
\hline																
$1^2\Sigma^+$	&	-15.191558	& 0.8  & 	23.6	&	2.4	&	1.3	& 0.1 &	&	1.1	&	0.6	&	0.2 & 0.1	\\
$2^2\Sigma^+$	&	-14.988419	& 12.8 &	28.6	&	3.5	&	1.8	& 0.1 &	&	6.0	&	1.7	&	0.4 & 0.2	\\
$3^2\Sigma^+$	&	-14.984730	& 27.2 &	35.4	&	5.9	&	3.9	& 0.1 &	&	9.9	&	3.8	&	0.7 & 0.3	\\
$4^2\Sigma^+$	&	-14.962559	& 28.2 &	34.6	&	2.6	&	1.4	& 0.1 &	&	8.6	&	1.8	&	0.2 & 0.1	\\
$5^2\Sigma^+$	&	-14.917953	& 25.5 &	41.8	&	6.2	&	4.9	& 0.1 &	&	24.7	&	4.7	&	0.6 & 0.2	\\
MAE	&		& 18.9 &	32.8	&	4.1	&	2.7	& 0.1 &	&	10.1	&	2.5	&	0.4 & 0.2	\\
\\																
$1^2\Pi$	&	-15.098974	& 26.3 &	37.4	&	5.4	&	1.8	& 0.1 &	&	23.7	&	0.7	&	0.6 & 0.1	\\
$1^4\Pi$	&	-14.980316	& 8.4  &	22.2	&	2.8	&	1.3	& 0.1 &	&	3.2	&	1.2	&	0.8 & 0.2	\\
$2^2\Pi$	&	-14.955116	& 29.8 &	42.0	&	4.1	&	0.8	& 0.1 &	&	26.7	&	0.3	&	0.3 & 0.1	\\
$3^2\Pi$	&	-14.921160	& 19.4 &	26.0	&	4.6	&	1.2	& 0.1 &	&	11.1	&	1.0	&	0.7 & 0.2	\\
MAE	&		& 21.0 &	31.9	&	4.2	&	1.3	& 0.1 &	&	16.2	&	0.8	&	0.6 & 0.2	\\
\hline\hline
\end{tabular}
\label{tab:behEx}
\end{table}

\section{Conclusions}\label{sec:conclusion}
The traditional routes to approach FCI accuracy include
utilizing all possible symmetries (antisymmetry, spin, and point group)
to make the FCI space compact, truncating the FCI space based on excitation rank,
or selecting important configurations based on perturbative estimates.
TNS based methods provide a fundamentally different way to
tackle the many-electron problem. By transforming the CI vector into a tensor,
it allows to explore the low entanglement among different physical dimensions.
Such low entanglement can either come from spatial or energetic locality.
 The redundancy of intermediate states can be reduced effectively through renormalization.
The present paper answered a fundamental question: Can  many-electron wavefunctions defined in a Hilbert space representation be approximated by
products of smaller pieces as the TNS in the Fock-space case? The key for this problem is the introduction
of prefix/suffix renormalization, which is a general  tool to synthesize many-body states from smaller pieces in Hilbert space.
The so-obtained renormalized states have their own orbital supports that can be identified by the prefix or suffix,
such that they can also be viewed as some special renormalized states in Fock space.
Consequently, the problems of dealing with antisymmetry and computing matrix elements
can be solved. The resulting HS-MPS ansatz is highly flexible. By defining
appropriate physical indices (nodes in the configuration graph) and coupling rules
(interconnections among nodes), various configuration spaces can be
represented, and the variational minimization of the energy in the linear
space is converted to a minimization on the HS-MPS manifold \eqref{GSopt}.
Several numerical examples have been studied to reveal the merits of HS-MPS and FS-MPS.
The results are useful for guiding the design of new kinds of wavefunction ansatz.
In particular, the ansatz obtained by limiting the maximal bond dimension but without
restricting the physical indices and coupling rules has more flexibility than
traditional truncated CI models. The important higher excitations can be sampled
efficiently with moderate bond dimensions, in a very distinct way from e.g. coupled-cluster theory\cite{shavitt2009many}.
The usefulness of HS-MPS as a reference, to efficiently cover most of the strong correlations,
and as a starting point for further including the residual dynamic correlation, deserves further exploration.
Hilbert-space formulations of tensor network states with other topologies, especially hierarchical tree (HT) structures\cite{shi2006classical},
may be useful in certain circumstances.
The HT structure
has seen application chemistry in the
multilayer formulation of the multiconfiguration time-dependent Hartree theory
(ML-MCTDH)\cite{MLMCTDH2009} in Fock space for identical particles. It can be
related to tree tensor networks by the removal of the physical indices of all tensors
except at the boundary of the tree\cite{murg2010simulating,nakatani2013efficient,ttns2015}. Its Hilbert-space
version can be formulated with the help of the prefix/suffix renormalization. In particular,
the spin-adapted HS-MPS proposed in Sec. \ref{sec:generalization} can be viewed as a simple realization of such a HT structure.

\section*{ACKNOWLEDGMENTS}
One of the authors (Z.L.) acknowledges helpful discussions with Sheng Guo, Sebastian Wounters, Qiming Sun, and Boxiao Zheng.
This work was primarily supported by the US National Science Foundation, through the award NSF:CHE-1265277.

%

\end{document}